\documentclass[11pt]{article}
\usepackage[utf8]{inputenc}
\usepackage[T1]{fontenc}
\pdfoutput = 1

\usepackage{amsmath}
\usepackage{amssymb}
\usepackage{bbm}
\usepackage{bbold}
\usepackage{braket}
\usepackage{caption}
\usepackage{color}
    \definecolor{darkgreen}{rgb}{0,0.5,0}
    \definecolor{darkred}{rgb}{0.5,0,0}
    \definecolor{darkblue}{rgb}{0,0,0.6}
    \definecolor{purple}{rgb}{0.4,.2,0.7}
\usepackage[mathcal]{eucal}
\usepackage{float}
\usepackage{graphicx}
\usepackage[hyperfootnotes = true, colorlinks = true, linkcolor = darkblue, citecolor = purple]{hyperref}
\usepackage{mathabx}
\usepackage{mathtools}
\usepackage{pdfsync}
\usepackage{slashed}
\usepackage[normalem]{ulem}
\usepackage{upgreek}
\usepackage{url}
\allowdisplaybreaks

\usepackage[margin = 2.2cm]{geometry}
\usepackage[ragged]{footmisc}
\def\({\left(}
\def\){\right)}
\def\[{\left[}
\def\]{\right]}

\newcommand{\qeq}{{\hbox{=\kern-2.3mm ? \kern.5mm }}}
\renewcommand{\qeq}{=}

\newcommand{\ve}{\varepsilon}

\newcommand{\BB}{{\cal B}}

\newcommand{\BBB}{{\cal B}}

\newcommand{\AAA}{{\cal A}}

\newcommand{\MM}{{\cal M}}
\newcommand{\CC}{{\cal C}}

\newcommand{\XX}{{\cal X}}
\newcommand{\YY}{{\cal Y}}

\newcommand{\wt}{\widetilde}

\newcommand{\NN}{{\cal N}}

\def\cl0{\tilde c_0}

\def\one{{\hbox{ 1\kern-.8mm l}}}
\def\zero{{\hbox{ 0\kern-1.5mm 0}}}

\newcommand{\NNP}{C}

\def\be{\begin{equation}}
\def\ee{\end{equation}}

\renewcommand{\d}{\mathrm{d}}
\renewcommand{\i}{\mathrm{i}}
\renewcommand{\tilde}{\widetilde}
\DeclareMathOperator{\Tr}{Tr}
\DeclareMathOperator{\arcsinh}{arcsinh}

\numberwithin{equation}{section}

\newcommand{\ben}{\begin{eqnarray}\displaystyle}
\newcommand{\een}{\end{eqnarray}}
\newcommand{\non}{\nonumber}
\newcommand{\refb}{\eqref}

\newcommand{\veff}{V_\text{eff}}

\begin{document}

\thispagestyle{empty}
\begin{center}
    ~\vspace{5mm}
    
    {\Large \bf 
    
    Multi-instantons in minimal string theory and in matrix integrals
    
    }
    
    \vspace{0.4in}
    
    {\bf 
    Dan Stefan Eniceicu,$^1$ 
    Raghu Mahajan,$^1$
    Chitraang Murdia,$^{2,3}$ and 
    Ashoke Sen$^{4}$
    }

    \vspace{0.4in}

    $^1$ Department of Physics, Stanford University, Stanford, CA 94305, USA \vskip1ex
    $^2$ Berkeley Center for Theoretical Physics, Department of Physics, University of California, Berkeley, CA 94720, USA \vskip1ex
    $^3$ Theoretical Physics Group, Lawrence Berkeley National Laboratory, Berkeley, CA 94720, USA \vskip1ex
    $^4$ International Centre for Theoretical Sciences, 
Bengaluru - 560089, India
    \vspace{0.1in}
    
    {\tt eniceicu@stanford.edu, raghumahajan@stanford.edu, murdia@berkeley.edu,
    ashoke.sen@icts.res.in}
\end{center}

\vspace{0.4in}

\begin{abstract}

We compute the normalization of the general multi-instanton contribution to the partition function of $(p',p)$ minimal string theory and also to the dual two-matrix integral.
We find perfect agreement between the two results.

\end{abstract}

\pagebreak

\tableofcontents

\section{Introduction and summary} \label{sintro}
Minimal string theories are toy models of string theory where the worldsheet CFT consists of a minimal model, the Liouville field and the $bc$ ghosts \cite{DiFrancesco:1993cyw, Seiberg:2004at}.
These theories admit a dual description in terms of double-scaled matrix integrals \cite{DiFrancesco:1993cyw}.
An insight from these toy models that generalizes even to critical superstring theory is the existence of ``D-instanton'' effects, which are nonperturbative effects of order $\exp(-A \,g_s^{-1})$ \cite{Shenker:1990uf, David:1990sk, PolchinskiCombinatorics}.  
Here $g_s$ is the closed string coupling and $A$ is some constant. This differs from the usual $\exp(-A\, g_s^{-2})$ that might be expected from field theory considerations. 
In minimal string theory, these objects were later identified as ZZ branes \cite{zz}.
From the perspective of the dual matrix integral, these nonperturbative effects correspond to one-eigenvalue instantons \cite{Shenker:1990uf}.

Motivated by the study of instanton effects in the $c=1$ string theory \cite{bryzz,Balthazar:2019ypi,SenNormalization,Balthazar:2022apu},  as well as critical (superstring) theories    \cite{Sen:2021tpp,Alexandrov:2021shf,Alexandrov:2021dyl,Alexandrov:2022mmy,Agmon:2022vdj}, in our previous work \cite{Eniceicu:2022nay} we used tools from string field theory to compute the one-loop normalization constant that multiplies these instanton contributions to various physical quantities.
In particular, the genus expansion of the free energy $F$ reads
\begin{align} \label{eseries}
    F = \sum_{g=0}^\infty F_g \, g_s^{-2+2g}  + \mathcal{N} \exp \left( - A \, g_s^{-1} \right) + \ldots
\end{align}
and we are interested in computing the constant $\mathcal{N}$.
We have indicated the perturbative contribution,  along with a correction from one particular instanton up to one-loop order.
The dots indicate other nonperturbative corrections.
Physically, $A g_s^{-1}$ is the tension of the ZZ brane and $\mathcal{N}$ is the exponential of the annulus diagram.
The annulus diagram between two ZZ branes is known \cite{zz, Martinec:2003ka, seibergannulus}, but, unfortunately, is ostensibly ill-defined when the two boundaries of the annulus lie on the same ZZ brane.
The dual matrix integral, on the other hand, predicts a finite value for $\mathcal{N}$ \cite{David:1990sk, David:1992za, Sato:2004tz, Hanada:2004im, Ishibashi:2005dh, Ishibashi:2005zf, Marino:2007te, Marino:2008vx, sss}.
In \cite{Eniceicu:2022nay}, we resolved this tension by noting that the exponentiated ZZ annulus is, in fact, finite. 
This is done by identifying the source of the divergence as a pair of Majorana zero modes, which are zero modes because Siegel gauge breaks down in the presence of D-instantons \cite{SenNormalization}.
In particular, this happens because the usual $U(1)$ gauge symmetry on the worldvolume of D-branes is a rigid $U(1)$ symmetry in the case of D-instantons.
Reverting back to a non gauge-fixed form of the path integral then yields meaningful results which agree with the matrix integral \cite{SenNormalization, Eniceicu:2022nay}.

Since the perturbation series given in \refb{eseries} is not Borel summable,  the significance of non-perturbative instanton corrections is not {\it a priori} clear. 
In \cite{Eniceicu:2022nay} we addressed this issue by noting that while the coefficients $F_g$ are all real, the normalization constant $\NN$ is purely imaginary. 
Therefore instanton corrections give the leading imaginary contribution to the free energy,  and this can be reliably computed using D-instanton physics. 
In this paper we shall take a somewhat different point of view that was already present in the early papers on this subject,  for example in \cite{David:1992za}. 
In the matrix model, once we express the integral over matrices as integration over eigenvalues, the effective potential of the eigenvalues develops various saddle points. 
We can then choose to integrate all the eigenvalues over the Lefschetz thimble (generalization of steepest descent contour for multi-dimensional integrals \cite{Marino:2012zq,Dunne:2015eaa,Aniceto:2018bis}) of the perturbative saddle point,  or we can choose to integrate all but a finite number of eigenvalues along the Lefschetz thimble of the perturbative saddle point and distribute the integration over the rest of the eigenvalues along the Lefschetz thimbles of various non-perturbative saddle points.
This gives a definition of the
general multi-instanton contribution to the matrix model partition function. 
Eventually, the complete non-perturbative result is obtained by expressing the actual
integration contour as sum over the Lefschetz thimbles of different saddle points with
appropriate weights, but this involves a separate analysis. 
From this view point, the analysis of \cite{Eniceicu:2022nay} involved the contribution
from the integration contour where all but one of the eigenvalues are integrated over the
Lefschetz thimble of the perturbative saddle point and one of the eigenvalues is integrated over the Lefschetz thimble of a non-perturbative saddle point. 
One minor difference between these two perspectives is that in the analysis of \cite{Eniceicu:2022nay},  since we focused on the full integration
contour that happened to contain only half of the Lefschetz thimble of the non-perturbative saddle point, the instanton contribution had an extra factor of half that will be absent in the new perspective.

We believe that a similar perspective should also exist in string (field) theory, but since at present we do not have an independent non-perturbative definition of string theory,  we cannot give a fully rigorous description of what the D-instanton contributions represent. 
Nevertheless we can offer the following limited perspective. 
Formally, D-branes with different boundary conditions, as well as the perturbative vacuum, are expected to be different classical solutions in a parent open string field theory that can be the open string field theory on a particular D-brane configuration \cite{Sen:1999xm,Sen:1999nx,Schnabl:2005gv,Erler:2019fye}.\footnote{
In the critical superstring theory, the role of the parent  theory could be played by the open string field theory on a set of unstable space-filling D-branes or brane-anti-brane systems. Various D-branes, including D-instantons,  as well as the perturbative vacuum, can be
regarded as classical solutions in this theory \cite{Sen:1998tt,Witten:1998cd,Berkovits:2000hf}.
} 
We can then regard the contribution from a given set of D-instantons as the result of the path integral in the open (+ closed) string field theory along the Lefschetz thimble of one of these saddle points. 
Irrespective of the details, the conformal field theory approach gives a systematic procedure for computing the contribution due to the D-instantons, and this is what is needed for comparison with the matrix integral results.

In this article, we continue our analysis of such nonperturbative effects in the partition
function of both the matrix integral and the minimal string theory,  and extend our results to include a general configuration of instantons.
By a general configuration, we mean that we can have $\ell_1$ instantons of one type, $\ell_2$ instantons of another type, and so on.
We present both the string theory and the matrix integral computations for completeness,
although various special cases of the matrix integral computation have been discussed
earlier.
For a single instanton in the one-matrix integral, the matrix integral computation has been considered by many authors \cite{David:1990sk, David:1992za, Sato:2004tz, Hanada:2004im, Ishibashi:2005dh, Marino:2007te, sss}.
For a single instanton in the two-matrix model, see \cite{Ishibashi:2005zf, Kazakov:2004du}.
For $\ell$ identical instantons in the one-matrix model, see \cite{Marino:2008vx, Schiappa:2013opa}.
We review and extend these results to a general configuration of instantons in the two-matrix integral. 

We find perfect agreement between the string theory result and the matrix model result.  The structure of the results, equation (\ref{emultistring}) in the string theory case and equation (\ref{eq:mainresult2}) in the matrix integral case, are identical.  Further, the quantities appearing in these formulas also match precisely.

To prevent the reader from getting lost in the technical details, 
we end this introduction by highlighting the key qualitative ideas in our computation. 
\begin{enumerate}
\item On the string theory side,  the cylinder between identical ZZ branes requires a string field theory analysis to get a finite meaningful answer \cite{SenNormalization,  Eniceicu:2022nay}.   Instead of using Siegel gauge, one needs to do the path integral over the fields with ghost number one, and explicitly divide by the volume of the gauge group.  The rigid gauge group on the worldvolume of $\ell$ ZZ branes is $U(\ell)$ and we need to carefully compute the proper volume of this group \cite{Sen:2021jbr}.
\item In the presence of non-identical instantons,  the result depends on the cylinder connecting two different ZZ branes.  This cylinder is finite \cite{Martinec:2003ka,  seibergannulus} and we do not need to resort to any string field theory analysis.  For the sake of completeness, we re-derive these results from a different perspective.
\item On the matrix integral side,  when we place $\ell$ eigenvalues at an extremum of the one-eigenvalue effective potential,  the result does not vanish despite the presence of the Vandermonde determinant. This is because we need to do an $\ell \times \ell$ Gaussian matrix integral exactly.
\item In the matrix computation of the normalization constant for the two-matrix integral, we need to take into account a slight subtlety that the expansion of $\log Z$ contains terms that are proportional to $N \log N$ \cite{Ishibashi:2005zf}.
\item We compute the matrix integral results before taking the double-scaling limit,  for which we need to take into account some $1/N$ corrections in the one-eigenvalue effective potential and the perturbative free energy \cite{Marino:2007te}.
\end{enumerate}

The rest of the paper is organized as follows.
Section \ref{sec:string} discusses the string theory computation of the multi-instanton
contribution to the partition function in $(p',p)$ minimal string theory.
Section \ref{sec:onematrix} discusses the computation in one-matrix integrals that are dual to $(2,p)$ minimal string theory.  
Section \ref{stwomatrix} discusses the same computation in two-matrix integrals that are dual
to general $(p',p)$ minimal string theory. 
The appendices contain various technical details of the computations as well as some background material.

\section{String theory computations}
\label{sec:string}

The worldsheet CFT for the minimal string theory of interest consists of the $(p',p)$ minimal model serving the role of a matter sector, the Liouville field and the $bc$-ghosts \cite{Seiberg:2004at}.
Here $p'$ and $p$ are two relatively prime integers with $p' < p$.
The central charges of the minimal model and Liouville sectors are
\begin{align}
    c_\text{matter} = 1 - \frac{6(p-p')^2}{pp'}\, , \quad 
    c_\text{Liouville} = 1 + \frac{6(p+p')^2}{pp'}\, ,
\end{align}
and the $b$ parameter of Liouville theory is given by $b = \sqrt{p'/p}$.

Minimal string theory contains ZZ-branes \cite{zz} which are akin to D-instantons.
These branes have the ZZ boundary conditions \cite{zz} for the Liouville sector and Cardy boundary conditions \cite{Cardy89} for the minimal model.
They give rise to nonperturbative effects proportional to $\exp(- A g_s^{-1})$, realizing the fact that closed string worldsheets can develop boundaries nonperturbatively \cite{PolchinskiCombinatorics}.
There are some equivalences between these boundary conditions, and an independent set of boundary conditions can be obtained by restricting to the $(m,n)$ ZZ boundary condition for the Liouville field \cite{zz} and the most basic Cardy state for the minimal model (the one that only contains the identity character in the open string channel) \cite{Kazakov:2004du,Seiberg:2003nm}. 
The integers $m,n$ are restricted to the range $1\le m\le p'-1$, $1\le n\le p-1$ with
further identification under $(m,n) \to (p'-m, p-n)$.

In general, minimal string theory is dual to the double-scaling limit 
of a two-matrix integral \cite{Douglas:1990pt, Daul:1993bg}.
When $p'=2$, one of the matrix potentials is Gaussian and the matrix integral can be reduced to a one-matrix integral.

\subsection{Cylinder with identical boundary conditions}
We begin by quoting our result from \cite{Eniceicu:2022nay} about the exponential of the ZZ annulus with $(1,1)$ boundary conditions on both ends.
This requires un-gauge fixing back out from Siegel gauge and using a form of the string field theory path integral with an explicit division by the volume of the rigid gauge group on the worldvolume of the ZZ instanton. The result takes the form:
\begin{equation} \label{e22.2}
    \mathcal{N}_{1,1}
    = \left(T_{1,1}\right)^{-\frac{1}{2}} \frac{\i}{\sqrt{8\pi}} \left( \frac{ \cot^2(\pi/p) - \cot^2(\pi/p') }{p^2-p'^2} \right)^{\frac{1}{2}}.
\end{equation}
As remarked in section \ref{sintro}, compared to \cite{Eniceicu:2022nay}, we have integrated over the full steepest descent contour, rather than only half of it.

The Liouville contribution to the annulus diagram between two identical ZZ branes,  both labelled by $(m,n)$, is given by \cite{zz}
\begin{align}
    Z_{m,n}^{\mathrm{Liouville}}(t) &= 
    \sum_{k=1}^{m} \sum_{l=1}^{n} \chi_{2k-1,2l-1}(t) \, , \quad \text{with }
    \label{zz_annulus_diagonal_original} \\
    \chi_{k,l}(t) &= \eta(\i t)^{-1}\, ( q^{-kl} -1 ) \, q^{-(kp-lp')^2/4pp'}, \qquad q := e^{-2\pi t}\, .
\end{align}
The contribution from the minimal model Cardy state is \cite{Cardy89,DiFrancesco:1997nk}
\begin{align}
Z^\text{matter}_{1,1}(t)  &= \eta(i t)^{-1} \sum_{j = - \infty}^\infty \left( 
q^{\frac{(2pp'j + p - p' )^2}{4pp'}}
-
q^{\frac{(2pp'j + p + p' )^2}{4pp'}} 
\right)\, , 
\end{align}
and the contribution from the ghosts is $\eta(\i t)^2$ \cite{Polchinski:1998rq}.
The net result is a contribution $\sum_{k=1}^m\sum_{\ell=1}^n
F_{2k-1,2l-1}(t)$ with
\begin{align}
    F_{k,l}(t) := (q^{-kl} - 1) \, q^{-(kp-lp')^2/4pp'}  \sum_{j=-\infty}^{\infty}\left[ q^{(2pp' j + p - p')^2/4pp'} - q^{(2pp' j + p + p')^2/4pp'} \right].
\end{align}
The exponentiated annulus thus becomes a product
\begin{equation}
    \mathcal{N}_{m,n} = \exp \left[ \sum_{k=1}^{m} \sum_{l=1}^{n} \int_0^{\infty} \frac{\d t}{2t} F_{2k-1,2l-1}(t) \right] = 
    \prod_{k=1}^{m} \prod_{l=1}^{n} \, \mathcal{M}_{2k-1,2l-1} 
    \label{n_using_m}
\end{equation}
where we have defined
\begin{equation}
    \mathcal{M}_{k,l} := 
    \exp \left[ \int_0^{\infty} \frac{\d t}{2t} F_{k,l}(t) \right].
\end{equation}
Note that $F_{1,1}$ has a small $q$ expansion that reads $q^{-1} - 2 +O(q)$, which causes $\mathcal{M}_{1,1}$ to be ill-defined.
There are two issues: The tachyon gives the $q^{-1}$ and the two fermionic zero modes give the $-2$.
These need to be dealt with using insights from string field theory, as was explained in detail for the case of minimal string theory in \cite{Eniceicu:2022nay}.
For $(k,l)\neq (1,1)$, we show in appendix \ref{app:mkl} that  we get the following simple formula for $\mathcal{M}_{2k-1,2l-1}$
\begin{align}
    \mathcal{M}_{2k-1,2l-1}
    = \left[ \frac{ \left( \sin^2(\frac{\pi l}{p}) - \sin^2(\frac{\pi k}{p'}) \right) \left( \sin^2(\frac{\pi (l-1)}{p}) - \sin^2( \frac{\pi (k-1)}{p'} ) \right) }{ \left( \sin^2(\frac{\pi (l-1)}{p}) - \sin^2(\frac{\pi k}{p'}) \right) \left( \sin^2(\frac{\pi l}{p}) - \sin^2( \frac{\pi (k-1)}{p'} ) \right)} \right]^{\frac{1}{2}}.
    \label{mklanswer}
\end{align}
With this formula for $\mathcal{M}_{k,l}$ in hand, we note that the product in equation (\ref{n_using_m}) telescopes, and we can reduce the computation of $\mathcal{N}_{m,n}$ to that of $\mathcal{M}_{1,1}$  (see appendix \ref{app:mkl} for details)
\begin{align}
    \mathcal{N}_{m,n} = \mathcal{M}_{1,1} 
    \left[ 
    \frac{ \cot^2(\frac{\pi m}{p'}) - \cot^2(\frac{\pi n}{p}) }
    { \cot^2(\frac{\pi}{p'}) - \cot^2(\frac{\pi}{p}) } 
    \right]^\frac{1}{2}\, .
    \label{full_telescope}
\end{align}

Now, we need to treat $\mathcal{M}_{1,1}$ using string field theory.
Each brane,  labeled by the integers $(m,n)$, has a worldvolume theory. 
There is an action for this theory, which is the standard cubic action of open string field theory \cite{Witten:1985cc}.
The overall coefficient in front of this action is 
$\left(g_o^{(m,n)}\right)^{-2}$, and the tension of the brane is related to this coupling via \cite{Sen:1999xm}
\begin{align} \label{etension}
    T_{m,n} = \frac{1}{2\pi^2} \left(g_o^{(m,n)}\right)^{-2}\, .
\end{align}
Minimal string theories are known to have
branes with negative tension. We expect that the relation \refb{etension} will continue to hold in that case as well if we demand that the universal `tachyon condensation' on those
branes continue to give the perturbative vacuum as usual. Note that for negative tension
branes the universal tachyon has positive mass-squared and `tachyon condensation' actually
raises the tension instead of lowering it.

We will not repeat the steps of the string field theory analysis, but the main insight is that the two zero modes arise due to the failure of Siegel gauge, which, in turn, is related to the fact that the usual gauge symmetry on a D-brane worldvolume is a rigid symmetry in the case of D-instantons.
So we need to work with the string field theory path integral over fields with ghost number one, and explicitly divide by the volume of the rigid symmetry group.
The proper volume of that group equals $2\pi/g_o^{(m,n)}=2\sqrt 2\pi^2 T_{m,n}^{1/2}$ \cite{SenNormalization}.
Following these steps gives us the result identical to \refb{e22.2}, except for the
replacement of $T_{1,1}$ by $T_{m,n}$:
\be
\MM_{1,1}= \left(T_{m,n}\right)^{-\frac{1}{2}} \, \frac{\i}{\sqrt{8\pi}} 
    \left( \frac{ \cot^2(\pi /p) - \cot^2(\pi /p') }{p^2-p'^2} \right)^{\frac{1}{2}}.
\ee
Equation \refb{full_telescope} now gives
\begin{equation}
    \mathcal{N}_{m,n}
    = \left(T_{m,n}\right)^{-\frac{1}{2}} \, \frac{\i}{\sqrt{8\pi}} 
    \left( \frac{ \cot^2(\pi n/p) - \cot^2(\pi m/p') }{p^2-p'^2} \right)^{\frac{1}{2}}.
    \label{nt_string}
\end{equation}

\subsection{Cylinder with non-identical boundary conditions}

Let us extend the result to the annulus diagram with the one boundary lying on an $(m,n)$ ZZ brane and the other on an $(m',n')$ ZZ brane.
In some sense, this computation is easier since the worldsheet computation yields a finite answer, and there is no need to resort to string field theory.
This computation has already been done \cite{Martinec:2003ka, seibergannulus}, but we reproduce the result here using a different method.

The Liouville contribution is \cite{zz}
\begin{equation}
    Z_{(m,n)(m',n')}^{\mathrm{Liouville}}(t) = \sum_{k=|m-m'|+1,2}^{m+m'-1}\; \sum_{l=|n-n'|+1,2}^{n+n'-1} \chi_{k,l} (t) \, ,
\end{equation}
where the $2$ in the subscript of the summation sign indicates that we sum over every other value of the indices $k$ and $l$.
The normalization constant is given by
\begin{equation}
\begin{split}
    \NNP_{(m,n), (m',n')}
    &= \exp \left[ \int_{0}^{\infty} \frac{\d t}{t}\, Z_{(m,n)(m',n')}^{\mathrm{Liouville}}(t) \,
    Z_{(1,1)}^{\mathrm{Matter}}(t) \, 
    \eta(\i t)^2 \right]\, .
\end{split}
\end{equation}
Note that the measure for the $t$-integral is now $\frac{\d t}{t}$ as opposed to $\frac{\d t}{2t}$ because the two sides of the annulus are distinct.
If $(m,n) \neq (m',n')$, this can be written as a product of the contributions $\mathcal{M}_{k,l}$ given in \refb{eq:M_kl} which telescopes
\begin{align} \label{eann_diff}
    \NNP_{(m,n), (m',n')}
    &= \prod_{k=|m-m'|+1,2}^{m+m'-1} \; 
    \prod_{l=|n-n'|+1,2}^{n+n'-1} \left( \mathcal{M}_{k,l}\right)^2 \nonumber \\
    &=  \prod_{k=|m-m'|+1,2}^{m+m'-1} \; 
    \prod_{l=|n-n'|+1,2}^{n+n'-1} 
    \frac{ \left(\sin^2(  \frac{\pi(l+1)}{2p}) - \sin^2( \frac{\pi(k+1)}{2p'}) \right) \left(\sin^2(  \frac{\pi(l-1)}{2p}) - \sin^2( \frac{\pi(k-1)}{2p'}) \right) }{\left(\sin^2(  \frac{\pi(l-1)}{2p}) - \sin^2( \frac{\pi(k+1)}{2p'}) \right) \left(\sin^2(  \frac{\pi(l+1)}{2p}) - \sin^2( \frac{\pi(k-1)}{2p'}) \right) } \nonumber\\ 
    &=   \frac{ \left(\sin^2(  \frac{\pi(n+n')}{2p}) - \sin^2( \frac{\pi(m+m')}{2p'}) \right) \left(\sin^2(  \frac{\pi|n-n'|}{2p}) - \sin^2( \frac{\pi|m-m'|}{2p'}) \right) }{\left(\sin^2(  \frac{\pi|n-n'|}{2p}) - \sin^2( \frac{\pi(m+m')}{2p'}) \right) \left(\sin^2(  \frac{\pi(n+n')}{2p}) - \sin^2( \frac{\pi|m-m'|}{2p'}) \right) }.
\end{align}
This agrees with the result of \cite{seibergannulus} for the annulus between two different ZZ branes after using some trigonometric identities;  see section \ref{sec:ds2matrix} for some more details.
An important point is that the annulus between two different ZZ branes is finite and completely well-defined.
As we will see, this is the case for matrix integrals as well.

\subsection{General multi-instanton contribution to the partition function}

We shall now determine the ratio of the contribution to the partition function from $\ell$ ZZ branes of type $(m,n)$ to the perturbative contribution to the
partition function  in the $(p',p)$ minimal string theory.
The $\ell=1$ case has already been discussed in \cite{Eniceicu:2022nay}, and our goal will be to express the result for general $\ell$ in terms of quantities that already appear in the result for $\ell=1$.
For this let us express the result for $\ell=1$ as
\begin{align} \label{e22.17}
\frac{Z^{(1)}}{Z^{(0)}} = 
\exp\left[ - T_{m,n} \right] \, \wt B_{m,n}\, \,  \frac{g_o^{(m,n)}}{2\pi} \, ,
\end{align}
where, from \refb{etension} and \refb{nt_string}, we have
\begin{align}
\wt B_{m,n} = {2\pi \over g_o^{(m,n)}} \NN_{m,n}= \i\, \pi^{3/2}\,
 \left( \frac{ \cot^2(\pi n/p) - \cot^2(\pi m/p') }{p^2-p'^2} \right)^{\frac{1}{2}}. 
\end{align}
An important point to recall is that in \refb{e22.17} the
$g_o^{(m,n)}/(2\pi)$ term comes from division by the volume of the $U(1)$ gauge group
and the $\wt B_{m,n}$ factor comes from integration over the tachyon, all the massive modes
and the out-of-Siegel-gauge mode.

For $\ell$ instantons, the action is $\ell \, T_{m,n}$. 
Furthermore, the open string spectrum gets repeated $\ell^2$ times. So the contribution from massive states, tachyon as well as the out-of-Siegel-gauge mode gets repeated $\ell^2$ times. This produces a net factor:
\begin{align}
\label{e1}
\exp\left[-\ell\, T_{m,n} \right] \, \left(\wt B_{m,n} \right)^{\ell^2} \, .
\end{align}

The slightly non-trivial part of the calculation is division by the volume of the gauge group. For this we follow the logic of \cite{Sen:2021jbr}.  
We denote by $\theta^a$ the string field theory gauge  transformation parameters and by $\tilde\theta^a$ the $U(\ell)$ gauge transformation parameters on the D-instanton worldvolume. 
Then we have the relation \cite{Sen:2021jbr}
\be\label{ethtth}
\theta^a = \tilde \theta^a/g_o^{(m,n)}\, .
\ee
Now,  using \refb{ethtth} we see that division by the gauge group volume generates a factor of
\be\label{e3}
(g_o^{(m,n)})^{\ell^2} / V_{U(\ell)} \, ,
\ee
where $V_{U(\ell)}$ denotes the volume of the group $U(\ell)$ as measured by the parameters $\tilde \theta^a$. 
The volume $V_{U(\ell)}$ in this normalization was found in \cite{Sen:2021jbr}.  
The result is\footnote{The relation between $\theta^a$
and $\tilde \theta^a$ in \cite{Sen:2021jbr}  had an extra factor of 2 compared to
\refb{ethtth}, but this can be traced to an extra
factor of 2 in the definition of the SFT gauge transformation parameters and in fact cancels
against a factor of 2 coming from the out-of-Siegel-gauge mode integral. 
So as far as the volume of $U(\ell)$ is concerned, there is no difference and we can directly
take the result of \cite{Sen:2021jbr}.}
\begin{align}
V_{U(\ell)} = \frac{(2\pi)^{\frac{1}{2}(\ell^2 + \ell)}}{G_2(\ell+1)}\, ,\label{volul}
\end{align}
where $G_2(\ell+1) = \prod_{i=1}^{\ell-1} i!$ is the Barnes-G  double gamma function.
In particular,  in the normalization convention for $\tilde\theta^a$ in which \refb{ethtth} is valid,
the volume of the $U(1)^\ell$ diagonal subgroup of $U(\ell)$ is $(2\pi)^\ell$.  
In appendix \ref{appgaussian} we have computed the volume of
$U(\ell)$  using the same normalization and reproduced the result in \refb{volul}.

Multiplying \refb{e1} by \refb{e3} and using \refb{volul}, we get the net normalization factor:
\begin{align}
\frac{Z^{(\ell)}}{Z^{(0)}}  &=
\exp\left[-\ell\, T_{m,n} \right] \, \left( {B}_{m,n} \right)^{\ell^2} \,  
\frac{G_2(\ell+1)}{(2\pi)^{\frac{1}{2}(\ell^2+\ell)}}\,  , \label{ezlzzratio} \\ 
{B}_{m,n}  &:= \wt B_{m,n} \, g_o^{(m,n)} = 2\pi\, \NN_{m,n} = \left(T_{m,n}\right)^{-\frac{1}{2}} \, \i\, \sqrt{\pi\over 2}
    \left( \frac{ \cot^2(\pi n/p) - \cot^2(\pi m/p') }{p^2-p'^2} \right)^{\frac{1}{2}}\, .
    \label{edefbalphamn}
\end{align}
We shall see in sections \ref{sec:onematrix} and \ref{stwomatrix} that this result agrees with the matrix model result for $\ell$ identical instantons. 
For $\ell=1$ and $(m,n)=(1,1)$ this agreement was observed in \cite{Eniceicu:2022nay}.

One can now easily generalize the result to the case where we have
$\ell_\alpha$ instantons of type $\alpha$, where $\alpha$ takes values over
different pairs $(m,n)$. 
The result is:
\begin{align} \label{emultistring}
\frac{Z^{\{\ell_\alpha\}}}{Z^{(0)}} = 
\prod_\alpha \left\{\exp\left[-\ell_\alpha T_\alpha\right] \, \({B}_\alpha\)^{\ell_\alpha^2} 
\frac{G_2(\ell_\alpha+1)}{(2\pi)^{\frac{1}{2} (\ell_\alpha^2 +\ell_\alpha)}} \right\}
\prod_{\alpha,\beta\atop\alpha<\beta} (\NNP_{\alpha,\beta})^{\ell_\alpha\ell_\beta}\, ,
\end{align}
where $B_\alpha$ is as in (\ref{edefbalphamn}) and $\NNP_{\alpha,\beta}$, given in \refb{eann_diff}, accounts for the contribution from the exponential of the annulus amplitude with
one boundary on the instanton of type 
$\alpha$ and the other boundary on the instanton of type 
$\beta$.
The exponent $\ell_\alpha\ell_\beta$ represents the trace over the Chan-Paton factors on the two boundaries.

\section{One-matrix integrals}
\label{sec:onematrix}
The one-matrix integral (in the double-scaling limit) is dual to the $(2,p)$ family of minimal string theories, where $p \geq 3$ is an odd integer. 
The computation of the normalization constant for the case of $\ell$ identical instantons in the one-matrix integral was done in \cite{Marino:2008vx} using a degeneration limit of the two-cut solution to the matrix integral.
We will present a simpler approach that is similar to \cite{Ishibashi:2005dh, Marino:2007te}.
Some useful results about the perturbative structure of one-matrix integrals are collected in appendix \ref{app:onematrix}. 
We follow the conventions of \cite{marino2005chern, Marino:2007te}.

The quantity of interest is the integral over $N\times  N$ Hermitian matrices $M$:
\begin{align}
\label{e2.1}
Z(N,t) &:= {1\over V_{U(N)}} \int \d M \exp\[- \frac{N}{t} \Tr V(M)\] \\
&= {1\over N!}\int \prod_{i=1}^N {\d x_i\over 2\pi} \prod_{i,j=1 \atop i<j}^{N} 
(x_i-x_j)^2 
\exp\[-{N \over t}\sum_{i=1}^N V(x_i)\] \, . \label{zdef_eig}
\end{align}
As usual, the large-$N$ limit is taken keeping the 't Hooft coupling $t$ fixed.\footnote{
Hopefully,  there is no confusion between using the same letter $t$ for the 't Hooft coupling in the matrix integral and the open string modulus in string theory.
}
The measure $\d M$ is defined as the volume measure that is induced by the metric $\d s^2 = \Tr (\d M^2)$ on the space of Hermitian matrices.
We diagonalize $M = U \text{diag}(x_i) U^\dagger$ and change variables to the eigenvalues $x_i$ and $U$.
Here $V_{U(N)}$ denotes the volume of the $U(N)$ group, with a local measure that is induced on the space of $U$'s by the above change of variables.
The factor $(N!\, (2\pi)^N)^{-1}$ on the right hand side denotes the volume of $U(1)^N \times S_N$ which corresponds to rotating the phase of each column of $U$ and the permutations of the eigenvalues. 
This subgroup is left ``unfixed'' when we change variables from $M$ to the $x_i$'s and $U$. 
Also note that even though we have chosen a convenient normalization of $Z(N,t)$
in \refb{e2.1}, this choice will not affect our final result since we shall be
computing the ratio of two different contributions to $Z(N,t)$.

For simplicity, we take the potential to be an even polynomial, with degree $p+1$.
\begin{align} \label{e33.3}
    V(x) = \frac{x^2}{2} + \sum_{k=1}^{(p-1)/2} \frac{g_{2k+2}}{2k+2} \, x^{2k+2}\, .
\end{align}
We focus on the case where the perturbative contribution to the free energy comes from the so-called one-cut saddle point. 
The one-cut saddle point is the one where the eigenvalues of the matrix are distributed on a single interval $[-b,b]$.
The cut end-point $b$ depends on $t$ and on all the coefficients $g_{2k+2}$ appearing in the potential.

It is expected that $\log Z(N,t)$ has an asymptotic expansion in powers of $N^{-2}$ \cite{Brezin:1977sv, Bessis:1980ss, DiFrancesco:1993cyw}:
\begin{align}
    \log Z(N,t) \stackrel{?}{=} \sum_{g=0}^\infty 
    \left(\frac{N}{t}\right)^{2-2g} F_g(t)\, .
    \label{znseries}
\end{align}
This is almost correct. 
However, a more precise statement is that, in order to get this asymptotic series,  one needs to divide by the Gaussian matrix integral (the matrix integral with all the coefficients $g_{2k+2}$ set to zero),
see the rigorous mathematical treatment in \cite{ercolani2003asymptotics}. 
\begin{align}
    \log \frac{Z(N,t)}{Z_G(N,t)} = \sum_{g=0}^\infty \left(\frac{N}{t}\right)^{2-2g} (F_g(t)- F_{g,G}(t))\, . \label{e33.5}
\end{align}
The subtlety has to do with the fact that the $\log Z_G$ contains terms proportional to $\log N$ (see equation (\ref{gaussian_one_asymp})), and every matrix integral contains these same terms. 
This has been discussed in more detail in appendix \ref{appgaussian}, but an intuitive reason for this is that if we compute the integral via perturbation theory in the couplings $g_{2k+2}$, we get the Gaussian matrix integral as an overall factor \cite{Brezin:1977sv}.

We want to go beyond the perturbative expansion (\ref{znseries}) and include effects from one-eigenvalue instantons \cite{Shenker:1990uf,David:1990sk}.
A single eigenvalue in the matrix integral (\ref{zdef_eig}) at position $x_i$ feels an effective potential
\begin{align}
    V(x_i) - \frac{2t}{N}  \sum_{j:j\neq i} \log \vert
    x_i - x_j
    \vert \, .
\end{align}
In the large-$N$ limit, it is useful to introduce a \emph{holomorphic} effective potential defined as \cite{Marino:2007te}
\begin{align} \label{edefveff}
    \veff(x,t) := V(x) - 2 t \int_{-b}^b \d y \, \rho(y) \log (y-x)\, ,
\end{align}
where $\rho(y)$ is the eigenvalue density normalized according to $\int_{-b}^b \d y \, \rho(y) = 1$.
The actual potential felt by an eigenvalue is the \emph{real part} of the holomorphic effective potential.
For a review of some basic properties of the effective potential and more features of the large-$N$ one-cut saddle point,  please see appendix \ref{app:onematrix}.
Here,  we just mention the following important relations \cite{marino2005chern},
reviewed in \refb{ecc8}, \refb{ecc9},
\begin{align}
\veff'(x) &= M(x) \sqrt{x^2 - b^2} \, ,   \label{e33.11}\\
\rho(x) &= \frac{1}{2\pi t} M(x) \sqrt{b^2 - x^2} \ \Theta(b -|x|)\, , \label{e33.12}
\end{align}
where $\Theta(x)$ is the Heaviside step function and
$M(x)$ is a polynomial determined from the potential by the requirement that the resolvent  $\omega_0(x) = \frac{1}{2t}(V'(x) - M(x)\sqrt{x^2-b^2})$ behaves as $1/x$ as $x \to \infty$ on the physical $x$-sheet.

The one-eigenvalue instantons correspond to the extrema of $\veff(x)$ with $x$ being outside the interval $[-b, b]$.
In other words,  the one-eigenvalue instantons are the zeroes of $M(x)$.
Once we have identified these extrema $\{x_\alpha\}$,  we can define a general $\ell$-instanton partition
function $Z^{\{\ell_\alpha\}}(N,t)$ as follows. 
Let $\{\ell_\alpha\}$ be a finite sequence of non-negative integers (all being $O(N^0)$, not all zero) satisfying $\sum_\alpha \ell_\alpha=\ell$. 
Then $Z^{\{\ell_\alpha\}}(N,t)$ is defined as the contribution to the integral \refb{zdef_eig} where a number $\ell_\alpha$ of eigenvalues are integrated along the steepest descent contour of the particular extremum $x_{\alpha}$,  and $(N-\ell)$ of the eigenvalues are integrated along the interval $[-b,b]$ corresponding to the perturbative regime.
The quantity $Z^{(0)}$ denotes the contribution to the matrix integral where the integration range of each eigenvalue extends over the perturbatively allowed region $[-b,b]$. 
Once we have chosen a particular integration contour for the $x_i$'s  in \refb{zdef_eig}, we can express this contour as a weighted sum of the steepest descent contours. 
Accordingly,  the partition function is given by a weighted sum of the $Z^{\{\ell_\alpha\}}$'s.

Unlike \cite{SenNormalization, Eniceicu:2022nay,sss}, in this paper we will not be careful about the defining contour for the eigenvalues for the full partition function, and what linear combination of the steepest descent contours is homologous to the defining contour.
In particular, the question of the existence of the double-scaling limit, at finite values of the double-scaled coupling constant, is beyond the scope of this work.
Hence, we will simply compute $Z^{\{\ell_\alpha\}}$ by integrating the eigenvalues along the \emph{full} steepest descent contours.

\subsection{Identical instantons}
\label{sec:onematrix_identical}
We will consider the most general instanton configuration in section \ref{sec:onematrix_non_identical}, but for now we consider $\ell$ identical instantons.
To be more precise, we integrate $\ell$ eigenvalues along the steepest descent contour corresponding to the extremum $x^\star$ of the one-eigenvalue effective potential, and we want to compute the ratio $\frac{Z^{(\ell)}(N,t)}{Z^{(0)}(N,t)}$.

The definition of $\veff(x,t)$ given in \refb{edefveff} was in the strict large-$N$ limit. 
Since we want to compute the answer including the one-loop correction, we need to carefully
keep track of $1/N$ corrections. 
For this we need to take into account the fact that the original cut now only contains $N - \ell$ eigenvalues.
Since the overall coefficient in front of the potential remains $N/t$, if we want to interpret the second term in \refb{edefveff} as an expectation value in the matrix model,  this integral over $N - \ell$ eigenvalues must be evaluated at a shifted value of the 't Hooft coupling, namely $t' = t - t\ell/N$. 
This is so that $N/t$ can be rewritten as $(N-\ell)/t'$.

An important remark is that we need to properly treat the Vandermonde repulsion between the $\ell$ eigenvalues. If $\ell>1$ and we naively substitute $x^\star$ for each of the $\ell$ eigenvalues, the result will vanish.
This means that there is an $\ell \times \ell$ Gaussian matrix integral that we need to compute exactly.

Separating out $\ell$ eigenvalues to be placed near $x^\star$ in the integral (\ref{zdef_eig}), we get the expression
\begin{align}
Z^{(\ell)}(N,t) &= {1\over N!} {N\choose \ell} \, 
 \int_{\CC_0} \prod_{i=1}^{N-\ell} {dx_i\over 2\pi}  
 \prod_{i<j\atop i,j=1}^{N-\ell} (x_i-x_j)^2 
 \exp\[-{N \over t }\sum_{i=1}^{N-\ell} V(x_i)\] \non\\*
& \times\
\int_{\CC_1} \prod_{i=N-\ell+1}^{N} {dx_i\over 2\pi}  
\prod_{i<j\atop i,j=N-\ell+1}^{N}
(x_i-x_j)^2  \exp\bigg[-{N\over t }\sum_{i=N-\ell+1}^{N} V_{\rm eff}(x_i,t-t\ell/N ) \nonumber \\*
 & \hspace{0.6in}+  2\sum_{i=N-\ell+1}^{N} \sum_{j=N-\ell+1}^{N} A_{0,2} (x_i,x_j,t-t\ell/N ) +\cdots
 \bigg]\, ,  \quad \text{where} \label{zell_first_form}\\
&V_{\rm eff}(x,t-t\ell /N )) = V(x) - 
\frac{2t}{N} \, \left\langle \sum_{j=1}^{N-\ell} \log (x_j-x) \right\rangle \, , \quad\text{ and } \\
&A_{0,2} (x,x',t-t\ell/N )  = \sum_{i=1}^{N-\ell} \sum_{j=1}^{N-\ell}
 \left\langle \log (x_i-x) \log (x_j-x')\right\rangle_c\, .  \label{e33.15}
\end{align}
We used the fact that $\langle e^X \rangle = e^{\langle X \rangle + \frac{1}{2} \langle X^2 \rangle_c + \ldots}$.
Here $\CC_0$ is the perturbatively allowed range for the eigenvalues and $\CC_1$ is the steepest descent contour of the effective potential around the non-perturbative saddle point. 
In the large-$N$ limit with $t$ fixed, the quantities $V_{\rm eff}$ and $A_{0,2}$ are of order unity with corrections of order $1/N^2$.
The quantity $A_{0,2}$ denotes the \emph{connected} two-point function of the Vandermonde potential exerted by the $N-\ell$ eigenvalues. 
It contributes at the same order as the one-loop Gaussian integral around the non-perturbative saddle point.

Let $x^\star$ denote the location of an extremum of $V_{\rm eff}(x,t)$ and let us denote the derivative with respect to $x$ with a prime.\footnote{
The quantity $x^\star$ depends on $t$ and so the shift in the argument $t\to t - \ell t/N$ causes a shift of order $1/N$ in the value of $x^\star$. However the effect of this is suppressed by inverse powers of $1/N$ compared to the terms we keep. 
Hence we shall ignore this effect.
}
We now approximate the term $\veff(x_i, t - \ell t/N)$ appearing in (\ref{zell_first_form}) as follows
\begin{align}
    \veff(x_i, t - \ell t/N) \approx
    \veff(x^\star, t) + \frac{1}{2} \veff''(x^\star,t) \left( 
    x_i - x^\star \right)^2 - \frac{\ell t}{N} \, 
    \partial_t \veff(x^\star, t) \, .
    \label{veff_shifted}
\end{align}
Since we are interested in the answer up to one-loop order, we need to keep the last term \cite{Marino:2007te}.
We also replaced $x_i$ with $x^\star$ in this last term, since we will be evaluating the $x_i$ integral by the steepest descent method.
As far as the term $A_{0,2} (x, x', t-t\ell/N)$ is concerned, 
we can replace it with $A_{0,2} (x^\star, x^\star, t)$.

Using these ingredients, we get
\begin{align}
\label{e2.6}
Z^{(\ell)}(N,t) &= 
Z^{(0)}(N-\ell, t - t \ell/N) \,
\exp \left[ 
- \frac{N}{t} \ell \, \veff(x^\star,t) + \ell^2 \partial_t \veff(x^\star,t)
+ 2 \ell^2 A_{0,2}(x^\star, x^\star, t)
\right]  \notag \\*
&\times \ {1\over \ell!} \int_{\CC_1} \prod_{i=N-\ell+1}^{N} {dx_i\over 2\pi}
\prod_{i<j\atop i,j=N-\ell+1}^{N} (x_i-x_j)^2
\exp\[- \frac{N}{t} V''_{\rm eff}(x^\star,t) 
\sum_{i=N-\ell+1}^{N} \frac{1}{2} (x_i-x^\star)^2
\].
\end{align}
We now recognize the second line as a Gaussian matrix integral over $\ell \times \ell$ matrices. 
The exact result for this is given in appendix \ref{appgaussian}, equation (\ref{gaussianonematrix}).
Writing $Z^{(0)}(N-\ell, t-t\ell/N) \approx \exp \left( \frac{N^2}{t^2} F_0(t-t\ell/N) \right)$ and Taylor expanding $F_0(t-t\ell/N)$ to second order we get\footnote{As discussed below
\refb{e33.5}, $\log Z^{(0)}(N,t)$ has logarithmic terms that invalidate the expansion in
power series in $N^{-2}$. However these logarithmic terms are the same as
those that appear in the gaussian matrix integral $Z_G(N,t)$. We show in 
\refb{zn_by_znminus1_onematrix} that
the effect of the logarithmic terms drops out in the ratio
$\frac{Z_{G}(N-\ell, t-t\ell/N)}{Z_{G}(N,t)}$, and hence it also drops out in the
ratio $\frac{Z^{(0)}(N-\ell, t-t\ell/N)}{Z^{(0)}(N,t)}$.
\label{fo3}}
\begin{align}
\label{eznlone2}
\frac{Z^{(\ell)}(N,t)}{Z^{(0)}(N,t)} &= 
\exp\left[ - \frac{N}{t} \ell \, \mathcal{A}\right] \, 
\mathcal{B}^{\ell^2} \,
\frac{G_2(\ell+1)}{(2\pi)^{\frac{1}{2}(\ell^2 + \ell)}} \, ,
\end{align}
where we have defined
\begin{align}
\mathcal{A} &:= \veff(x^\star,t)+\partial_t F_0(t)\, ,  \label{defa}\\
\mathcal{B} &:= \exp \left[ \frac{1}{2}\partial_t^2 F_0(t)  
+ \partial_t \veff(x^\star,t) + 2 A_{0,2} (x^\star, x^\star,t) \right] 
\left( \frac{2\pi t}{N V''_{\rm eff}(x^\star)} \right)^\frac{1}{2}\, ,
\label{defb}
\end{align}
and $G_2(\ell+1) = \prod_{i=1}^{\ell} i^{\ell-i}$ is the Barnes-G  double gamma function.
The quantity $N\mathcal{A}/t$ is interpreted as the tension of a single instanton.
The exponential expression appearing in the quantity $\mathcal{B}$ turns out to have a simple formula in terms of the perturbative one-cut saddle point, see (\ref{largeN-B-onematrix}).
As described in \refb{F0prime_one-matrix},  the quantity $\partial_t F_0(t)$ is equal to minus the real part of the effective potential on the cut.
This implies that
\begin{align}
    \mathcal{A} = \int_{-b}^{x^\star} \d x \, \veff'(x)\, .
\end{align}
We shall see that the quantities $N \mathcal{A}/t$ and $\mathcal{B}$ stay finite in the double-scaling limit.

\subsection{General multi-instanton configuration}
\label{sec:onematrix_non_identical}
Suppose we integrate $\ell_1$ of the eigenvalues along the steepest
descent contour of the extremum $x^\star_{1}$ (of the one-eigenvalue effective potential), 
$\ell_2$ of the eigenvalues along the steepest descent contour of the extremum $x^\star_{2}$, and so on.
Let $\ell = \sum_\alpha \ell_\alpha$ denote the total number of eigenvalues that have been pulled out of the original cut, which now contains $N - \ell$ eigenvalues.
First of all, we get a factor like
the one on the right hand side of \refb{eznlone2} for each $\alpha$. 
Besides this, we get four new types of contributions involving each pair $(\alpha,\beta)$ for $\alpha\ne \beta$:
\begin{enumerate}
    \item The term $\ell^2 \, \partial_t^2 F_0$ now contains terms proportional to $\ell_\alpha \ell_\beta\, \partial_t^2 F_0$.
    \item There will be a contribution proportional to $\ell_\alpha \ell_\beta \, A_{0,2}(x^\star_\alpha,x^\star_\beta,t)$ with $A_{0,2}$ given by the connected correlator \refb{e33.15}
of the Vandermonde potential exerted by the $N - \ell$ eigenvalues in the cut $[-b,b]$. 
    \item For each $x^\star_\alpha$, the last term of (\ref{veff_shifted}) is still proportional to $\ell$.  Therefore, in \refb{e2.6} it generates a term proportional to 
    $ \ell\, \ell_\alpha  \, \partial_t \veff(x^\star_\alpha,t)$. After writing
    $\ell=\sum_\beta\ell_\beta$ this leads to terms of the form $\ell_\alpha \ell_\beta \, \partial_t \veff(x^\star_\alpha,t)$.
    \item There are an $\ell_\alpha\ell_\beta$ number of factors of $(x^\star_\alpha - x^\star_\beta)^2$  in the Vandermonde determinant.
\end{enumerate}
Keeping these four things in mind and repeating the steps in section \ref{sec:onematrix_identical}, we get
\begin{align}
    \frac{Z^{\{\ell_\alpha\}}(N,t)}{Z^{(0)}(N,t)} &= 
    \exp\left[ - \frac{N}{t} \sum_\alpha \ell_\alpha \mathcal{A}_\alpha \right] \, 
    \prod_\alpha \left\{\left(\mathcal{B}_\alpha\right)^{\, \ell_\alpha^2} \,
    \frac{G_2(\ell_\alpha+1)}{(2\pi)^{\frac{1}{2}(\ell_\alpha^2 + \ell_\alpha)}}\right\}\, 
    \prod_{\alpha < \beta} \left(C_{\alpha, \beta}\right)^{\ell_\alpha \ell_\beta}\, ,
    \label{eq:mainresult1}
\end{align}
where
\begin{align}
    \mathcal{A}_\alpha &:= 
    \veff(x^\star_\alpha,t)+\partial_t F_0(t) =  
    \int_{-b}^{x^\star_\alpha} \d x \, \veff'(x)
    \, , \label{defa2} \\
    \mathcal{B}_\alpha &:= \exp \left[\frac{1}{2} \partial_t^2 F_0(t)  
    + \partial_t \veff(x^\star_\alpha,t) + 2 A_{0,2} (x^\star_{\alpha}, x^\star_\alpha,t) \right] 
    \left(\frac{2\pi t}{N V''_{\rm eff}(x^\star_{\alpha})} \right)^\frac{1}{2}\, , \label{defb2} \\
    \mathcal{C}_{\alpha,\beta} &:= (x^\star_\alpha - x^\star_\beta)^2\, 
    \exp \left[ \partial_t^2 F_0(t) 
    + \partial_t \veff(x^\star_\alpha,t) + \partial_t \veff(x^\star_\beta,t) + 
    4 A_{0,2} (x^\star_\alpha, x^\star_\beta,t) \right] \, . \label{defc2}
\end{align}
The quantities $\mathcal{A}_\alpha$ and $\mathcal{B}_\alpha$ are the same as in (\ref{defa}) and (\ref{defb}) with $x^\star$ replaced by $x^\star_{\alpha}$, but we have reproduced them here for completeness.

The quantities $\mathcal{B}_\alpha$ and $\mathcal{C}_{\alpha,\beta}$ turn out to have a simple formula in terms of the perturbative one-cut saddle point, even outside of the double-scaling limit \cite{Ishibashi:2005dh, Marino:2007te}. 
Using (\ref{largeN-B-onematrix}) we get,
\begin{align}
    \mathcal{B}_\alpha &= 
        \frac{b}{2((x^\star_\alpha)^2-b^2)} 
  \left(  \frac{2\pi t}{N V''_{\rm eff}(x^\star_{\alpha})} \right)^\frac{1}{2}\, , \label{defb3}\\
    \mathcal{C}_{\alpha,\beta} &= (x^\star_\alpha - x^\star_\beta)^2\, 
    \left(
    \frac{b}{x^\star_\alpha x^\star_\beta - b^2 +     \sqrt{(x^\star_\alpha)^2-b^2 }
    \sqrt{(x^\star_\beta)^2-b^2 }}
     \right)^2\, . 
    \label{defc3}
\end{align}

\subsection{The double-scaling limit}
\label{sec:onematrixds}
The double-scaling limit refers to a procedure where we tune the parameters of the potential and simultaneously zoom in near an edge of the eigenvalue spectrum, say the left one,  such that the Feynman diagrams dominating the partition sum resemble continuum surfaces \cite{DiFrancesco:1993cyw}.
We define the energy variable $E$ via $x = - b + \varepsilon E$, and also introduce the variable $z$ via $E = - z^2$. The double-scaling limit is taken by sending
$\ve$ to zero and $N$ to infinity keeping the combination\footnote{In the $z$ coordinate introduced in \refb{ec.25}, the double-scaling limit 
is defined by zooming in near $x=-2\gamma$ or $z=-1$.
Using the relation $x = -2 \gamma + \gamma(z+1)^2 + \ldots$ in the neighborhood of $z=-1$,  and $x = -2\gamma + \varepsilon E$,
we see that in this limit $z \simeq -1 + \sqrt{-\varepsilon E/\gamma}$.
}
\be
e^{S_0} := {N} \ve^{{p\over 2}+1}
\ee
fixed. 
At the same time,  we tune the coefficients of the polynomial potential $V$ in a suitably analytic fashion such that $N \rho(x) \, \d x$ approaches $e^{S_0} \, \d E$ times a finite function of $E$ in this limit.
This function of $E$ is supported on the entire positive real axis.
There is some freedom in this process.
We want to focus on the so-called ``conformal background'' \cite{Moore:1991ir} of minimal string theory,  where only the cosmological constant operator is turned on. 
In this case the density of states $\rho(E)$ takes the form:\footnote{See,  for example,
\cite{sss} for a recent discussion on this and
\cite{msy} for explicit potentials that lead to the above density of states in the double-scaling limit.  Compared to the conventions of \cite{Eniceicu:2022nay},  we have set the constant $\kappa$ appearing there to be $\kappa = 1/2$.  However,  if one is interested in taking the JT gravity limit $p \to \infty$ \cite{sss},  one should scale $\kappa \sim p^2$. }
\begin{align}
\rho(E)= \ve^{{p\over 2}+1} \, \frac{1}{\pi}\, 
\sinh \left( p\, \arcsinh \sqrt{E} \right) + O\left(\varepsilon^{\frac{p}{2}+2}\right)\, .
\label{rhoconformal}
\end{align}
In this limit, $e^{-S_0}$ becomes the genus
expansion parameter and the perturbative contribution to the partition function has an
expansion in even powers of $e^{-S_0}$. 
The precise relation between $e^{-S_0}$ and the
string coupling $g_s$ may be found by comparing the matrix model results with the string
theory results.
From now on, we shall use $E$  instead of $x$ as the argument of $\veff$ and $\rho$, and by an abuse of notation,  we will denote derivatives with respect to $E$ also by a prime.

\begin{figure}
    \centering
    \includegraphics[width=0.6\textwidth]{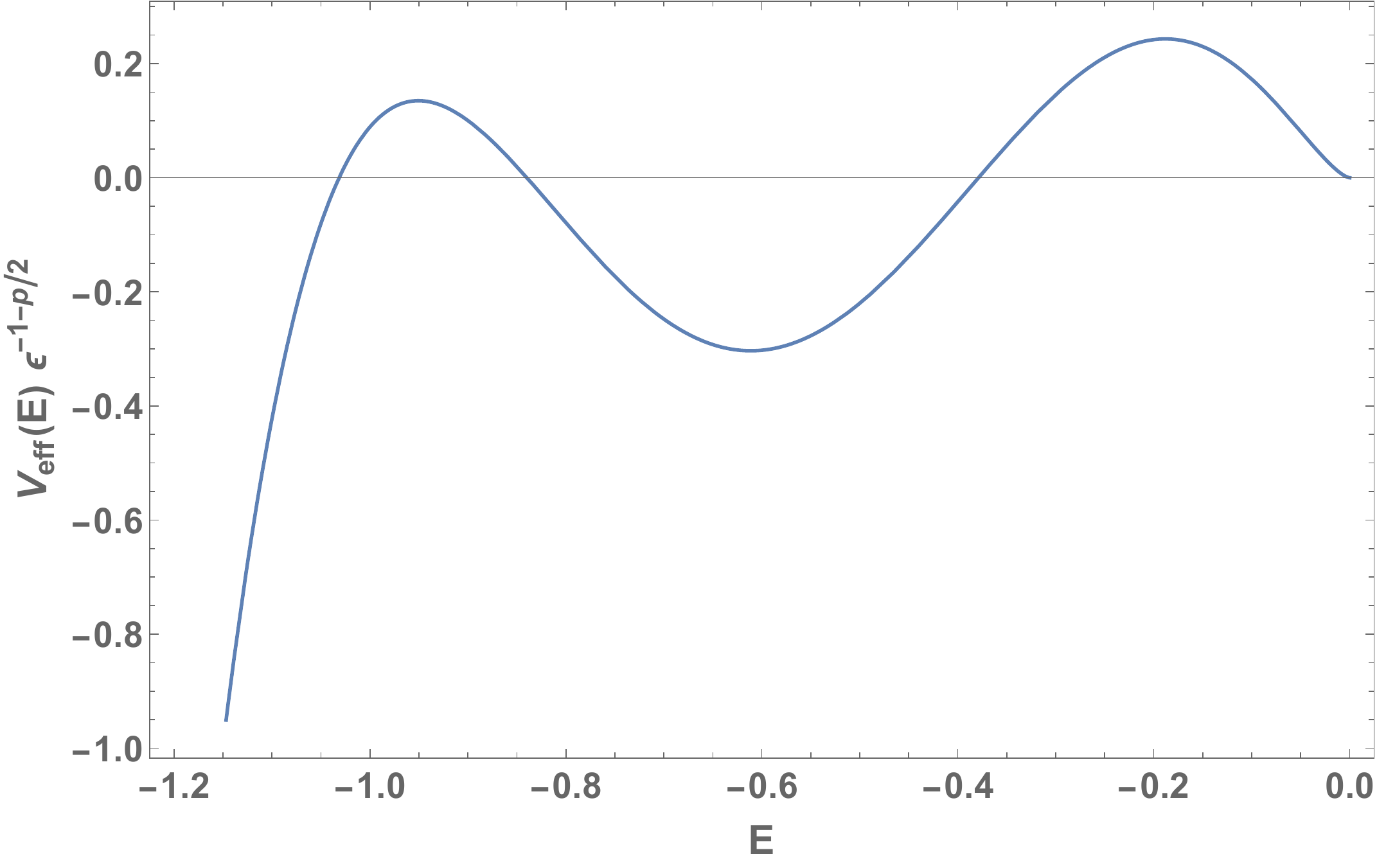}
    \caption{A plot of $\veff(E) \, \varepsilon^{-1-\frac{p}{2}}$ in equation (\ref{veffexplicit}) for the case $p=7$. We have set $t=1$. 
    In general there are $\frac{p-1}{2}$ extrema in the forbidden region, given by (\ref{enstar}). These extrema are in a one-to-one correspondence with the $(1,n)$ ZZ branes in the $(2,p)$ minimal string theory. The extremum closest to $E=0$ is always a maximum and corresponds to the simplest $(1,1)$ ZZ brane.}
    \label{fig:veff_p7}
\end{figure}

From (\ref{e33.11}) and \refb{e33.12}  we see that the analytic continuation
of $2\pi\i t\rho(E)$ from the interval $[-b,b]$ on the real line to the complex plane gives
$\veff'(E)$.
Using the form of $\rho(E)$ given in \refb{rhoconformal} we get the effective potential
\begin{align}
    V_\text{eff}(E) &= - t\, \ve^{{p\over 2}+1}\, 
\left[ {1\over p+2} \sin\left( (p+2) \arcsin\sqrt{{-E}}\right) - {1\over p-2} \sin\left( (p-2) \arcsin\sqrt{{-E}}\right)\right] \label{veffexplicit} \, .
\end{align}
We have chosen the additive constant in the potential such that $\veff$ vanishes at $E=0$.
Note that $V_\text{eff}(E)$ introduced here differs from that of 
\cite{Eniceicu:2022nay} by an overall normalization factor.
See figure \ref{fig:veff_p7} for a plot of $\veff(E)$ for the case $p=7$.

Since the zeros of
$\veff'(E)$ on the negative real axis give the locations of the instanton, we conclude that
the instantons are located at
\begin{align}
    E^\star_n = -   \left(\sin\frac{n\pi}{p}\right)^2 \, ,
    \quad n \in \left\{1, \ldots, \frac{p-1}{2}\right\}\, .
    \label{enstar}
\end{align}
The index $n$ is the same as the one that appears in the label $(1,n)$ for the ZZ branes in the $(2,p)$ minimal string (see more on this below).
From \refb{enstar} and \refb{veffexplicit} we conclude that
\begin{align}
    \veff(E^\star_n) &= (-1)^{n+1} \, t\, \ve^{{p\over 2}+1} \frac{2p\sin(2\pi n/p)}{p^2 -4}  \, , \label{veffestar} \\
    V_\text{eff}''(E^\star_n) &= (-1)^n \, t\, \ve^{{p\over 2}+1} \frac{2p}{\sin (2\pi n /p)}\, .\label{veffpp}
\end{align}

It follows from \refb{eq:mainresult1}, \refb{defa2} and the relation $\veff(0)=0$, that the tension
$T_\alpha$ of the $\alpha$-th ZZ brane is given by,
\be\label{einsttens}
T_\alpha= {N\over t} \AAA_\alpha = {N\over t} \veff(E_\alpha^\star ) = (-1)^{n+1} \,  e^{S_0}\, \frac{2p\sin(2\pi n/p)}{p^2 -4} \quad \hbox{for $\alpha=(1,n)$}\, . 
\ee
Since the right hand side does not involve $N$ or $\ve$, this has finite double-scaling limit.
In order to compare this with the string theory result, we need to find the explicit
relation between $e^{S_0}$ and $g_s$ via perturbative computation. We shall not
attempt to do this here. Instead we shall follow  \cite{Eniceicu:2022nay} and
express our result for the other quantities
in terms of the tension of the
instanton given in \refb{einsttens}.

Next, we need to work out the quantities $\mathcal{B}_{\alpha}$ and $\mathcal{C}_{\alpha,\beta}$ in the double-scaling limit. 
The expression for $\mathcal{B}_\alpha$ before taking this limit is given in (\ref{defb3}).
In the double-scaling limit,  the first factor in (\ref{defb3}) equals $\frac{1}{-4 \varepsilon E}$. 
This $\varepsilon$ in the denominator combines with the $\partial_x^2 \veff$ appearing in (\ref{defb3}) to convert the $\partial_x^2$ into an $\partial_E^2$.
Thus, the combination $\frac{1}{\varepsilon^2} \frac{t}{N \, \partial_x^2 \veff(x^\star)}$ equals $\frac{t}{N \partial_E^2 \veff(E^\star)}$. 
This gives
\be \label{ebalphafin}
 \mathcal{B}_\alpha = \frac{1}{-4 E^\star_\alpha} \, 
    \left(\frac{2\pi t}{N\veff''(E^\star_\alpha)} \right)^\frac{1}{2} 
    =\frac{1}{4\sin^2(n\pi/p)} \,
\({(-1)^{n} \, \pi\, \sin(2\pi n/p)\over e^{S_0} p}\)^{1/2} \quad \hbox{for $\alpha=(1,n)$.}
\ee
There are no subtleties in taking the double-scaling limit for $\mathcal{C}_{\alpha,\beta}$, since
the expression given in \refb{defc3} remains finite in this limit. 
The final result takes the form:
\be\label{ecalphafin}
        \mathcal{C}_{\alpha,\beta} = \left( \frac{\sqrt{-E^\star_\alpha} -\sqrt{-E^\star_\beta}}{\sqrt{-E^\star_\alpha} + \sqrt{-E^\star_\beta}} \right)^2 \quad \Rightarrow \quad
        \CC_{(1,n),(1,n')} = \({\sin{n\pi\over p} - \sin{n'\pi\over p}\over \sin{n\pi\over p} + \sin{n'\pi\over p}}\)^2\, .
\ee

To compare these with the string theory results, we note that \refb{eq:mainresult1} has the same
structure as \refb{emultistring} with $B_\alpha$ replaced by $\BB_\alpha$ 
and $\NNP_{\alpha,\beta}$
replaced by $\CC_{\alpha,\beta}$. Therefore we need to compare $B_\alpha$ with 
$\BB_\alpha$ and $\NNP_{\alpha,\beta}$ with $\CC_{\alpha,\beta}$.
First, we see from \refb{eann_diff} with $p'=2$, $m=m'=1$ that,
\be
\NNP_{(1,n),(1,n')}=\frac{ \cos^2\(  \frac{\pi(n+n')}{2p}\) \ \sin^2\(  \frac{\pi|n-n'|}{2p}\) }
{\cos^2\(  \frac{\pi|n-n'|}{2p}\) \ \sin^2\(  \frac{\pi(n+n')}{2p}\)} =\CC_{(1,n),(1,n')} .
\ee
Next,  using \refb{einsttens}, we can express \refb{ebalphafin} as
\be
\BB_{1,n}=
\left(T_{1,n}\right)^{-\frac{1}{2}} \, \i\, \sqrt{\pi\over 2}
     \frac{ \cot(\pi n/p)  } {\sqrt{p^2-4}} \, .
\ee
On the other hand, from \refb{edefbalphamn} with $p'=2$, $m=1$, we get
\be
B_{1,n}= \left(T_{1,n}\right)^{-\frac{1}{2}} \, \i\, \sqrt{\pi\over 2}
    \left( \frac{ \cot^2(\pi n/p)  } {p^2-4} \right)^{\frac{1}{2}}\, .
\ee
Therefore, we see that there is perfect agreement between the result in $(2,p)$ minimal
string theory and the double-scaled one-matrix model.

We remind the reader that the result presented here for $\mathcal{B} T^\frac{1}{2}$ is twice that quoted in our previous work \cite{Eniceicu:2022nay}, which only dealt with the case $n=1$. 
This is because we are integrating over the full steepest descent contour of the saddle point, and we are not worrying about what linear combination of the steepest descent contours is homologous to the defining contour.

Let us remark that we can perform a consistency check of (\ref{einsttens}) by comparing the dependence on $n$ to that obtained from the string theory computation. 
In string theory, the tension of the ZZ branes can be obtained, up to an overall proportionality constant, using the formulas for the boundary state wavefunction $\Psi_{1,n}(P)$ \cite{zz}.
Indeed, the entire dependence on $n$ is given by a multiplicative factor $\sinh (2\pi n b P)$ (see equation (5.15) of \cite{zz}).
Here $P$ represents the Liouville momentum,  which labels the exponential Liouville operators $e^{2\alpha \phi}$ with $\alpha = (b+b^{-1})/2+\i P$. 
Therefore we set $P$ to $-\i(b-b^{-1})/2$ in order to get the one-point function of the cosmological constant operator $e^{2b \phi}$ \cite{akk, seibergannulus}.
Since $\langle e^{2b \phi} \rangle = \partial_\mu Z_\text{disk}$ and $Z_\text{disk}$ is proportional to the tension of the ZZ brane, we see that $T_{1,n} \, \propto \, \sinh(2\pi n b P)$ with $P = -\i(b-b^{-1})/2$.
Using the fact that $b = \sqrt{2/p}$, we get $T_{1,n} \, \propto \, (-1)^{n-1}\sin (2\pi n/p)$. 
This reproduces the $n$-dependence of the matrix integral formula (\ref{einsttens}).

\section{Two-matrix integrals} \label{stwomatrix}
The general $(p',p)$ minimal string theory is dual to a matrix integral over two matrices in the double-scaling limit \cite{Douglas:1990pt, Daul:1993bg}.
Denoting the matrices by $M_1$ and $M_2$, the action is $\frac{N}{t} \Tr (V_1(M_1) + V_2(M_2) - M_1 M_2)$.
Here $M_1$ and $M_2$ are Hermitian matrices, both of size $N \times N$.
Using the Harishchandra-Itzykson-Zuber formula \cite{Harish-Chandra:1957dhy, Itzykson:1979fi, Zinn-Justin:2002rai}, it is possible to reduce this integral, up to an overall constant, to the following integral over the eigenvalues
\begin{align}
    Z(N,t):= \frac{1}{N!} \int \prod_{i=1}^N \frac{\d x_i \d y_i}{2\pi}
    \Delta(x)\Delta(y)
    \exp \left[ - \frac{N}{t} \sum_{i=1}^N (V_1(x_i) + V_2(y_i) - x_i y_i) \right]\, . 
    \label{zntwodef}
\end{align}
Here, the $x_i$ are the eigenvalues of $M_1$ and $y_i$ are the eigenvalues of $M_2$.
Note that there is only one power of the Vandermonde determinant for each set of eigenvalues.

In appendix \ref{appgaussian} we present the results for this integral when both potentials are Gaussian.
When we vary the potentials to get different integrals, we choose to keep the coefficients of the quadratic terms fixed.
For much the same reasons as discussed in section \ref{sec:onematrix} and appendix \ref{appgaussian},  we obtain a nice asymptotic expansion in powers of $1/N^2$ after dividing by the Gaussian matrix integral
\begin{align} \label{e44.2}
    \log \frac{Z(N,t)}{Z_G(N,t)} = \sum_{g=0}^{\infty} N^{2-2g} (F_g(t) - F_{g,G}(t)) \, .
\end{align}
We are interested in corrections to $\log Z$ that are of order $e^{-N}$.
A single instanton would correspond to placing one pair $(x_i, y_i)$ at an extremum of the effective potential that this pair feels.
Explicitly, from (\ref{zntwodef}),  this effective potential is
\begin{align} \label{e44.3}
    \veff(x_i, y_i) := 
    V_1(x_i) - \frac{t}{N} \sum_{j:j \neq i} \log (x_i - x_j) + 
    V_2(y_i) - \frac{t}{N} \sum_{j:j \neq i} \log (y_i - y_j)
    - x_i y_i\, .
\end{align}

Reference \cite{Ishibashi:2005zf} obtained the normalization constant for a single instanton in the two-matrix integral.
We shall now generalize this to the case of multiple instantons, possibly of different types.
Our results at the intermediate stages will differ from those of \cite{Ishibashi:2005zf} since we express our results in terms of correlation functions in the theory with $N$ eigenvalues while \cite{Ishibashi:2005zf} expresses the results in terms of correlation functions in the theory with $(N-1)$ eigenvalues \cite{Marino:2007te}.

\subsection{Identical instantons}

In this section, we shall analyze the contribution due to $\ell$ instantons  of the same type.
The main idea is similar to that in section \ref{sec:onematrix_identical}.
We pull out $\ell$ pairs $(x_i, y_i)$ and integrate them along the Lefschetz thimble of one particular extremum of $\veff(x_i, y_i)$.
Let us denote this extremum by $(x^\star, y^\star)$.

We define the following quantities
\begin{align}
    V_{1, \text{eff}}(x,t-t\ell/N )) &= 
    V_1(x) - 
    \frac{t}{N} \left\langle \sum_{j=1}^{N-\ell} \log (x-x_j) \right\rangle \, , \label{def:v1eff} \\ 
    V_{2,\text{eff}}(y,t-t\ell/N )) &= V_2(y) - 
    \frac{t}{N} \left\langle \sum_{j=1}^{N-\ell} \log (y-y_j) \right\rangle \, , \label{def:v2eff}
\end{align}    
and also the connected correlators of the Vandermonde potentials
\begin{align}
    A_{0,2}^{(1)} (x,x',t-t\ell/N )  &= 
    \sum_{i=1}^{N-\ell} \sum_{j=1}^{N-\ell}
    \left\langle \log (x-x_i) \log (x'-x_j)\right\rangle_c\,, \label{defa021}\\
    A_{0,2}^{(2)} (y,y',t-t\ell/N)  &= 
    \sum_{i=1}^{N-\ell} \sum_{j=1}^{N-\ell}
    \left\langle \log (y-y_i) \log (y'-y_j)\right\rangle_c\, ,  \label{defa022}\\
    A_{0,2}^{(3)} (x,y,t-t\ell/N)  &= 
    \sum_{i=1}^{N-\ell} \sum_{j=1}^{N-\ell}
    \left\langle \log (x-x_i) \log (y-y_j)\right\rangle_c\, . \label{defa023}
\end{align}
It follows from \refb{e44.3}-\refb{def:v2eff} that $(x^\star, y^\star)$ are determined from the equations
\be\label{exstarystar}
V_{1, \text{eff}}'(x^\star,t)=y^\star, \qquad V_{2, \text{eff}}'(y^\star,t)=x^\star\, ,
\ee
up to corrections that do not affect the result to the order of the $1/N$ expansion that we
are interested in.

Qualitatively, we have the same terms as in the one-matrix case: 
since there are $N-\ell$ pairs of eigenvalues that are integrated on the perturbative contour, and the coefficient $N/t$ does not change, the 't Hooft coupling is shifted to $t-\ell t/N$.
We also get the connected correlators of the Vandermonde potentials.
There is an $\ell\times \ell$ two-matrix Gaussian integral that one needs to do exactly.
So we have
\begin{align}
    Z^{(\ell)}(N,t) &= {1\over N!} {N\choose \ell} \, \int_{\CC_0} \prod_{i=1}^{N-\ell} \frac{\d x_i \d y_i}{2\pi}
    \prod_{i<j\atop i,j=1}^{N-\ell} (x_i-x_j) (y_i-y_j) 
    \exp\[-\frac{N}{t}\sum_{i=1}^{N-\ell} 
    \{V_1(x_i) + V_2(y_i)-x_iy_i\}
    \] \non\\* 
    & \times\
    \int_{\CC_1} \prod_{i=N-\ell+1}^{N} 
    \frac{\d x_i \d y_i}{2\pi}
    \prod_{i<j\atop i,j=N-\ell+1}^{N}
    (x_i-x_j) (y_i-y_j) \non\\* 
    & \exp\bigg[-\frac{N}{t}\sum_{i=N-\ell+1}^{N} 
    \{  V_{1,\text{eff}}(x_i,t-t\ell/N)+
        V_{2,\text{eff}}(y_i,t-t\ell/N) -x_iy_i\}\non\\* &
    \quad + {1\over 2}
    \sum_{i=N-\ell+1}^{N} \sum_{j=N-\ell+1}^{N} 
    \{ A_{0,2}^{(1)}(x_i, x_j, t)+
    A_{0,2}^{(2)}(y_i, y_j, t)+
    2A_{0,2}^{(3)}(x_i, y_j, t) \}
    +\cdots
     \bigg]\, . 
\end{align}
Here $\CC_0$ is the perturbatively allowed range for the eigenvalues 
and $\CC_1$ is the Lefschetz thimble of the effective potential around the
non-perturbative saddle point. 
Just as in the one-matrix case (\ref{veff_shifted}), we need to Taylor expand $V_{1,\text{eff}}(x_i,t-t\ell/N)$ and $V_{2,\text{eff}}(y_i,t-t\ell/N)$
to first order in $\ell/N$, while the shift in $t$ can be ignored in the connected correlators; indeed we have already replaced $t-t\ell/N$ with $t$ for these terms.
Also, we need to Taylor expand $V_{1,\text{eff}}(x_i,t-t\ell/N)$ and $V_{2,\text{eff}}(y_i,t-t\ell/N)$ to second order in $x_i - x^\star$ and $y_i - y^\star$, while we can replace $x_i$ and $y_i$ by their saddle point values in the connected correlators.
The integral over $\mathcal{C}_1$ now forms a two-matrix Gaussian integral that is given in (\ref{twomatrixgaussian}).
Thus, we get
\begin{align}
    Z^{(\ell)}(N,t) &= Z^{(0)}(N-\ell, t-t\ell/N)\, 
    \exp\left[ \frac{\ell^2}{2} \left( 
    A_{0,2}^{(1)}(x^\star, x^\star, t)+
    A_{0,2}^{(2)}(y^\star, y^\star, t)+
    2A_{0,2}^{(3)}(x^\star, y^\star, t) \right) \right] \nonumber \\*
    & \times 
    \exp \left[ \ell^2 \left( \partial_t V_{1,\text{eff}}(x^\star,t) + \partial_t V_{2,\text{eff}}(y^\star,t)  \right)  \right] 
    \times 
    \exp \left[ - \frac{N}{t} \, \ell\, \veff(x^\star, y^\star) \right]
    \nonumber \\*
    &\times G_2(\ell+1) \left( \frac{t}{N} \right)^{\frac{1}{2}(\ell^2 + \ell)}
    \left( V_{1,\text{eff}}''(x^\star,t) V_{2,\text{eff}}''(y^\star,t)  -1\right)^{-\ell^2/2}\, . \label{zell_two_intermediate}
\end{align}
Now we would like to compute the ratio $\frac{Z^{(\ell)}(N,t)}{Z^{(0)}(N,t)}$.
From the above equation we see that this involves the ratio $\frac{Z^{(0)}(N-\ell,t-t\ell/N)}{Z^{(0)}(N,t)}$.
The main novelty as compared to the one-matrix case is that we need to worry about the division by the Gaussian matrix integral. 
As we remarked in \refb{e44.2}, it is $\frac{Z^{(0)}(N,t)}{Z_{G,2}(N,t)}$ that has a nice asymptotic expansion, and from \refb{ebb9} we see that
\begin{align}
 \log \, \frac{Z_{G,2}(N-\ell, t-t\ell/N)}{Z_{G,2}(N,t)}
    = \frac{N^2}{t^2} \left( F_{0,G}(t-t\ell/N) - F_{0,G}(t) \right)
     - \frac{\ell}{2} \log \frac{2\pi t }{N} + O\left(\frac{1}{N} \right)\, .
\end{align}
We want to emphasize the term $- \frac{\ell}{2} \log \frac{2\pi t }{N}$ in this equation, which is novel in the two-matrix case.\footnote{We can see from (\ref{zn_by_znminus1_onematrix}) that the correction is order $1/N$ in the one-matrix case,  and hence not important to the order that we are working at. This was discussed in footnote \ref{fo3}.}
This means that in the large-$N$ limit we should write
\begin{align}
    \frac{Z^{(0)}(N-\ell,t-t\ell/N)}{Z^{(0)}(N,t)}
    = \exp \left[ \frac{N^2}{t^2} \left( - \frac{\ell t}{N} \partial_t F_0(t) 
    + \frac{1}{2} \frac{\ell^2 t^2}{N^2} \partial_t^2 F_0(t)
    \right) \right] \, \left( \frac{N}{2\pi t} \right)^{\frac{\ell}{2}}\, . \label{zshiftedbyz}
\end{align}
The multiplicative power of $N^{\ell/2}$ is important since it combines with $N^{-\frac{1}{2}(\ell^2 + \ell)}$ in (\ref{zell_two_intermediate}) to give
$N^{-\ell^2/2}$.
It is important to get this power; otherwise,  the answer would not agree with the string theory result.

Using (\ref{zell_two_intermediate}) and (\ref{zshiftedbyz}),  we see that the final result is
\begin{align} \label{e44.14}
    \frac{Z^{(\ell)}(N,t)}{Z^{(0)}(N,t)} = 
    \exp \left[ - \frac{N}{t} \, \ell\, \mathcal{A} \right] 
    \mathcal{B}^{\ell^2} \, \frac{G_2(\ell+1)}{(2\pi)^{\frac{1}{2}(\ell^2 + \ell)}}
\end{align}
with the quantities $\mathcal{A}$ and $\mathcal{B}$ defined as
\begin{align}
    \mathcal{A} &:= \veff(x^\star, y^\star) + \partial_t F_0(t)  \, , \quad \text{ and}\\
    \mathcal{B} &:= \left( \frac{2\pi t}{N} \, \frac{1}{V_{1,\text{eff}}''(x^\star,t) V_{2,\text{eff}}''(y^\star,t)  -1}  \right)^{\frac{1}{2}}
    \exp \left[ \frac{1}{2}\partial_t^2 F_0(t) + \partial_t V_{1,\text{eff}}(x^\star,t) + \partial_t V_{2,\text{eff}}(y^\star,t)  \,+  
    \right. \nonumber \\*
    & \hspace{2.2in}+ \left. 
    \frac{1}{2} A_{0,2}^{(1)}(x^\star, x^\star, t)+
    \frac{1}{2} A_{0,2}^{(2)}(y^\star, y^\star, t)+
    A_{0,2}^{(3)}(x^\star, y^\star, t)
    \right]\, .
\end{align}
These quantities are similar to (\ref{defa}) and (\ref{defb}) in the one-matrix case.
They represent the on-shell action and the total one-loop contribution about the instanton configuration. 
Furthermore,  we shall see in section \ref{sec:ds2matrix} that $N \mathcal{A}/t$ and $\mathcal{B}$ are finite in the double-scaling limit, representing the tension of the ZZ brane and the exponential of the annulus between a ZZ brane and itself. 

\subsection{General multi-instanton configuration}
We now follow the logic of section \ref{sec:onematrix_non_identical} and generalize to an arbitrary configuration of instantons.
Let us integrate
a number $\ell_1$ of $(x, y)$ pairs along the Lefschetz thimble of
the saddle point $(x_1^\star, y_1^\star)$,
a number $\ell_2$ of $(x, y)$ pairs along the Lefschetz thimble of
the saddle point $(x_2^\star, y_2^\star)$, and so on.
Let $\ell = \sum_\alpha \ell_\alpha$ be the total number of instantons.
For each $\alpha$ we shall get a factor of the form \refb{e44.14}. 
Besides this, there are four types of contributions that give rise to a multiplicative factor $\mathcal{C}_{\alpha, \beta}^{\ell_\alpha\ell_\beta}$.
They are similar to the ones we enumerated in section \ref{sec:onematrix_non_identical}, except that we have more functions to keep track of. 
Also, the Vandermonde contribution is now a power of 
$(x^\star_\alpha -x^\star_\beta)(y^\star_\alpha -y^\star_\beta)$.

After a straightforward, though perhaps slightly tedious calculation, we arrive at the result
\begin{align}
    \frac{Z^{(\ell_1, \ell_2, \ldots)}(N,t)}{Z^{(0)}(N,t)} &= 
    \exp\left[ - \frac{N}{t} \sum_\alpha \ell_\alpha \mathcal{A}_\alpha \right] \, 
    \prod_\alpha \left\{ \left(\mathcal{B}_\alpha \right)^{\, \ell_\alpha^2} \,
    \frac{G_2(\ell_\alpha+1)}{(2\pi)^{\frac{1}{2}(\ell_\alpha^2 + \ell_\alpha)}}\right\}\, 
    \prod_{\alpha < \beta} \CC_{\alpha, \beta}^{\ell_\alpha \ell_\beta}\, ,
    \label{eq:mainresult2}
\end{align}
with the definitions
\begin{align}
 \mathcal{A}_\alpha &:= \veff(x^\star_\alpha, y^\star_\alpha) + \partial_t F_0(t)     \, , \label{defa5}  \\
    \mathcal{B}_\alpha &:= 
    \left( \frac{2\pi t}{N} \, \frac{1}{V_{1,\text{eff}}''(x^\star_\alpha) V_{2,\text{eff}}''(y^\star_\alpha)  -1}  \right)^\frac{1}{2}
    \exp \left[\frac{1}{2}\partial_t^2 F_0(t) + 
    \partial_t V_{1,\text{eff}}(x^\star_\alpha) + 
    \partial_t V_{2,\text{eff}}(y^\star_\alpha)  \,+  
    \right. \nonumber \\
    & \hspace{2.5in}+ \left. 
    \frac{1}{2} A_{0,2}^{(1)}(x^\star_\alpha, x^\star_\alpha)+
    \frac{1}{2} A_{0,2}^{(2)}(y^\star_\alpha, y^\star_\alpha)+
    A_{0,2}^{(3)}(x^\star_\alpha, y^\star_\alpha)
    \right] \, ,  \label{defb5} \\
    \mathcal{C}_{\alpha,\beta} &:= (x^\star_\alpha-x^\star_\beta)(y^\star_\alpha - y^\star_\beta) \exp \left[ 
    \partial_t^2 F_0(t) + 
    \partial_t V_{1, \text{eff}}(x_\alpha^\star) +
    \partial_t V_{2, \text{eff}}(y_\alpha^\star) +
    \partial_t V_{1, \text{eff}}(x_\beta^\star) +
    \partial_t V_{2, \text{eff}}(y_\beta^\star) \, + \right. \nonumber \\
    &\hspace{1.5in}\left. + \, A_{0,2}^{(1)} (x_\alpha^\star, x_\beta^\star)+
    A_{0,2}^{(2)} (y_\alpha^\star, y_\beta^\star) + 
    A_{0,2}^{(3)} (x_\alpha^\star, y_\beta^\star) + 
    A_{0,2}^{(3)} (x_\beta^\star, y_\alpha^\star) \right]\, . \label{defc5}
\end{align}
The quantities $\mathcal{A}_\alpha$ and $\mathcal{B}_\alpha$ are the same as what we derived in the previous section, but we have included them in this result for the sake of convenience.
Comparing \refb{eq:mainresult2} with \refb{ezlzzratio} we see that the matrix model
results agree with the string theory results provided we identify $T_\alpha$ with
$N\AAA_{\alpha}/t$, $B_\alpha$ with $\BBB_\alpha$ and $C_{\alpha,\beta}$ with
$\CC_{\alpha,\beta}$. We shall verify these in section \ref{sec:ds2matrix}.

\subsection{The double-scaling limit}
\label{sec:ds2matrix}
The details about the saddle point structure of the two-matrix integral and the double-scaling limit are presented in appendix \ref{app:twomatrixsaddle}.
Here, we just present the final results.

The instantons are labeled by two integers $(m,n)$ with $m \in \{1, \ldots,p'-1\}$ and $n \in \{1, \ldots,p-1\}$ subject to the identification $(m,n) \equiv (p'-m,p-n)$ \cite{Seiberg:2003nm}.
One finds that, in the double-scaling limit, the quantities $T_{m,n} = \frac{N}{t}\, \AAA_{m,n}$ and $\mathcal{B}_{m,n}$ remain finite.
Using (\ref{amn}) and (\ref{bmn}),  we see that
\begin{align}
   \BBB_{m,n}
    = \left(T_{m,n}\right)^{-\frac{1}{2}} \, \i\sqrt{\pi\over 2}
    \left( \frac{ \cot^2(\pi n/p) - \cot^2(\pi m/p') }{p^2-p'^2} \right)^{\frac{1}{2}}\, ,
    \label{nt_matrixmaintext}
\end{align}
agreeing precisely with the string theory result (\ref{edefbalphamn}).

Finally,  we check that the string theory expression for $\NNP_{\alpha,\beta}$
agrees with the matrix model result for $\mathcal{C}_{\alpha,\beta}$.
The result for $\mathcal{C}_{\alpha,\beta}$ from (\ref{cmnmpnp}) is
\begin{align}
\CC_{(m,n),(m',n')} =  {\cos \left( \frac{\pi m}{p'} + \frac{\pi n}{p} \right) -  \cos \left( \frac{\pi m'}{p'} + \frac{\pi n'}{p} \right) \over \cos \left( \frac{\pi m}{p'} + \frac{\pi n}{p} \right) -  \cos \left( \frac{\pi m'}{p'} - \frac{\pi n'}{p} \right)} \times
{\cos \left( \frac{\pi m}{p'} - \frac{\pi n}{p} \right) -  \cos \left( \frac{\pi m'}{p'} - \frac{\pi n'}{p} \right) \over \cos \left( \frac{\pi m}{p'} - \frac{\pi n}{p} \right) -  \cos \left( \frac{\pi m'}{p'} + \frac{\pi n'}{p} \right)}\, .
\label{cab2matrixzMainText}
\end{align}
After some simplification using basic trigonometric identities, we see that this agrees with the string theory result (\ref{eann_diff}).

Finally, let us make a remark about the ratio of the tensions of the various ZZ branes.
From (\ref{amn}) we see that the matrix integral predicts that
\begin{align}
\frac{T_{m,n}}{T_{1,1}} = (-1)^{m+n}  \,  \frac{\sin \frac{\pi m p}{p'} \sin \frac{\pi n p'}{p}}{\sin \frac{\pi p}{p'} \sin \frac{\pi p'}{p}}\, .
\end{align}
Similarly to the remarks in section \ref{sec:onematrixds},  this agrees with the results from the string theory side \cite{zz,  seibergannulus, akk}. 
The boundary state wavefunction $\Psi_{m,n}(P)\, \propto \, \sinh (2\pi m P b^{-1}) \sinh (2\pi n P b)$ which for $P = -\i(b-b^{-1})/2$ equals $(-1)^{m+n} \sin \frac{\pi m p}{p'} \sin \frac{\pi n p'}{p}$.

\paragraph{Acknowledgments.} 
C.M.  would like to thank Mykhaylo Usatyuk for helpful conversations.
D.S.E. would like to acknowledge the Shoucheng Zhang Graduate Fellowship for support.
R.M.  is supported in part by AFOSR grant FA9550-16-0092. 
C.M. is supported in part by the U.S. Department of Energy, Office of Science, Office of High Energy Physics under QuantISED Award DE-SC0019380 and contract DE-AC02-05CH11231.
A.S. is supported by ICTS-Infosys Madhava Chair Professorship and the J. C. Bose fellowship of the Department of Science and Technology, India.

\appendix

\section{String theory computation of the exponentiated annulus}
\label{app:mkl}
The goal of this appendix is to derive the formula (\ref{mklanswer}) for $\mathcal{M}_{2k-1,2l-1}$ and also to present some details of the derivation of (\ref{full_telescope}).
Recall the definitions
\begin{align}
    F_{k,l}(t) &:= (q^{-kl} - 1) \, q^{-(kp-lp')^2/4pp'}  \sum_{j=-\infty}^{\infty}\left[ q^{(2pp' j + p - p')^2/4pp'} - q^{(2pp' j + p + p')^2/4pp'} \right], \quad q:= e^{-2\pi t} \, ,  \label{efklexp}\\
    \mathcal{M}_{k,l} &:= 
    \exp \left[ \int_0^{\infty} \frac{\d t}{2t} F_{k,l}(t) \right] \, .
\end{align}
We want to derive a simple explicit formula for $\mathcal{M}_{k,l}$.
The main identity that we will need is
\begin{equation}
\label{eq:int_log}
    \int_{0}^{\infty} \frac{\d t}{2t} \left( e^{-2\pi h_1 t} - e^{-2\pi h_2 t} \right) = 
    \frac{1}{2} \log \frac{h_2}{h_1}.
\end{equation}
This identity is valid for $h_1,h_2>0$. 
If either $h_1$ or $h_2$ is negative,  we can use the right hand side to compute the analytic continuation of the left hand side. 
The use of analytic continuation may be justified by noting that in string field theory the steepest descent contour for a mode with negative $h$ runs along the imaginary axis instead of the real axis.

When $(k,l)\neq(1,1)$, we can use the analytically continued version of \eqref{eq:int_log} to perform the integral over $t$, which gives us\footnote{
As discussed in appendix \ref{sintegral}, for special values of $(p',p)$ we get some vanishing exponents of $q$ even for $(k,\ell)\ne (1,1)$.
For now we shall ignore this problem and simply use the fact that the problematic terms cancel pairwise. 
However we should keep in mind that for these special values of $(p',p)$, our string
theory results remain somewhat formal.}
\begin{align}
    \mathcal{M}_{k,l}
    =& \prod_{j \in \mathbb{Z}} \left[ \left\{\frac{(2pp'j + p + p')^2 - (kp + lp')^2}{(2pp'j + p + p')^2 - (kp - lp')^2}\right\} \left\{ \frac{(2pp'j + p - p')^2 - (kp - lp')^2}{(2pp'j + p - p')^2 - (kp + lp')^2} \right\}\right]^{\frac{1}{2}} \nonumber \\
    =& \prod_{j \in \mathbb{Z}} \left[ \frac{(2pp'j + (k+1)p + (l+1)p') (2pp'j - (k-1)p - (l-1)p') }{(2pp'j + (k+1)p - (l-1)p') (2pp'j - (k-1)p + (l+1)p')} \right.\nonumber \\
    & \quad\quad\times \left. \frac{(2pp'j + (k+1)p - (l+1)p') (2pp'j - (k-1)p + (l-1)p') }{(2pp'j + (k+1)p + (l-1)p') (2pp'j - (k-1)p - (l+1)p')} \right]^{\frac{1}{2}} \nonumber \\
    =& \prod_{j \in \mathbb{Z}} \left[ \frac{(j+\frac{k+1}{2p'} + \frac{l+1}{2p}) (j - \frac{k-1}{2p'} - \frac{l-1}{2p}) (j + \frac{k+1}{2p'} - \frac{l+1}{2p}) (j - \frac{k-1}{2p'} + \frac{l-1}{2p}) }{(j + \frac{k+1}{2p'} - \frac{l-1}{2p}) (j - \frac{k-1}{2p'} + \frac{l+1}{2p}) (j + \frac{k+1}{2p'} + \frac{l-1}{2p}) (j - \frac{k-1}{2p'} - \frac{l+1}{2p})} \right]^{\frac{1}{2}} \nonumber \\
    =& \prod_{j \in \mathbb{Z}} \left[ \frac{(j + \frac{l+1}{2p} +\frac{k+1}{2p'} ) (j + \frac{l-1}{2p} + \frac{k-1}{2p'}) (j + \frac{l+1}{2p} - \frac{k+1}{2p'}) (j + \frac{l-1}{2p} - \frac{k-1}{2p'} ) }{(j + \frac{l-1}{2p} - \frac{k+1}{2p'} ) (j + \frac{l+1}{2p} - \frac{k-1}{2p'} ) (j + \frac{l-1}{2p} + \frac{k+1}{2p'} ) (j + \frac{l+1}{2p} + \frac{k-1}{2p'} )} \right]^{\frac{1}{2}}.
\end{align}
In the last step we have made a $j\to -j$ transformation in the second and third factors
in the numerator, and in the first and fourth factors in the denominator.
Now we can use the infinite product identity
\begin{equation}
    \prod_{j \in \mathbb{Z}}  \frac{j + a}{j + b} = \frac{a}{b} \prod_{j \in \mathbb{Z}^*}  \frac{1 + \frac{a}{j}}{1 + \frac{b}{j}} = \frac{a}{b} \prod_{j \in \mathbb{Z}^+}  \frac{1 - \frac{a^2}{j^2}}{1 - \frac{b^2}{j^2}} = \frac{\sin(\pi a) }{\sin(\pi b)}
\end{equation}
to get
\begin{align}
    \mathcal{M}_{k,l} 
    &= \left[ \frac{\sin( \pi( \frac{l+1}{2p} +\frac{k+1}{2p'} )) \sin( \pi(\frac{l-1}{2p} + \frac{k-1}{2p'} )) \sin( \pi(\frac{l+1}{2p} - \frac{k+1}{2p'} )) \sin( \pi(\frac{l-1}{2p} - \frac{k-1}{2p'} )) }{\sin( \pi(\frac{l-1}{2p} - \frac{k+1}{2p'} )) \sin( \pi( \frac{l+1}{2p} - \frac{k-1}{2p'} )) \sin( \pi(\frac{l-1}{2p} + \frac{k+1}{2p'} )) \sin( \pi(\frac{l+1}{2p} + \frac{k-1}{2p'} )) } \right]^{\frac{1}{2}} \nonumber \\
    &= \left[ \frac{ \left(\sin^2(  \frac{\pi(l+1)}{2p}) - \sin^2( \frac{\pi(k+1)}{2p'}) \right) \left(\sin^2(  \frac{\pi(l-1)}{2p}) - \sin^2( \frac{\pi(k-1)}{2p'}) \right) }{\left(\sin^2(  \frac{\pi(l-1)}{2p}) - \sin^2( \frac{\pi(k+1)}{2p'}) \right) \left(\sin^2(  \frac{\pi(l+1)}{2p}) - \sin^2( \frac{\pi(k-1)}{2p'}) \right) } \right]^{\frac{1}{2}}.
\label{eq:M_kl}
\end{align}
This gives us the result (\ref{mklanswer}) that we needed in the main text.

Armed with this formula for $\mathcal{M}_{k,l}$, we can see that the product formula for $\mathcal{N}_{m,n}$ in equation (\ref{n_using_m}) telescopes.
We first do the product over all values of $(2k-1,2l-1)$ except $(1,l)$.
\begin{align}
    \mathcal{N}_{m,n} 
    &= \left( \prod_{l=1}^{n} \mathcal{M}_{1,2l-1} \right)
    \prod_{k=2}^{m} \prod_{l=1}^{n} \mathcal{M}_{2k-1,2l-1}  \nonumber \\
    &= \left( \prod_{l=1}^{n} \mathcal{M}_{1,2l-1}\right)
    \prod_{k=2}^{m} \prod_{l=1}^{n} 
    \left[ \frac{ \left( \sin^2(\frac{\pi l}{p}) - \sin^2(\frac{\pi k}{p'}) \right) \left( \sin^2(\frac{\pi (l-1)}{p}) - \sin^2( \frac{\pi (k-1)}{p'} ) \right) }{ \left( \sin^2(\frac{\pi (l-1)}{p}) - \sin^2(\frac{\pi k}{p'}) \right) \left( \sin^2(\frac{\pi l}{p}) - \sin^2( \frac{\pi (k-1)}{p'} ) \right)} \right]^{\frac{1}{2}} \nonumber \\
    &= \left(\prod_{l=1}^{n} \mathcal{M}_{1,2l-1}\right)
    \left[ \frac{ \left( \sin^2(\frac{\pi n}{p}) - \sin^2(\frac{\pi m}{p'}) \right) \left( \sin^2( \frac{\pi}{p'} ) \right) }{ \left( \sin^2(\frac{\pi m}{p'}) \right) \left( \sin^2(\frac{\pi n}{p}) - \sin^2( \frac{\pi}{p'} ) \right)} \right]^{\frac{1}{2}} \nonumber\\
    &= \left(\prod_{l=1}^{n} \mathcal{M}_{1,2l-1} \right)
    \left[ 
    \frac{ \cot^2(\frac{\pi n}{p}) - \cot^2(\frac{\pi m}{p'}) }
            { \cot^2(\frac{\pi n}{p}) - \cot^2( \frac{\pi}{p'} ) }
    \right]^{\frac{1}{2}}\, .
    \label{firsttelescope}
\end{align}
Now we compute the remaining product over $l$, but leaving out $\mathcal{M}_{1,1}$.
\begin{align}
   \prod_{l=1}^{n} \mathcal{M}_{1,2l-1}
    &= \mathcal{M}_{1,1} \prod_{l=2}^{n} \left[ \frac{ \left( \sin^2(\frac{\pi}{p'}) - \sin^2(\frac{\pi l}{p}) \right) \left( \sin^2( \frac{\pi (l-1)}{p} ) \right) }{ \left( \sin^2(\frac{\pi l}{p}) \right) \left( \sin^2(\frac{\pi}{p'}) - \sin^2( \frac{\pi (l-1)}{p} ) \right)} \right]^{\frac{1}{2}} \nonumber\\
    &= \mathcal{M}_{1,1} \prod_{l=2}^{n} \left[ 
    \frac{ \cot^2(\frac{\pi}{p'}) - \cot^2(\frac{\pi l}{p}) }
    { \cot^2(\frac{\pi}{p'}) - \cot^2(\frac{\pi (l-1)}{p})} \right]^{\frac{1}{2}}
    = \mathcal{M}_{1,1} \left[ 
    \frac{ \cot^2(\frac{\pi}{p'}) - \cot^2(\frac{\pi n}{p}) }
    { \cot^2(\frac{\pi}{p'}) - \cot^2(\frac{\pi}{p}) } 
    \right]^\frac{1}{2}\, .
    \label{secondtelescope}
\end{align}
Combining (\ref{firsttelescope}) and (\ref{secondtelescope}), we get
\begin{align}
    \mathcal{N}_{m,n} = \mathcal{M}_{1,1} 
    \left[ 
    \frac{ \cot^2(\frac{\pi m}{p'}) - \cot^2(\frac{\pi n}{p}) }
    { \cot^2(\frac{\pi}{p'}) - \cot^2(\frac{\pi}{p}) } 
    \right]^\frac{1}{2}\, ,
\end{align}
which is the relation (\ref{full_telescope}) that was used in the main text.

\section{Gaussian matrix integrals}
\label{appgaussian}
\paragraph{The one-matrix gaussian integral.}
In the main text we need the following integral, which is the Gaussian one-matrix integral written as an integral over the eigenvalues \cite{marino2005chern}
\begin{align}
    Z_G(\ell) 
    := \frac{1}{\ell !}\int \prod_{i=1}^\ell \frac{\d x_i}{2\pi} \,
    \Delta(x)^2 \, \exp \left( - \frac{N}{t} \sum_{i=1}^\ell 
    \frac{x_i^2}{2} \right)
    = \frac{G_2(\ell+1)}{(2\pi)^{\ell/2}} \, \left( \frac{t}{N} \right)^{\ell^2/2}\, .
    \label{gaussianonematrix}
\end{align}
Here $G_2$ denotes the Barnes-G double gamma function.
This equation is exact, and, in particular, is true even if $\ell$ is of order $N$.
This can be derived using the fact that Hermite polynomials are orthogonal with respect to the Gaussian measure \cite{marino2005chern}.
We note that the asymptotic expansion of $Z_G(N)$ reads \cite{marino2005chern}
\begin{align}
    \log Z_G(N) = N^2 \left( \frac{1}{2} \log t - \frac{3}{4} \right) + O(\log N)\, .
    \label{gaussian_one_asymp}
\end{align}
We identify the first term as $N^2 F_{0,G}(t)/t^2$.
In particular, note that terms of order $N \log N$ or $N$ are absent.
This will no longer be the case in the two-matrix integral.
The above expression implies the following result that we need for our calculations
\begin{align}
    \log \, \frac{Z_{G}(N-\ell, t-t\ell/N)}{Z_{G}(N,t)}
    = \frac{N^2}{t^2} \left( F_{0,G}(t-t\ell/N) - F_{0,G}(t) \right)
     + O\left(\frac{1}{N} \right)\, . \label{zn_by_znminus1_onematrix}
\end{align}

\paragraph{Volume of $U(\ell)$.}
We can use the result (\ref{gaussianonematrix}) and the equality between the two integrals (\ref{e2.1}) and (\ref{zdef_eig}) to check the expression \refb{volul} for the  volume of the group $U(\ell)$.
Picking $V(x)=x^2/2$, the Gaussian integral over the matrix in (\ref{e2.1}) can be done in a trivial fashion by doing $\ell^2$ separate Gaussian integrals over the individual matrix elements of $M$.
The upshot is that
\begin{align}
Z_G(\ell) = \frac{1}{V_{U(\ell)}} \left( \frac{2\pi t}{N} \right)^{\ell^2/2}\, .
\end{align}
Equating the right hand side of this equation to (\ref{gaussianonematrix}), we get
\begin{align}
V_{U(\ell)} = \frac{(2\pi)^{\frac{1}{2}(\ell^2 + \ell)}}{G_2(\ell+1)}\, .
\label{volulapp}
\end{align}

\paragraph{Universality of the logarithmic terms.}
If we consider a general potential $V(M)$ of the form given in
\refb{e33.3},  then $F_0(t)$ differs from the result for quadratic potential,
but the logarithmic terms remain unchanged. 
One way to see this is as follows. 
Introducing the function $r(\xi)$ satisfying the ``string equation'' 
\cite{Bessis:1980ss, marino2005chern, DiFrancesco:1993cyw}
\begin{align}
    t \, \xi = r(\xi) + \sum_{k=1}^{(p-1)/2} g_{2k+2} 
    {\binom{2k+1}{k+1}} \, r(\xi)^{k+1}\, ,
\end{align}
the planar part of the free energy can be written as
\begin{align}
    F_0(t) = t^2 \int_0^1 \d \xi \, (1-\xi) \log r(\xi)\, .
    \label{f0general}
\end{align}
This integral is obtained as a continuum approximation to a discrete sum \cite{Bessis:1980ss, marino2005chern, DiFrancesco:1993cyw}, with errors potentially given by the Euler-Maclaurin formula.
Note that $r(\xi) \approx t \xi$ as $\xi \to 0$. 
The relation $r(\xi) = t \xi$ would be exact for the Gaussian matrix integral.
The integrand in (\ref{f0general}) thus behaves as $\log \xi$ near the lower limit $\xi = 0$.
It is for this reason that the combination
\begin{align}
    F_0(t) - F_{0,G}(t) = t^2 \int_0^1 \d \xi \, (1-\xi) \log \frac{r(\xi)}{t \xi}
\end{align}
is better behaved and $\log \frac{Z(N,t)}{Z_G(N,t)}$ has a good asymptotic expansion without logarithmic terms \cite{Bessis:1980ss, marino2005chern, ercolani2003asymptotics}.

\paragraph{The two-matrix gaussian integral.}
The two-matrix Gaussian integral is given by \cite{Mehta:1981xt}
\begin{align}
    Z_{G,2}(\ell)
    &:= \frac{1}{\ell !}\int \prod_{i=1}^\ell 
    \frac{\d x_i \d y_i}{2\pi} \,
    \Delta(x)\Delta(y) \, \exp \left( - \frac{N}{t} \sum_{i=1}^\ell 
    \left( 
    \frac{c_1 x_i^2}{2} + \frac{c_2 y_i^2}{2} - c_3 x_i y_i
    \right)
    \right) \\
    &= G_2(\ell+1) \, \left(\frac{t}{N} \right)^{\frac{1}{2}(\ell^2 + \ell)} \, (c_1 c_2 - c_3^2)^{-\frac{1}{2}\ell^2}
    c_3^{\frac{1}{2}(\ell^2 - \ell)}\, . \label{twomatrixgaussian}
\end{align}
The lemma in the appendix of \cite{Mehta:1981xt} allows us to reduce this integral to the one-matrix Gaussian integral.
Alternatively, one can derive it using two-matrix orthogonal polynomial technology; the orthogonal polynomials are still the Hermite polynomials. 
Again, this expression is exact and can be used for $\ell$ of order $N$.
The asymptotic expansion of $Z_{G,2}(N)$ reads
\begin{align} 
    \log Z_{G,2}(N) = N^2 \left(\frac{1}{2} \log t - \frac{1}{2} \log 
    \frac{c_1 c_2 - c_3^2}{c_3} - \frac{3}{4}\right)
    - \frac{1}{2} N \log N 
    + \frac{1}{2} N \, \log \frac{2\pi t}{c_3} + O(\log N)\, .
\end{align}
We identify the first term on the right hand side as $N^2 F_{0,G}(t) / t^2$.
Note the presence of terms of order $N \log N$ and $N$.
Using the above expression we see that
\begin{align} \label{ebb9}
    \log \, \frac{Z_{G,2}(N-\ell, t-t\ell/N)}{Z_{G,2}(N,t)}
    = \frac{N^2}{t^2} \left( F_{0,G}(t-t\ell/N) - F_{0,G}(t) \right)
     - \frac{\ell}{2} \log \frac{2\pi t }{c_3 N} + O\left(\frac{1}{N} \right)\, .
\end{align}
Comparing this to (\ref{zn_by_znminus1_onematrix}), we see the presence of an extra logarithmic term \cite{Ishibashi:2005zf}, which will be important in our analysis.

\section{Saddle point technology for the one-matrix integral}
In this appendix we shall review some  results in the theory of one-matrix integrals that we need.
We refer to \cite{marino2005chern, Marino:2007te} for more details, whose notations we also use.
\subsection{The large-$N$ limit}
\label{app:onematrix}
We consider integrals over an $N\times N$ Hermitian matrix $M$; in terms of the eigenvalues, the matrix integral is defined via
\begin{align}
    Z(N,t) := \frac{1}{N!}\int \prod_{i=1}^N\frac{\d x_i}{2\pi} \,
    \Delta(x)^2 \, \exp \left( - \frac{N}{t} \sum_i V(x_i) \right)\, . \label{znt_def_one_matrix}
\end{align}
We take the potential to be an even polynomial, for simplicity.
The planar free energy $F_0$ is defined as
\begin{align}
    F_0(t) := \lim_{N \to \infty} \frac{1}{N^2}\, t^2 \log Z(N,t)\, .
\end{align}

We work with the one-cut solution in which the resolvent takes the form
\begin{align}
    \omega_0(x) &:= \lim_{N \to \infty}\frac{1}{N} \, \left\langle \Tr \frac{1}{x-M} \right\rangle = \int_{-b}^b \d y \, \frac{\rho(y)}{x-y}\, , \label{ecc3}
\end{align}
The support of the eigenvalue density $\rho(y)$ is on the interval $[-b,b]$. One can show 
that \cite{Marino:2007te}
\begin{align}
    2t \, \omega_0(x) &= V'(x) - M(x) \sqrt{x^2 - b^2}\, , \label{ecc4}
\end{align}
where, 
if $V(x)$ is a polynomial of degree $d+1$, then $M(x)$ is a polynomial of degree $d-1$.
It will be understood that $\sqrt{x^2-b^2}\simeq x$ for large $|x|$ on the physical sheet.
Both $M(x)$ and $b$ can be determined using the fact that $\omega_0(x) = 1/x +O(x^0)$ as $x\to\infty$ on the physical sheet; they depend on $t$.
An important relation that will be useful for us is \cite{marino2005chern}
\begin{align}
    \frac{\partial}{\partial t} (t\omega_0(x)) = \frac{1}{\sqrt{x^2 - b^2}}\, .
    \label{dttomega}
\end{align}

The \emph{holomorphic} effective potential is defined via \cite{Marino:2007te}
\begin{align}
    \veff(x) &:= V(x) - 2 t \int_{-b}^b \d y \, \rho(y) \, \log(y - x) 
    = V(x) - 2 t \int_{-\Lambda}^x \d x' \, \omega_0(x') - 2t\log \Lambda \, , \label{ecc6}
\end{align}
where the limit $\Lambda\to \infty$ is understood in the last expression. Since we shall
define the double-scaling limit by zooming in near the region $x\simeq -b$, we have 
defined $\veff(x)$ such that it is real on the negative $x$-axis.
A consequence of the large-$N$ saddle point equation is that the \emph{real part} of the effective potential is constant on the interval $[-b,b]$.
Using (\ref{dttomega}) we get the derivative of the effective potential with respect to $t$:
\begin{align}
    \partial_t \veff(x) &= - 2 \int_{-\Lambda}^x \d x' \, \frac{1}{\sqrt{x'^2 - b^2}} - 2 \log \Lambda
    = -2 \log \left( \frac{-x - \sqrt{x^2-b^2}}{2} \right) \, .
    \label{dt_veff} 
\end{align}

It follows from \refb{ecc3} that the imaginary part of $\omega_0$ in the interval $[-b,b]$ is
given by $\pi\rho(x)$. 
On the other hand, we see from \refb{ecc4} that the imaginary part of $2t\omega_0$ in the same interval is given by $M(x)\sqrt{b^2-x^2}$. 
This gives
\be\label{ecc8}
\rho(x) = {1\over 2\pi t} M(x)\sqrt{b^2-x^2}\  \Theta(b-|x|)\, .
\ee
From \refb{ecc6} and \refb{ecc4} we also have
\be \label{ecc9}
\veff'(x) = V'(x)-2t\omega_0(x) = M(x)\sqrt{x^2-b^2} \, .
\ee

Next, we collect the following results for the planar free energy and its $t$-derivatives \cite{Marino:2007te, Brezin:1977sv}
\begin{align}
    F_0(t) &= - \frac{t}{2} \int_{-b}^b \d x \, \rho(x) V(x) - \frac{t}{2} \veff(-b)\, , \label{F0_one-matrix} \\
    \partial_t F_0(t) &= - \veff(-b) \, , \label{F0prime_one-matrix} \\
    \partial_t^2 F_0(t) &= 2\log \frac{b}{2} \, . \label{F0doubleprime_one-matrix}
\end{align}
To derive (\ref{F0_one-matrix}), we need to use the fact that the real part of $\veff$ is constant on the cut, and so $V(x) - 2 t \int \d y \, \rho(y) \log \vert x - y \vert = \veff(-b)$ for $x \in [-b,b]$. 
Using this to simplify the Coulomb-repulsion term in the on-shell action for the defining integral (\ref{znt_def_one_matrix}), we get
\begin{align}
    \frac{1}{t^2}\, F_0(t) &= - \frac{1}{t} \int_{-b}^b \d x \rho(x) V(x) + \int_{-b}^b \int_{-b}^b \d x \d y\,  \rho(x) \rho(y) \log \vert x - y \vert \non\\
    &=- \frac{1}{2t} \int_{-b}^b \d x \rho(x) V(x) - \frac{1}{2t} \veff(-b)\, ,
\end{align}
as desired.
To derive (\ref{F0prime_one-matrix}), we use the relation $\partial_t \log Z = \frac{N^2}{t^2} \int_{-b}^b \d x \, \rho(x) V(x)$ which follows directly by taking a $t$-derivative in the definition (\ref{znt_def_one_matrix}).
Using the large-$N$ approximation $\log Z = N^2 F_0 /t^2$ and (\ref{F0_one-matrix}), we get (\ref{F0prime_one-matrix}).
To derive (\ref{F0doubleprime_one-matrix}), we take a $t$-derivative in (\ref{F0prime_one-matrix}) and use (\ref{dt_veff}), together with the fact that $\veff'(-b) = 0 $ so that $\partial_t b$ does not contribute.

We also need the connected correlator of two resolvent operators \cite{Ambjorn:1992gw, Ambjorn:1990ji} (for a recent exposition,  see \cite{sss})
\begin{align}
    R_{0,2}(x_1, x_2) := \left \langle 
    \Tr \frac{1}{x_1 - M} \Tr \frac{1}{x_2 - M} 
    \right \rangle_\text{c} = \frac{1}{2(x_1 - x_2)^2} \left( 
    \frac{x_1 x_2 - b^2}{\sqrt{x_1^2 - b^2}\sqrt{x_2^2 - b^2}}
     - 1
    \right) \, .
\end{align}
We integrate this twice to get the connected correlator
\begin{align}
    A_{0,2}(x_1, x_2) := \left \langle 
    \Tr \log(M-x_1) \Tr \log (M-x_2)
    \right \rangle_\text{c}\, .
\end{align}
We first integrate over $x_2$ and get 
\begin{align}
\int_{-\infty}^{x_2} \d x_2' \, R_{0,2}(x_1, x_2')  
= \left( \frac{1}{2(x_2-x_1)} - \frac{1}{2(x_2-x_1)} \sqrt{\frac{x_2^2-b^2}{x_1^2-b^2}} \right) + \frac{1}{2\sqrt{x_1^2-b^2}}\, .
\end{align}
At large $\vert x_1 \vert$, this expression behaves as $x_1^{-2}$ so we should not have any problems integrating it.
The final answer is
\begin{align}
A_{0,2}(x_1, x_2) &= - \frac{1}{2} \log \left[ \frac{2(x_1 x_2-b^2 + \sqrt{x_1^2-b^2}\sqrt{x_2^2-b^2})}{(x_1+\sqrt{x_1^2-b^2})( x_2+\sqrt{x_2^2-b^2})} \right]\, .
\label{a02_one_matrix}
\end{align}
Eventually, we will need the following combination, which we get using (\ref{dt_veff}), (\ref{F0doubleprime_one-matrix}) and (\ref{a02_one_matrix})
\begin{align}
    \exp \left( 
    \partial_t^2 F_0 + \partial_t \veff(x_1) + \partial_t \veff(x_2) +
    4 A_{0,2} (x_1, x_2)  \right)
    = \left(
    \frac{b}{x_1 x_2 - b^2 + \sqrt{x_1^2-b^2}\sqrt{x_2^2-b^2}}\right)^2\, .
    \label{largeN-B-onematrix}
\end{align}

\subsection{A more geometrical perspective} \label{appgeometric}
We will now recast some of the above formulas in terms of a more formal perspective,  using the language of algebraic curves.
This will help us  in understanding the generalization to the two-matrix case.
We follow the notations of \cite{Eynard:2015aea}.

The starting point is to define a function $Y(x)$ via
\begin{align}
Y(x) := V'(x) - \frac{2t}{N} \left \langle \Tr \frac{1}{x-M} \right\rangle\, . \label{def:bigYonematrix}
\end{align}
Using (\ref{ecc6}), we note that $Y(x) = \veff'(x)$.
We can use a Schwinger-Dyson equation in the defining matrix integral to show that 
$Y(x)$ satisfies a polynomial equation
\begin{align}
Y(x)^2 &= V'(x)^2 - 4 t P(x) \, , \quad \text{where}
\label{schwinger-dyson1} \\
P(x)&:= \frac{1}{N} \left \langle \Tr \(\frac{V'(x)-V'(M)}{x-M} \)\right \rangle\, .
\end{align}
Let us define the spectral curve $\Sigma$ as
\begin{align}
\Sigma := \{(u,v) \in \mathbb{C}^2 \, \vert \, v^2 - V'(u)^2 + 4 t P(u) = 0  \}\, .
\end{align}
We can restate the Schwinger-Dyson equation as the statement that the point $(x,Y(x))$ lies on $\Sigma$,  with $Y(x)$ defined in (\ref{def:bigYonematrix}).

If $V(x)$ is a polynomial of degree $d+1$,  then $P(x)$ is a polynomial of degree $d-1$, which makes the right hand side of (\ref{schwinger-dyson1}) a polynomial of degree $2d$.
Note,  in particular, that if $V(x)=x^2/2$, then $P(x)=1$.
If we do not insist on using an even polynomial, then we can take $d=p-1$ in order to describe the $(2,p)$ minimal string.
The ``one-cut assumption'' can be rephrased as the condition that $d-1$ of the roots of the polynomial $V'(x)^2 - 4 t P(x)$ have multiplicity two, so that $V'(x)^2 - 4 t P(x) = M(x)^2 (x-a)(x-b)$.
Here $a$ and $b$ are the cut endpoints.
The roots of $M(x)$, which are $d-1$ in number,  are precisely the locations of the one-eigenvalue instantons.\footnote{
Not all of these one-eigenvalue instantons survive in the double-scaling limit.  
To get the conformal background of $(2,p)$ minimal string theory,  we can take a polynomial of degree $d+1 = p$,  or we could also take an even polynomial of degree $d+1 = p+1$.
In either case,  only $(p-1)/2$ of the $d-1$ one-eigenvalue instantons will survive in the double-scaling limit.}
Note also that, in the immediate neighborhood of a one-eigenvalue instanton,  the equation defining the spectral curve looks like $y^2 - c\, x^2 = 0$ (for some constant $c$), clearly exhibiting the ``double-point'' singularity \cite{Seiberg:2003nm}.

If all the roots of $V'(x)^2 - 4 t P(x)$ were distinct, the spectral curve $\Sigma$ would have genus $d-1$, and we would have an eigenvalue density supported on $d$ distinct arcs in the complex-$x$ plane.
In the one-cut solution,  $d-1$ of these arcs have length zero,  giving us $d-1$ distinct one-eigenvalue instantons,  as also argued in the previous paragraph.

Let us now specialize to the one-cut case. 
The curve defined by (\ref{schwinger-dyson1}) admits a uniformization parameter $z$ with the projection to the $x$-coordinate given by
\begin{align}
x =  \frac{a+b}{2}+ \gamma \left(z + z^{-1} \right)\,  , \quad \gamma := \frac{b-a}{4}\, .
\end{align}
The special property of the above map is that the cut end-points $x=a$ and $x=b$ get mapped to $z=-1$ and $z=1$.
These are also the points where $\d x/\d z = 0$.
We need two copies or sheets of the $x$-plane to cover the $z$-plane, the ``physical sheet'' is the one whose $z$-image contains $z = \infty$,  the second sheet is the one whose $z$-image contains $z=0$.
The boundary dividing the $z$-image of these two sheets is the unit circle in the complex-$z$ plane.

Further specializing to an even potential, so that $b = - a > 0$ and $b =2\gamma$,  we see that,  on the physical sheet we have
\begin{align} \label{ec.25}
x &= \gamma \left(z + z^{-1} \right) \, , \quad 
\sqrt{x^2 - 4\gamma^2} =\gamma \left( z - z^{-1}\right)\, .
\end{align}
Thus,  we see that many of the formulas in the previous section would be somewhat simpler when written in terms of $z$.
For instance,   we can rewrite (\ref{dt_veff}), (\ref{a02_one_matrix}) and (\ref{largeN-B-onematrix}) as
\begin{align}
&\partial_t \veff = -2 \log (-\gamma z)\, , \label{dtveff_simple_z} \\
&A_{0,2}(x_1, x_2) = - \log \frac{x_1 - x_2}{\gamma z_1 - \gamma z_2} =\log {z_1 z_2\over z_1 z_2-1} \, , \label{a02_simple_z}\\
& \exp \left( 
    \partial_t^2 F_0 + \partial_t \veff(x_1) + \partial_t \veff(x_2) +
    4 A_{0,2} (x_1, x_2)  \right) = \left( \frac{z_1 z_2}{\gamma(z_1 z_2 - 1)^2} \right)^2\, .
    \label{calphabeta_z1z2}
\end{align}

\section{Saddle-point structure of the two-matrix integral}
\label{app:twomatrixsaddle}

In this appendix we shall review some relevant formulas in the large-$N$ and the double-scaling limit of two-matrix integrals.
We follow \cite{Eynard:2002kg} and \cite{Bertola:2003rp}.

\subsection{The large-$N$ limit}
Recall that we have defined the two-matrix integral via
\begin{align}
    Z(N,t):= \frac{1}{N!} \int \prod_{i=1}^N \frac{\d x_i \d y_i}{2\pi}
    \Delta(x)\Delta(y)
    \exp \left[ - \frac{N}{t} \sum_{i=1}^N (V_1(x_i) + V_2(y_i) - x_i y_i) \right]\, . 
    \label{zntwodef_repeat}
\end{align}
Up to an overall normalization,  this integral is proportional to
\begin{align}
Z(N,t) \, \propto \,  \int \d M_1 \d M_2 \, \exp \left[ - \frac{N}{t} \Tr \left(V_1(M_1) + V_2(M_2) - M_1 M_2 \right) \right] \, ,
\end{align}
via the Harishchandra-Itzykson-Zuber formula \cite{Harish-Chandra:1957dhy,  Itzykson:1979fi, Zinn-Justin:2002rai}.
In order to get to the $(p',p)$ minimal string, we can take $V_1$ to be a polynomial of degree $p$ and $V_2$ to be a polynomial of degree $p'$ \cite{Daul:1993bg}.
To get back the one-matrix case,  one can set $p'=2$ and integrate out the matrix $M_2$.

We define two functions $Y(x)$ and $X(y)$ as follows \cite{Eynard:2002kg} 
\begin{align}
    Y(x) := V_1'(x) - \frac{t}{N} \left\langle \Tr \frac{1}{x-M_1} \right\rangle
    \, , \\
    X(y) := V_2'(y) - \frac{t}{N} \left\langle \Tr \frac{1}{y-M_2} \right\rangle 
    \, .
\end{align}
Note that in terms of $V_{1,\text{eff}}(x)$ and $V_{2,\text{eff}}(y)$ defined in (\ref{def:v1eff}) and (\ref{def:v2eff}),  we have 
\be\label{eveffxyrel}
V_{1,\text{eff}}'(x) = Y(x), \qquad V_{2,\text{eff}}'(y)= X(y)\, .
\ee
We have suppressed the dependence on $t$, with the understanding that both sides will have
the same dependence on $t$.
Let us define the spectral curve $\Sigma$ via
\begin{align}
\Sigma &:= \{ (u,v) \in\mathbb{C}^2 \, \vert \,  (V_1'(u)-v)(V_2'(v)-u) - P(u,v) + t = 0  \} \,  , \quad \text{where} \\
P(u,v) &:= \frac{t}{N} \left \langle \Tr \left(\frac{V_1'(u)- V_1'(M_1)}{u-M_1} \frac{V_2'(v)- V_2'(M_2)}{v-M_2} \right)\right\rangle\, .
\end{align}
It can be shown,  via Schwinger-Dyson equations,  that \emph{both} the points $(x,Y(x))$ and $(X(y),y)$ lie on the spectral curve $\Sigma$ \cite{Eynard:2002kg}.
Generically,  we can use either $x$ or $y$ as the local coordinate on $\Sigma$.
The projection to $x$ ceases to be a good coordinate when $\d Y(x)/\d x = \infty$. 
Typically,  this will happen when $Y(x)$ has a square root behavior near some $x$.
A similar remark holds for projection to $y$.

Like in the one-matrix case,  we work with the case when $\Sigma$ has genus zero (apart from the singular points to be discussed below).
Denoting the uniformization parameter by $z \in \mathbb{C}\cup \{\infty\}$,  we can coordinatize $\Sigma$ as \cite{Eynard:2002kg, Bertola:2003rp}
\begin{align}
\left( \mathcal{X}(z), \mathcal{Y}(z) \right) &\in \Sigma \, , \\
\mathcal{X}(z) = \gamma z + \sum_{k=0}^{p'-1} \alpha_k \, z^{-k} \,  , & \quad
\mathcal{Y}(z) = \gamma z^{-1} + \sum_{k=0}^{p-1} \beta_k \, z^{k}\, .
\label{curlyXz}
\end{align}
This means that for given $x$, we can find a $z$ such that $x=\XX(z)$, $Y(x)=\YY(z)$, and
for given $y$, we can find a $z'$ such that $X(y)=\XX(z')$, $y=\YY(z')$.
The map $\XX: z \mapsto x$ is,  generically,  a $p'$-to-$1$ map except at $p'$ values of $z$
where $\d \mathcal{X}(z)/\d z = 0$.
This means that we need $p'$ number of $x$-sheets to cover the $z$-plane,  or,  equivalently, to cover $\Sigma$.
The ``physical'' $x$-sheet contains the point $z=\infty$, near which the resolvent $\frac{1}{N} \langle \Tr\frac{1}{x-M_1} \rangle$ behaves as $\frac{1}{x}$.
The boundary of the physical $x$-sheet on $\Sigma$ is where the eigenvalues of $M_1$ are distributed.
Of course,  analogous comments apply to the map $\YY(z):z \mapsto y$, with $p'$ replaced by $p$.
The important distinction is that the physical $y$-sheet contains the point $z=0$, near which the resolvent $\frac{1}{N} \langle \Tr\frac{1}{y-M_2} \rangle$ behaves as $\frac{1}{y}$.

If $y=Y(x)$, then $(x,y)\in\Sigma$. On the other hand, $(X(y),y)$ is also on $\Sigma$.
However it does not follow from this that $x=X(y)$, since in general $(x,y)$ and
$(X(y),y)$ belong to different Riemann sheets. Exceptions are the
``one-eigenvalue instantons'' since
it follows from \refb{exstarystar} and \refb{eveffxyrel} that  they  are located at the points $(x^\star, y^\star)$ satisfying
\be\label{ed.10}
x^\star = X(y^\star), \quad y^\star=Y(x^\star)\, .
\ee
It then follows that if $\XX(z^\star)=x^\star$, then $\YY(z^\star)= Y(x^\star)=y^\star$.
These represent ZZ branes in minimal string theory after taking the double-scaling limit \cite{Seiberg:2003nm}. However, in the neighborhood of $(x^\star,y^\star)$ the points
$(x, Y(x))$ and $(X(y), y)$ belong to different branches and 
there will exist \emph{two distinct values} of $z$,  call them $z^{\star(1)}$ and $z^{\star(2)}$, such that
\begin{align} \label{ed.11}
(x^\star, y^\star) = (\mathcal{X}(z^{\star(1)})\, , \mathcal{Y}(z^{\star(1)})) = 
(\mathcal{X}(z^{\star(2)})\, , \mathcal{Y}(z^{\star(2)})) \, .
\end{align}In the neighborhood of these points the equation defining $\Sigma$ looks like $\alpha (x-x^\star)^2 - \beta (y-y^\star)^2 = 0$ and the surface is singular.\footnote{
As a simple example, consider the ``figure-8'' curve defined by $x^4 = x^2 - y^2$, embedded in $\mathbb{R}^2$ and parametrized as $(x,y) = (\sin t, \sin t \cos t)$.  The point $(0,0)$ is a double point,  and corresponds to both $t=0$ and $t = \pi$.  
In the discussion of the one-matrix case in appendix \ref{appgeometric},  the two values of the uniformizing coordinate at the instanton locations are related as $z^{\star(2)} = 1/z^{\star(1)}$.}

Now we come to the calculation of the objects that we need,  namely those that appear in the formulas (\ref{defa5}), (\ref{defb5}) and (\ref{defc5}).
Using the expressions in \cite{Bertola:2003rp},  it was shown in \cite{Ishibashi:2005zf} that
\begin{align}
\mathcal{A_\alpha} = \veff(x^\star,y^\star) + \partial_t F_0 = \int_{z^{\star(2)}}^{z^{\star(1)}} \d z \,  \mathcal{Y}(z) \, \frac{\d \mathcal{X}(z)}{\d z} \, .
\label{mathcalAtwomatrix}
\end{align} 
It also holds that \cite{Bertola:2003rp}
\begin{align}
\partial_t^2 F_0 = 2 \log \gamma\, ,
\end{align}
which is directly analogous to the equation (\ref{F0doubleprime_one-matrix}) in the one-matrix case.
Further, we have the following equality of one-forms  \cite{Bertola:2003rp}
\begin{align}
\left. \partial_t Y(x)\right|_{x=\XX(z)} \, \d \XX(z) = - \left. \partial_t X(y)\right|_{y=\YY(z)} \, 
\d \YY(z) = -\frac{\d z}{z} \label{three_one_forms}
\end{align}
In the one-matrix case, the analog of this equation would be (\ref{dttomega}).
To get an expression for the $t$-derivatives of the effective potentials, note that the definition (\ref{def:v1eff}) implies that
\begin{align}
V_{1,\text{eff}} (x) = \int_{\Lambda_x}^x \d x' \, Y(x') + \left( V_1(\Lambda_x) - t \log \Lambda_x \right)\, ,
\end{align}
where the limit $\Lambda_x \to \infty$ on the physical $x$-sheet is understood. 
Using (\ref{three_one_forms}) we see that if we introduce the variable $z'$ via $x'=\XX(z')$
then,
\begin{align}
\partial_t V_{1,\text{eff}} (x) &= -\int_{\Lambda_x}^x \d x' \,  \frac{1}{z'} \, \frac{\d z'}{\d x'} - \log \Lambda_x 
=-\int_{\Lambda_x/\gamma}^z \frac{\d z'}{z'} - \log \Lambda_x  = -\log (z \gamma)\, ,
\end{align}
where we used the fact that $x \approx \gamma z$ near $x = \infty$ on the physical $x$-sheet, see (\ref{curlyXz}). 
This equation is the direct analog of (\ref{dtveff_simple_z}) in the two-matrix case.
Similarly, we have
\begin{align}
\partial_t V_{2,\text{eff}} (y) &= \int_{\Lambda_y}^y \d y' \,  \frac{1}{z'} \, \frac{\d z'}{\d y'} - \log \Lambda_y
= \int_{\gamma/\Lambda_y}^z \frac{\d z'}{z'} - \log \Lambda_y  = -\log (\gamma/z)\, ,
\end{align}
where we used the fact that $y \approx \gamma/z$ near $y = \infty$ on the physical $y$-sheet, see (\ref{curlyXz}).
Finally,  the connected two-point correlators of the Vandermonde potential, defined in (\ref{defa021}), (\ref{defa022}) and (\ref{defa023}) are given by \cite{Daul:1993bg, Ishibashi:2005zf}
\begin{align}
A_{0,2}^{(1)}(x_1, x_2) &= 
- \log \frac{\mathcal{X}(z_1)-\mathcal{X}(z_2)}{\gamma z_1 - \gamma z_2}, \quad
\hbox{for $x_1=\XX(z_1)$, \ $x_2=\XX(z_2)$}  \,,\\
A_{0,2}^{(2)}(y_1, y_2) &= - \log \frac{\mathcal{Y}(z_1)-\mathcal{Y}(z_2)}{\gamma/z_1 - \gamma/z_2}, \quad
\hbox{for $y_1=\YY(z_1)$, \ $y_2=\YY(z_2)$} \,,\\
A_{0,2}^{(3)}(x_1, y_2) &= - \log \left( 1 - \frac{z_2}{z_1} \right), \quad
\hbox{for $x_1=\XX(z_1)$, \ $y_2=\YY(z_2)$}\, .
\end{align}
These formulas generalize (\ref{a02_simple_z}) to the two-matrix case.
From now on we shall drop the $\star$'s and add a subscript $\alpha$ to a variable to
denote its value at the $\alpha$-th saddle point.

Using these formulas, let us compute the quantity $\mathcal{C}_{\alpha,\beta}$ defined in (\ref{defc5}) that appears in the general multi-instanton contribution to the partition function:
\begin{align}
\log \mathcal{C}_{\alpha,\beta} 
&= \log[(x_\alpha - x_\beta)(y_\alpha-y_\beta)] + 2 \log \gamma - \log (z_\alpha^{(1)} \gamma ) - \log (z_\beta^{(1)} \gamma )
- \log (\gamma/z_\alpha^{(2)}) - \log (\gamma/z_\beta^{(2)}) \nonumber \\
&
- \log \frac{x_\alpha - x_\beta}{\gamma z_\alpha^{(1)} - \gamma z_\beta^{(1)}}
- \log \frac{y_\alpha - y_\beta}{\gamma/z_\alpha^{(2)} - \gamma/z_\beta^{(2)}}
- \log \left( 1 - \frac{z_\beta^{(2)}}{z_\alpha^{(1)}} \right)
- \log \left( 1 - \frac{z_\alpha^{(2)}}{z_\beta^{(1)}} \right) \, .
\end{align}
In writing these formulas, we have to pick the branch $z^{(1)}_\alpha$ for $x_\alpha$ and the branch $z^{(2)}_\alpha$ for $y_\alpha$.
We now see that the $\log \gamma$ terms cancel and the contribution $ \log[(x_\alpha - x_\beta)(y_\alpha-y_\beta)]$ from the Vandermonde factors also cancels with the corresponding factors from $A_{0,2}^{(1)}$ and $A_{0,2}^{(2)}$.
Simplifying a bit, we find\footnote{As a consistency check,  we can see that this expression is consistent with the one-matrix results where $z_\alpha^{(2)} = 1/z_\alpha^{(1)}$. To see this,  we multiply (\ref{calphabeta_z1z2}) with the contribution from the Vandermonde $(x_\alpha-x_\beta)^2$ and use $x_\alpha = \gamma(z_\alpha+1/z_\alpha)$ and the corresponding relation for $x_\beta$. This gives $\mathcal{C}_{\alpha,\beta} = \frac{z_\alpha-z_\beta}{z_\alpha-1/z_\beta} \frac{1/z_\alpha-1/z_\beta}{1/z_\alpha - z_\beta}$ in agreement with (\ref{calphabetatwomatrixz}).}
\begin{align}
\log \mathcal{C}_{\alpha,\beta} = \log 
\frac{z_\alpha^{(1)} - z_\beta^{(1)}}{z_\alpha^{(1)} - z_\beta^{(2)}}
\frac{z_\alpha^{(2)} - z_\beta^{(2)}}{z_\alpha^{(2)} - z_\beta^{(1)}}\,. \label{calphabetatwomatrixz}
\end{align}

Now let us come to the computation of $\mathcal{B}_\alpha$ defined in (\ref{defb5}).
Again,  we have to pick the branch $z^{(1)}_\alpha$ for $x_\alpha$ and the branch $z^{(2)}_\alpha$ for $y_\alpha$.
We first simplify the exponential piece appearing in (\ref{defb5}):
\begin{align}
&\frac{1}{2} \partial_t^2 F_0 + \partial_t V_{1,\text{eff}} +  \partial_t V_{2,\text{eff}}
+ \frac{1}{2} A_{0,2}^{(1)}+ \frac{1}{2} A_{0,2}^{(2)} + A_{0,2}^{(3)} \nonumber \\*
&= \log \gamma - \log (\gamma z_\alpha^{(1)}) - \log (\gamma/z_\alpha^{(2)})
- \frac{1}{2} \log \left(\frac{1}{\gamma}\frac{\d \mathcal{X}}{\d z} (z_\alpha^{(1)}) \right)
 \nonumber \\*
&\hspace{0.1in} - \frac{1}{2} \log \left( -\frac{1}{\gamma}\,
(z_\alpha^{(2)})^2\, \frac{\d \mathcal{Y}}{\d z} (z_\alpha^{(2)}) \right)
- \log \left(1 - \frac{z_\alpha^{(2)}}{z_\alpha^{(1)}} \right)\, .
\end{align}
We see again that the $\log \gamma$ cancels out and the expression simplifies to
\begin{align}
\exp\left[ 
\frac{1}{2} \partial_t^2 F_0 + \partial_t V_{1,\text{eff}} +  \partial_t V_{2,\text{eff}}
+ \frac{1}{2} A_{0,2}^{(1)}+ \frac{1}{2} A_{0,2}^{(2)} + A_{0,2}^{(3)}
\right]
= \frac{1}{z_\alpha^{(1)}-z_\alpha^{(2)}} \,  \left( 
- \frac{\d \mathcal{X}}{\d z} (z_\alpha^{(1)}) 
\frac{\d \mathcal{Y}}{\d z} (z_\alpha^{(2)})
\right)^{-\frac{1}{2}}\, . \nonumber
\end{align}
The prefactor term in (\ref{defb5}) combines nicely with the second term in the above equation. To see this note that $V_{1,\text{eff}}''(x) = \frac{\d Y}{\d x}$ and $V_{2,\text{eff}}''(y) = \frac{\d X}{\d y}$.  Thus,  we get
\begin{align}
\mathcal{B}_\alpha = \sqrt{\frac{2\pi t}{N}} \frac{1}{z_\alpha^{(1)}-z_\alpha^{(2)}}
\left( 
\frac{\d \mathcal{X}}{\d z} (z_\alpha^{(1)}) 
\frac{\d \mathcal{Y}}{\d z} (z_\alpha^{(2)})
- 
\frac{\d \mathcal{X}}{\d z} (z_\alpha^{(2)}) 
\frac{\d \mathcal{Y}}{\d z} (z_\alpha^{(1)})
\right)^{-\frac{1}{2}} \, .
\label{balpha2matrixz}
\end{align}

\subsection{The double-scaling limit}
Finally,  we discuss the double-scaling limit.
The double-scaling limit is defined by zooming in near a point on $\Sigma$ that corresponds to an endpoint of the distribution of the eigenvalues of $M_1$ (the symmetry between $M_1$ and $M_2$ is broken by which potential has a higher degree).
This is a point that lies on the boundary of the $z$-image of the physical $x$-sheet.
We appropriately choose the parameters in the potentials $V_1$ and $V_2$ and  
introduce new variables $x,y,z$ and new functions $\wt X$, $\wt Y$, $\wt\XX$ and
$\wt\YY$ as,
\begin{align}
x = c_x + d_x \varepsilon^\frac{p'}{2}\, \tilde x, \qquad & 
y = c_y + d_y 
\varepsilon^\frac{p}{2} \, \tilde y, \qquad z =  c_z + d_z  \varepsilon^\frac{1}{2} \tilde z\, ,\\
X(y) =  c_x+ d_x^{} \varepsilon^{\frac{p'}{2}} \wt X(\tilde y) \, , \qquad 
& Y(x) = c_y+ d_y^{}
\varepsilon^{\frac{p}{2}} \, \wt Y(\tilde x), \non\\
\XX(z) =  c_x+ d_x^{} \varepsilon^{\frac{p'}{2}} \wt \XX(\tilde z) \, , \qquad 
& \YY(z) = c_y+d_y^{}
\varepsilon^{\frac{p}{2}} \, \wt \YY(\tilde z) \, , \label{ed26}
\end{align}
for appropriate constants $c_x,c_y,c_z,d_x, d_y,d_z$.  
We now take the limit $N\to\infty$, $\ve\to 0$,  while keeping fixed the combination
\begin{align} \label{ed26.5}
e^{S_0} := \frac{N}{t} \,  d_x d_y \,\varepsilon^{\frac{p}{2} + \frac{p'}{2}}\,  .
\end{align}
The analog of the conformal background for the one-matrix case is a special choice of the
parameters of the potential $V_1$, $V_2$ and the parameters $c_x,c_y,c_z,d_x,d_y,d_z$ such that \cite{Seiberg:2003nm, seibergannulus}
\be
\wt \XX(\tilde z) = T_{p'}(\tilde z), \qquad \wt\YY(\tilde z) = T_p(\tilde z)\, .\label{ed27}
\ee
Here $T_p$ denotes the Chebyshev polynomial of the first kind which is defined by the relation $\cos p \theta = T_p(\cos \theta)$. 
Equation \refb{ed27} also implicitly defines the functions 
$\wt X$ and $\wt Y$ after eliminating $\tilde z$. In the new variables
the spectral curve of the conformal background of the $(p',p)$ minimal string is the following curve \cite{Seiberg:2003nm}
\begin{align}
\Sigma = \{ (\tilde x,\tilde y)\in \mathbb{C}^2 \, \vert \, T_{p}(\tilde x) - T_{p'}(\tilde y) = 0 \} \, ,
\end{align}
with $\tilde z$ being the uniformization parameter of this surface.

The relations \refb{ed.10}, \refb{ed.11} defining the instanton locations now take the form:
\be 
\tilde x^\star = \wt X(\tilde y^\star), \quad \tilde y^\star = \wt Y(\tilde x^\star)\, , 
\ee 
\be
(\tilde x^\star, \tilde y^\star) = (\, \wt\XX(\tilde z^{\star(1)})\,,  \wt \YY(\tilde z^{\star(1)})\,)
=
(\, \wt\XX(\tilde z^{\star(2)})\,,\,  \wt \YY(\tilde z^{\star(2)})\,)\, .
\ee
These represent the singular points of $\Sigma$ and correspond to the instanton locations.
Explicitly, there are $(p'-1)(p-1)/2$ singular points on $\Sigma$ given by
\begin{align}
(\tilde x_{m,n},  \tilde y_{m,n}) &= \left( (-1)^m \cos \frac{\pi n p'}{p} \,  ,  \, (-1)^n \cos \frac{\pi m p}{p'}  \right)\,  , \quad \text{with} \\
m & \in \{ 1, \ldots, p'-1\}\, , \quad  n  \in \{ 1, \ldots, p-1\} \,,
\end{align}
and subject to the identification $(m,n) \equiv (p'-m,p-n)$ since these two labels give the same point on the curve $\Sigma$.
We have omitted the stars from the notation to reduce the clutter,  since
the subscripts $m,n$ make it clear that these values refer to the singular points.
Each of these singular points corresponds to two distinct values of the uniformizing coordinate $z$.
Explicitly,  these are
\begin{align}
\tilde z^{(1)}_{m,n} = \cos \left( \frac{\pi m}{p'} + \frac{\pi n}{p} \right)\, , \quad
\tilde z^{(2)}_{m,n} = \cos \left( \frac{\pi m}{p'} - \frac{\pi n}{p} \right)\, . \label{zstar12}
\end{align}
From the analysis given above, we cannot determine which of the two values $\cos ( \frac{\pi m}{p'} \pm \frac{\pi n}{p} ) $ corresponds to $\tilde z^{(1)}_{m,n}$ and which to $\tilde z^{(2)}_{m,n}$.
However,  exchanging them changes the signs of $T_\alpha=N\AAA_\alpha/t$ and $(\BB_\alpha)^2$ computed from \refb{mathcalAtwomatrix} and \refb{balpha2matrixz},  respectively,  and leaves
$\CC_{\alpha,\beta}$ computed from \refb{calphabetatwomatrixz} unchanged.
Since we only compare the combinations $\BB_\alpha T_\alpha^{1/2}$ and
$\CC_{\alpha,\beta}$ with the string theory results, this ambiguity does not affect
our analysis. 
Ref. \cite{Ishibashi:2005zf} resolves this ambiguity using a physical input.

Let us now compute the on-shell action of the instanton labeled by $(m,n)$.
Using (\ref{ed26}) and \refb{ed26.5},  we can recast (\ref{mathcalAtwomatrix}) as
\begin{align}
\frac{N}{t} \mathcal{A}_{m,n} = e^{S_0}\, \int_{\tilde z_{m,n}^{(2)}}^{\tilde z_{m,n}^{(1)}} \d \tilde z \,  {\wt \YY}(\tilde z) \, \frac{\d {\wt \XX}(\tilde z)}{\d \tilde z}\, .
\end{align}
In particular, since $\wt\XX(\tilde z_{m,n}^{(1)})=\wt\XX(\tilde z_{m,n}^{(2)})=\tilde x_{m,n}$,
the constant terms $c_x$ and $c_y$ in \refb{ed26} drop out of this equation.
To evaluate this, we first compute the indefinite integral of $T_p(\tilde z) \frac{\d }{\d \tilde z} T_{p'}(\tilde z)$:
\begin{align}
\int^{\tilde z} \d u \, T_p(u) \frac{\d }{\d u} T_{p'}(u) 
= \frac{p'}{2} \left( 
\frac{1}{p+p'} \, T_{p+p'}(\tilde z) - \frac{1}{p-p'} \, T_{p-p'}(\tilde z) 
\right)+C\, ,
\end{align}
where $C$ is the constant of integration. 
Taking the difference between this indefinite integral expression evaluated at the two values of $\tilde z$ in (\ref{zstar12}), we get
\begin{align}
T_{m,n}:= \frac{N}{t}\,\mathcal{A}_{m,n} = e^{S_0}\, (-1)^{m+n} \frac{2p p'}{p'^2-p^2} \, \sin \frac{\pi m p}{p'}\, \sin \frac{\pi n p'}{p}\, .
\label{amn}
\end{align}

We now take the double-scaling limit in the formula (\ref{balpha2matrixz}) for $\mathcal{B}_\alpha$. 
The result is
\begin{align}
\mathcal{B}_{m,n}  &=
e^{-S_0/2}\, 
\frac{\sqrt{2\pi}}{\tilde z_\alpha^{(1)}-\tilde z_\alpha^{(2)}}
\left( 
\frac{\d \wt\XX}{\d \tilde z} (\tilde z_\alpha^{(1)}) 
\frac{\d \wt\YY}{\d \tilde z} (\tilde z_\alpha^{(2)})
- 
\frac{\d \wt\XX}{\d \tilde z} (\tilde z_\alpha^{(2)}) 
\frac{\d \wt\YY}{\d \tilde z} (\tilde z_\alpha^{(1)})
\right)^{-\frac{1}{2}}
\non\\
&= e^{-S_0/2}\, \frac{\sqrt{2\pi}}{2 \sin \frac{\pi m}{p'} \sin \frac{\pi n}{p}} 
\left( \frac{2 p'p \, (-1)^{m+n} \sin \frac{\pi m p}{p'}\sin \frac{\pi n p'}{p} }{\sin^2 \frac{\pi m}{p'} -\sin^2 \frac{\pi n}{p} } \right)^{-\frac{1}{2}}\, .
\label{bmn}
\end{align}
Note that the overall sign of $\BB_{m,n}$ can be changed by changing the orientation of the
steepest descent integration contour in the complex eigenvalue plane.  
This sign is not significant since the string theory computation also has a similar ambiguity.

Finally,  we discuss the double-scaling limit of $\mathcal{C}_{\alpha,\beta}$. 
Since $z$ just undergoes a shift and rescaling by $\varepsilon^{\frac{1}{2}}$, 
we get from \refb{calphabetatwomatrixz} that
\begin{align}
\mathcal{C}_{\alpha,\beta} =
\frac{\tilde z_\alpha^{(1)} - \tilde z_\beta^{(1)}}{\tilde z_\alpha^{(1)} - \tilde z_\beta^{(2)}}
\times \frac{\tilde z_\alpha^{(2)} - \tilde z_\beta^{(2)}}{\tilde z_\alpha^{(2)} - \tilde z_\beta^{(1)}}\,,\label{cab2matrixz}
\end{align}
with $\tilde z_\alpha$ given in (\ref{zstar12}).
This yields
\be
\CC_{(m,n)(m',n')} =  {\cos \left( \frac{\pi m}{p'} + \frac{\pi n}{p} \right) -  \cos \left( \frac{\pi m'}{p'} + \frac{\pi n'}{p} \right) \over \cos \left( \frac{\pi m}{p'} + \frac{\pi n}{p} \right) -  \cos \left( \frac{\pi m'}{p'} - \frac{\pi n'}{p} \right)} \times
{\cos \left( \frac{\pi m}{p'} - \frac{\pi n}{p} \right) -  \cos \left( \frac{\pi m'}{p'} - \frac{\pi n'}{p} \right) \over \cos \left( \frac{\pi m}{p'} - \frac{\pi n}{p} \right) -  \cos \left( \frac{\pi m'}{p'} + \frac{\pi n'}{p} \right)}\, .\label{cmnmpnp}
\ee
As expected, $T_\alpha$, $\BB_\alpha$ and $\CC_{\alpha,\beta}$ all have finite expressions
in the double-scaling limit.

\section{Integer dimensions in the open string spectrum} 
\label{sintegral}

Our analysis in the main text shows perfect agreement between the matrix model results and
string theory results for the instanton partition function. 
However we shall now argue that for some class of instantons this agreement is somewhat formal. 

For $(m,n)$ type ZZ branes, the spectrum of open strings is built by the action of ghost oscillators and matter
and Liouville Virasoro generators on Liouville $\times$ matter primaries $\times$ ghost vacuum
$c_1|0\rangle$
of dimension:
\begin{align}
h_{k\ell} &= -{1\over 4 pp'} \{(2m-2k-1)p+ (2n-2\ell-1)p'\}^2 + {1\over 4pp'} (p-p')^2, \non\\
0 & \le k\le m-1, \quad 0\le \ell\le n-1\, .
\end{align}
If $h_{k\ell}$ is a negative integer for any $k,\ell$, then by acting with ghost oscillators and matter Virasoro generators we shall produce zero modes. This means that in \refb{efklexp}, the power series expansion in $q$ will have some constant terms. 
This in turn will produce  logarithmic divergence in the integration over $t$.
The finiteness of the final result \refb{eq:M_kl} shows that the net coefficient of the logarithmically  divergent term vanishes, but in the spirit of string field theory the correct procedure is to interpret the divergences caused by individual terms as arising from integration over zero modes and to carry out the integration over these zero modes carefully. 
Since our analysis ignores this subtlety,  our results remain somewhat formal.
We shall now determine under what condition we get such additional zero modes.

The expression for $h_{k,\ell}$ may be written as
\begin{align}
h_{k,\ell} &= -{1\over 4pp'} \left[ (2m-2k-2)p + (2n-2\ell)p'\right] \left[ (2m-2k)p+(2n-2\ell-2)p'\right]\non\\
&= - {1\over pp'} [ap + (b+1)p'] [(a+1)p+ bp'] \non\\
&= -ab-(a+1)(b+1) - a(a+1) {p\over p'} - b (b+1) {p'\over p}\non \,, \quad \text{with }\\
& a := m-k-1, \quad b := n-\ell-1, \quad 0\le a\le m-1, \quad 0\le b\le n-1\, .
\end{align}
Since $p,p'$ are relatively prime, in order that $h_{k,\ell}$ be a negative integer, we must have
\be
p' \ |\ a(a+1), \qquad p\ | \ b(b+1)\, .
\ee
A simple solution is $a=b=0$ with $h_{k,\ell}=-1$.
This corresponds to the product of the identity fields from the matter and Liouville
sector and the ghost vacuum $c_1|0\rangle$. 
The zero modes produced from this sector correspond to ghost zero modes associated with the breakdown of the Siegel gauge.
These have already been taken into account in our analysis.
But there are other solutions. 
Here are some examples:
\ben
&& p'=2, \quad p=15, \quad a=0, \quad b=5, \non\\
&& p'=6, \quad p=55, \quad a=2, \quad b = 10,\non\\
&& p'=6, \quad p=35, \quad a=2, \quad b = 14\, .
\een
For such pairs $(p',p)$, the string theory results remain formal for $(m,n)$ type ZZ instantons
with $m\ge a+1$, $n\ge b+1$.

\bibliographystyle{apsrev4-1long}
\bibliography{multiple_zz_instantons}

\begin{thebibliography}{10}%
\makeatletter
\providecommand \@ifxundefined [1]{%
 \ifx #1\undefined \expandafter \@firstoftwo
 \else \expandafter \@secondoftwo
\fi
}%
\providecommand \@ifnum [1]{%
 \ifnum #1\expandafter \@firstoftwo
 \else \expandafter \@secondoftwo
\fi
}%
\providecommand \enquote [1]{``#1''}%
\providecommand \bibnamefont  [1]{#1}%
\providecommand \bibfnamefont [1]{#1}%
\providecommand \citenamefont [1]{#1}%
\providecommand\href[0]{\@sanitize\@href}%
\providecommand\@href[1]{\endgroup\@@startlink{#1}\endgroup\@@href}%
\providecommand\@@href[1]{#1\@@endlink}%
\providecommand \@sanitize [0]{\begingroup\catcode`\&12\catcode`\#12\relax}%
\@ifxundefined \pdfoutput {\@firstoftwo}{%
 \@ifnum{\z@=\pdfoutput}{\@firstoftwo}{\@secondoftwo}%
}{%
 \providecommand\@@startlink[1]{\leavevmode\special{html:<a href="#1">}}%
 \providecommand\@@endlink[0]{\special{html:</a>}}%
}{%
 \providecommand\@@startlink[1]{%
  \leavevmode
  \pdfstartlink
   attr{/Border[0 0 1 ]/H/I/C[0 1 1]}%
   user{/Subtype/Link/A<</Type/Action/S/URI/URI(#1)>>}%
  \relax
 }%
 \providecommand\@@endlink[0]{\pdfendlink}%
}%
\providecommand \url  [0]{\begingroup\@sanitize \@url }%
\providecommand \@url [1]{\endgroup\@href {#1}{\urlprefix}}%
\providecommand \urlprefix [0]{URL }%
\providecommand \Eprint[0]{\href }%
\@ifxundefined \urlstyle {%
  \providecommand \doi [1]{doi:\discretionary{}{}{}#1}%
}{%
  \providecommand \doi [0]{doi:\discretionary{}{}{}\begingroup
  \urlstyle{rm}\Url }%
}%
\providecommand \doibase [0]{http://dx.doi.org/}%
\providecommand \Doi[1]{\href{\doibase#1}}%
\providecommand \bibAnnote [3]{%
  \BibitemShut{#1}%
  \begin{quotation}\noindent
    \textsc{Key:}\ #2\\\textsc{Annotation:}\ #3%
  \end{quotation}%
}%
\providecommand \bibAnnoteFile [2]{%
  \IfFileExists{#2}{\bibAnnote {#1} {#2} {\input{#2}}}{}%
}%
\providecommand \typeout [0]{\immediate \write \m@ne }%
\providecommand \selectlanguage [0]{\@gobble}%
\providecommand \bibinfo [0]{\@secondoftwo}%
\providecommand \bibfield [0]{\@secondoftwo}%
\providecommand \translation [1]{[#1]}%
\providecommand \BibitemOpen[0]{}%
\providecommand \bibitemStop [0]{}%
\providecommand \bibitemNoStop [0]{.\EOS\space}%
\providecommand \EOS [0]{\spacefactor3000\relax}%
\providecommand \BibitemShut [1]{\csname bibitem#1\endcsname}%
\bibitem{DiFrancesco:1993cyw}%
  \BibitemOpen
  \bibfield{author}{%
  \bibinfo {author} {\bibfnamefont{P.}~\bibnamefont{Di~Francesco}}, \bibinfo
  {author} {\bibfnamefont{Paul~H.}\ \bibnamefont{Ginsparg}},\ and\ \bibinfo
  {author} {\bibfnamefont{Jean}\ \bibnamefont{Zinn-Justin}},\ }%
  \bibfield{title}{%
  \enquote{\bibinfo {title} {{2-D Gravity and random matrices}},}\ }%
  \bibfield{journal}{%
  \Doi{10.1016/0370-1573(94)00084-G}{\bibinfo {journal} {Phys. Rept.}}\ }%
  \textbf{\bibinfo {volume} {254}},\ \bibinfo {pages} {1--133} (\bibinfo {year}
  {1995}),\ \Eprint{http://arxiv.org/abs/hep-th/9306153}{arXiv:hep-th/9306153}%
  \bibAnnoteFile{NoStop}{DiFrancesco:1993cyw}%
\bibitem{Seiberg:2004at}%
  \BibitemOpen
  \bibfield{author}{%
  \bibinfo {author} {\bibfnamefont{Nathan}\ \bibnamefont{Seiberg}}\ and\
  \bibinfo {author} {\bibfnamefont{David}\ \bibnamefont{Shih}},\ }%
  \bibfield{title}{%
  \enquote{\bibinfo {title} {{Minimal string theory}},}\ }%
  \bibfield{journal}{%
  \Doi{10.1016/j.crhy.2004.12.007}{\bibinfo {journal} {Comptes Rendus
  Physique}}\ }%
  \textbf{\bibinfo {volume} {6}},\ \bibinfo {pages} {165--174} (\bibinfo {year}
  {2005}),\ \Eprint{http://arxiv.org/abs/hep-th/0409306}{arXiv:hep-th/0409306}%
  \bibAnnoteFile{NoStop}{Seiberg:2004at}%
\bibitem{Shenker:1990uf}%
  \BibitemOpen
  \bibfield{author}{%
  \bibinfo {author} {\bibfnamefont{Stephen~H.}\ \bibnamefont{Shenker}},\ }%
  \enquote{\bibinfo {title} {{The Strength of nonperturbative effects in string
  theory}},}\ in\ \emph{\bibinfo {booktitle} {{Cargese Study Institute: Random
  Surfaces, Quantum Gravity and Strings}}}\ (\bibinfo {year} {1990})\ pp.\
  \bibinfo {pages} {809--819}%
  \bibAnnoteFile{NoStop}{Shenker:1990uf}%
\bibitem{David:1990sk}%
  \BibitemOpen
  \bibfield{author}{%
  \bibinfo {author} {\bibfnamefont{Francois}\ \bibnamefont{David}},\ }%
  \bibfield{title}{%
  \enquote{\bibinfo {title} {{Phases of the large N matrix model and
  nonperturbative effects in 2-d gravity}},}\ }%
  \bibfield{journal}{%
  \Doi{10.1016/0550-3213(91)90202-9}{\bibinfo {journal} {Nucl. Phys. B}}\ }%
  \textbf{\bibinfo {volume} {348}},\ \bibinfo {pages} {507--524} (\bibinfo
  {year} {1991})%
  \bibAnnoteFile{NoStop}{David:1990sk}%
\bibitem{PolchinskiCombinatorics}%
  \BibitemOpen
  \bibfield{author}{%
  \bibinfo {author} {\bibfnamefont{Joseph}\ \bibnamefont{Polchinski}},\ }%
  \bibfield{title}{%
  \enquote{\bibinfo {title} {{Combinatorics of boundaries in string theory}},}\
  }%
  \bibfield{journal}{%
  \Doi{10.1103/PhysRevD.50.R6041}{\bibinfo {journal} {Phys. Rev. D}}\ }%
  \textbf{\bibinfo {volume} {50}},\ \bibinfo {pages} {R6041--R6045} (\bibinfo
  {year} {1994}),\
  \Eprint{http://arxiv.org/abs/hep-th/9407031}{arXiv:hep-th/9407031}%
  \bibAnnoteFile{NoStop}{PolchinskiCombinatorics}%
\bibitem{zz}%
  \BibitemOpen
  \bibfield{author}{%
  \bibinfo {author} {\bibfnamefont{Alexander~B.}\ \bibnamefont{Zamolodchikov}}\
  and\ \bibinfo {author} {\bibfnamefont{Alexei~B.}\
  \bibnamefont{Zamolodchikov}},\ }%
  \bibfield{title}{%
  \enquote{\bibinfo {title} {{Liouville field theory on a pseudosphere}},}\ }%
   (\bibinfo {month} {1}\ \bibinfo {year} {2001}),\
  \Eprint{http://arxiv.org/abs/hep-th/0101152}{arXiv:hep-th/0101152}%
  \bibAnnoteFile{NoStop}{zz}%
\bibitem{bryzz}%
  \BibitemOpen
  \bibfield{author}{%
  \bibinfo {author} {\bibfnamefont{Bruno}\ \bibnamefont{Balthazar}}, \bibinfo
  {author} {\bibfnamefont{Victor~A.}\ \bibnamefont{Rodriguez}},\ and\ \bibinfo
  {author} {\bibfnamefont{Xi}~\bibnamefont{Yin}},\ }%
  \bibfield{title}{%
  \enquote{\bibinfo {title} {{ZZ Instantons and the Non-Perturbative Dual of c
  = 1 String Theory}},}\ }%
   (\bibinfo {month} {7}\ \bibinfo {year} {2019}),\
  \Eprint{http://arxiv.org/abs/1907.07688}{arXiv:1907.07688 [hep-th]}%
  \bibAnnoteFile{NoStop}{bryzz}%
\bibitem{Balthazar:2019ypi}%
  \BibitemOpen
  \bibfield{author}{%
  \bibinfo {author} {\bibfnamefont{Bruno}\ \bibnamefont{Balthazar}}, \bibinfo
  {author} {\bibfnamefont{Victor~A.}\ \bibnamefont{Rodriguez}},\ and\ \bibinfo
  {author} {\bibfnamefont{Xi}~\bibnamefont{Yin}},\ }%
  \bibfield{title}{%
  \enquote{\bibinfo {title} {{Multi-Instanton Calculus in $c = 1$ String
  Theory}},}\ }%
   (\bibinfo {month} {12}\ \bibinfo {year} {2019}),\
  \Eprint{http://arxiv.org/abs/1912.07170}{arXiv:1912.07170 [hep-th]}%
  \bibAnnoteFile{NoStop}{Balthazar:2019ypi}%
\bibitem{SenNormalization}%
  \BibitemOpen
  \bibfield{author}{%
  \bibinfo {author} {\bibfnamefont{Ashoke}\ \bibnamefont{Sen}},\ }%
  \bibfield{title}{%
  \enquote{\bibinfo {title} {{Normalization of D-instanton amplitudes}},}\ }%
  \bibfield{journal}{%
  \Doi{10.1007/JHEP11(2021)077}{\bibinfo {journal} {JHEP}}\ }%
  \textbf{\bibinfo {volume} {11}},\ \bibinfo {pages} {077} (\bibinfo {year}
  {2021}),\ \Eprint{http://arxiv.org/abs/2101.08566}{arXiv:2101.08566
  [hep-th]}%
  \bibAnnoteFile{NoStop}{SenNormalization}%
\bibitem{Balthazar:2022apu}%
  \BibitemOpen
  \bibfield{author}{%
  \bibinfo {author} {\bibfnamefont{Bruno}\ \bibnamefont{Balthazar}}, \bibinfo
  {author} {\bibfnamefont{Victor~A.}\ \bibnamefont{Rodriguez}},\ and\ \bibinfo
  {author} {\bibfnamefont{Xi}~\bibnamefont{Yin}},\ }%
  \bibfield{title}{%
  \enquote{\bibinfo {title} {{The S-Matrix of 2D Type 0B String Theory Part 2:
  D-Instanton Effects}},}\ }%
   (\bibinfo {month} {4}\ \bibinfo {year} {2022}),\
  \Eprint{http://arxiv.org/abs/2204.01747}{arXiv:2204.01747 [hep-th]}%
  \bibAnnoteFile{NoStop}{Balthazar:2022apu}%
\bibitem{Sen:2021tpp}%
  \BibitemOpen
  \bibfield{author}{%
  \bibinfo {author} {\bibfnamefont{Ashoke}\ \bibnamefont{Sen}},\ }%
  \bibfield{title}{%
  \enquote{\bibinfo {title} {{Normalization of type IIB D-instanton
  amplitudes}},}\ }%
  \bibfield{journal}{%
  \Doi{10.1007/JHEP12(2021)146}{\bibinfo {journal} {JHEP}}\ }%
  \textbf{\bibinfo {volume} {12}},\ \bibinfo {pages} {146} (\bibinfo {year}
  {2021}),\ \Eprint{http://arxiv.org/abs/2104.11109}{arXiv:2104.11109
  [hep-th]}%
  \bibAnnoteFile{NoStop}{Sen:2021tpp}%
\bibitem{Alexandrov:2021shf}%
  \BibitemOpen
  \bibfield{author}{%
  \bibinfo {author} {\bibfnamefont{Sergei}\ \bibnamefont{Alexandrov}}, \bibinfo
  {author} {\bibfnamefont{Ashoke}\ \bibnamefont{Sen}},\ and\ \bibinfo {author}
  {\bibfnamefont{Bogdan}\ \bibnamefont{Stefanski}},\ }%
  \bibfield{title}{%
  \enquote{\bibinfo {title} {{D-instantons in Type IIA string theory on
  Calabi-Yau threefolds}},}\ }%
  \bibfield{journal}{%
  \Doi{10.1007/JHEP11(2021)018}{\bibinfo {journal} {JHEP}}\ }%
  \textbf{\bibinfo {volume} {11}},\ \bibinfo {pages} {018} (\bibinfo {year}
  {2021}),\ \Eprint{http://arxiv.org/abs/2108.04265}{arXiv:2108.04265
  [hep-th]}%
  \bibAnnoteFile{NoStop}{Alexandrov:2021shf}%
\bibitem{Alexandrov:2021dyl}%
  \BibitemOpen
  \bibfield{author}{%
  \bibinfo {author} {\bibfnamefont{Sergei}\ \bibnamefont{Alexandrov}}, \bibinfo
  {author} {\bibfnamefont{Ashoke}\ \bibnamefont{Sen}},\ and\ \bibinfo {author}
  {\bibfnamefont{Bogdan}\ \bibnamefont{Stefanski}},\ }%
  \bibfield{title}{%
  \enquote{\bibinfo {title} {{Euclidean D-branes in type IIB string theory on
  Calabi-Yau threefolds}},}\ }%
  \bibfield{journal}{%
  \Doi{10.1007/JHEP12(2021)044}{\bibinfo {journal} {JHEP}}\ }%
  \textbf{\bibinfo {volume} {12}},\ \bibinfo {pages} {044} (\bibinfo {year}
  {2021}),\ \Eprint{http://arxiv.org/abs/2110.06949}{arXiv:2110.06949
  [hep-th]}%
  \bibAnnoteFile{NoStop}{Alexandrov:2021dyl}%
\bibitem{Alexandrov:2022mmy}%
  \BibitemOpen
  \bibfield{author}{%
  \bibinfo {author} {\bibfnamefont{Sergei}\ \bibnamefont{Alexandrov}}, \bibinfo
  {author} {\bibfnamefont{Atakan~Hilmi}\ \bibnamefont{Firat}}, \bibinfo
  {author} {\bibfnamefont{Manki}\ \bibnamefont{Kim}}, \bibinfo {author}
  {\bibfnamefont{Ashoke}\ \bibnamefont{Sen}},\ and\ \bibinfo {author}
  {\bibfnamefont{Bogdan}\ \bibnamefont{Stefanski}},\ }%
  \bibfield{title}{%
  \enquote{\bibinfo {title} {{D-instanton Induced Superpotential}},}\ }%
   (\bibinfo {month} {4}\ \bibinfo {year} {2022}),\
  \Eprint{http://arxiv.org/abs/2204.02981}{arXiv:2204.02981 [hep-th]}%
  \bibAnnoteFile{NoStop}{Alexandrov:2022mmy}%
\bibitem{Agmon:2022vdj}%
  \BibitemOpen
  \bibfield{author}{%
  \bibinfo {author} {\bibfnamefont{Nathan~B.}\ \bibnamefont{Agmon}}, \bibinfo
  {author} {\bibfnamefont{Bruno}\ \bibnamefont{Balthazar}}, \bibinfo {author}
  {\bibfnamefont{Minjae}\ \bibnamefont{Cho}}, \bibinfo {author}
  {\bibfnamefont{Victor~A.}\ \bibnamefont{Rodriguez}},\ and\ \bibinfo {author}
  {\bibfnamefont{Xi}~\bibnamefont{Yin}},\ }%
  \bibfield{title}{%
  \enquote{\bibinfo {title} {{D-instanton Effects in Type IIB String
  Theory}},}\ }%
   (\bibinfo {month} {5}\ \bibinfo {year} {2022}),\
  \Eprint{http://arxiv.org/abs/2205.00609}{arXiv:2205.00609 [hep-th]}%
  \bibAnnoteFile{NoStop}{Agmon:2022vdj}%
\bibitem{Eniceicu:2022nay}%
  \BibitemOpen
  \bibfield{author}{%
  \bibinfo {author} {\bibfnamefont{Dan~Stefan}\ \bibnamefont{Eniceicu}},
  \bibinfo {author} {\bibfnamefont{Raghu}\ \bibnamefont{Mahajan}}, \bibinfo
  {author} {\bibfnamefont{Chitraang}\ \bibnamefont{Murdia}},\ and\ \bibinfo
  {author} {\bibfnamefont{Ashoke}\ \bibnamefont{Sen}},\ }%
  \bibfield{title}{%
  \enquote{\bibinfo {title} {{Normalization of ZZ instanton amplitudes in
  minimal string theory}},}\ }%
   (\bibinfo {month} {2}\ \bibinfo {year} {2022}),\
  \Eprint{http://arxiv.org/abs/2202.03448}{arXiv:2202.03448 [hep-th]}%
  \bibAnnoteFile{NoStop}{Eniceicu:2022nay}%
\bibitem{Martinec:2003ka}%
  \BibitemOpen
  \bibfield{author}{%
  \bibinfo {author} {\bibfnamefont{Emil~J.}\ \bibnamefont{Martinec}},\ }%
  \bibfield{title}{%
  \enquote{\bibinfo {title} {{The Annular report on noncritical string
  theory}},}\ }%
   (\bibinfo {month} {5}\ \bibinfo {year} {2003}),\
  \Eprint{http://arxiv.org/abs/hep-th/0305148}{arXiv:hep-th/0305148}%
  \bibAnnoteFile{NoStop}{Martinec:2003ka}%
\bibitem{seibergannulus}%
  \BibitemOpen
  \bibfield{author}{%
  \bibinfo {author} {\bibfnamefont{David}\ \bibnamefont{Kutasov}}, \bibinfo
  {author} {\bibfnamefont{Kazumi}\ \bibnamefont{Okuyama}}, \bibinfo {author}
  {\bibfnamefont{Jong-won}\ \bibnamefont{Park}}, \bibinfo {author}
  {\bibfnamefont{Nathan}\ \bibnamefont{Seiberg}},\ and\ \bibinfo {author}
  {\bibfnamefont{David}\ \bibnamefont{Shih}},\ }%
  \bibfield{title}{%
  \enquote{\bibinfo {title} {{Annulus amplitudes and ZZ branes in minimal
  string theory}},}\ }%
  \bibfield{journal}{%
  \Doi{10.1088/1126-6708/2004/08/026}{\bibinfo {journal} {JHEP}}\ }%
  \textbf{\bibinfo {volume} {08}},\ \bibinfo {pages} {026} (\bibinfo {year}
  {2004}),\ \Eprint{http://arxiv.org/abs/hep-th/0406030}{arXiv:hep-th/0406030}%
  \bibAnnoteFile{NoStop}{seibergannulus}%
\bibitem{David:1992za}%
  \BibitemOpen
  \bibfield{author}{%
  \bibinfo {author} {\bibfnamefont{Francois}\ \bibnamefont{David}},\ }%
  \bibfield{title}{%
  \enquote{\bibinfo {title} {{Nonperturbative effects in matrix models and
  vacua of two-dimensional gravity}},}\ }%
  \bibfield{journal}{%
  \Doi{10.1016/0370-2693(93)90417-G}{\bibinfo {journal} {Phys. Lett. B}}\ }%
  \textbf{\bibinfo {volume} {302}},\ \bibinfo {pages} {403--410} (\bibinfo
  {year} {1993}),\
  \Eprint{http://arxiv.org/abs/hep-th/9212106}{arXiv:hep-th/9212106}%
  \bibAnnoteFile{NoStop}{David:1992za}%
\bibitem{Sato:2004tz}%
  \BibitemOpen
  \bibfield{author}{%
  \bibinfo {author} {\bibfnamefont{Akira}\ \bibnamefont{Sato}}\ and\ \bibinfo
  {author} {\bibfnamefont{Asato}\ \bibnamefont{Tsuchiya}},\ }%
  \bibfield{title}{%
  \enquote{\bibinfo {title} {{ZZ brane amplitudes from matrix models}},}\ }%
  \bibfield{journal}{%
  \Doi{10.1088/1126-6708/2005/02/032}{\bibinfo {journal} {JHEP}}\ }%
  \textbf{\bibinfo {volume} {02}},\ \bibinfo {pages} {032} (\bibinfo {year}
  {2005}),\ \Eprint{http://arxiv.org/abs/hep-th/0412201}{arXiv:hep-th/0412201}%
  \bibAnnoteFile{NoStop}{Sato:2004tz}%
\bibitem{Hanada:2004im}%
  \BibitemOpen
  \bibfield{author}{%
  \bibinfo {author} {\bibfnamefont{Masanori}\ \bibnamefont{Hanada}}, \bibinfo
  {author} {\bibfnamefont{Masashi}\ \bibnamefont{Hayakawa}}, \bibinfo {author}
  {\bibfnamefont{Nobuyuki}\ \bibnamefont{Ishibashi}}, \bibinfo {author}
  {\bibfnamefont{Hikaru}\ \bibnamefont{Kawai}}, \bibinfo {author}
  {\bibfnamefont{Tsunehide}\ \bibnamefont{Kuroki}}, \bibinfo {author}
  {\bibfnamefont{Yoshinori}\ \bibnamefont{Matsuo}},\ and\ \bibinfo {author}
  {\bibfnamefont{Tsukasa}\ \bibnamefont{Tada}},\ }%
  \bibfield{title}{%
  \enquote{\bibinfo {title} {{Loops versus matrices: The Nonperturbative
  aspects of noncritical string}},}\ }%
  \bibfield{journal}{%
  \Doi{10.1143/PTP.112.131}{\bibinfo {journal} {Prog. Theor. Phys.}}\ }%
  \textbf{\bibinfo {volume} {112}},\ \bibinfo {pages} {131--181} (\bibinfo
  {year} {2004}),\
  \Eprint{http://arxiv.org/abs/hep-th/0405076}{arXiv:hep-th/0405076}%
  \bibAnnoteFile{NoStop}{Hanada:2004im}%
\bibitem{Ishibashi:2005dh}%
  \BibitemOpen
  \bibfield{author}{%
  \bibinfo {author} {\bibfnamefont{Nobuyuki}\ \bibnamefont{Ishibashi}}\ and\
  \bibinfo {author} {\bibfnamefont{Atsushi}\ \bibnamefont{Yamaguchi}},\ }%
  \bibfield{title}{%
  \enquote{\bibinfo {title} {{On the chemical potential of D-instantons in c=0
  noncritical string theory}},}\ }%
  \bibfield{journal}{%
  \Doi{10.1088/1126-6708/2005/06/082}{\bibinfo {journal} {JHEP}}\ }%
  \textbf{\bibinfo {volume} {06}},\ \bibinfo {pages} {082} (\bibinfo {year}
  {2005}),\ \Eprint{http://arxiv.org/abs/hep-th/0503199}{arXiv:hep-th/0503199}%
  \bibAnnoteFile{NoStop}{Ishibashi:2005dh}%
\bibitem{Ishibashi:2005zf}%
  \BibitemOpen
  \bibfield{author}{%
  \bibinfo {author} {\bibfnamefont{Nobuyuki}\ \bibnamefont{Ishibashi}},
  \bibinfo {author} {\bibfnamefont{Tsunehide}\ \bibnamefont{Kuroki}},\ and\
  \bibinfo {author} {\bibfnamefont{Atsushi}\ \bibnamefont{Yamaguchi}},\ }%
  \bibfield{title}{%
  \enquote{\bibinfo {title} {{Universality of nonperturbative effects in
  c\ensuremath{<}1 noncritical string theory}},}\ }%
  \bibfield{journal}{%
  \Doi{10.1088/1126-6708/2005/09/043}{\bibinfo {journal} {JHEP}}\ }%
  \textbf{\bibinfo {volume} {09}},\ \bibinfo {pages} {043} (\bibinfo {year}
  {2005}),\ \Eprint{http://arxiv.org/abs/hep-th/0507263}{arXiv:hep-th/0507263}%
  \bibAnnoteFile{NoStop}{Ishibashi:2005zf}%
\bibitem{Marino:2007te}%
  \BibitemOpen
  \bibfield{author}{%
  \bibinfo {author} {\bibfnamefont{Marcos}\ \bibnamefont{Marino}}, \bibinfo
  {author} {\bibfnamefont{Ricardo}\ \bibnamefont{Schiappa}},\ and\ \bibinfo
  {author} {\bibfnamefont{Marlene}\ \bibnamefont{Weiss}},\ }%
  \bibfield{title}{%
  \enquote{\bibinfo {title} {{Nonperturbative Effects and the Large-Order
  Behavior of Matrix Models and Topological Strings}},}\ }%
  \bibfield{journal}{%
  \Doi{10.4310/CNTP.2008.v2.n2.a3}{\bibinfo {journal} {Commun. Num. Theor.
  Phys.}}\ }%
  \textbf{\bibinfo {volume} {2}},\ \bibinfo {pages} {349--419} (\bibinfo {year}
  {2008}),\ \Eprint{http://arxiv.org/abs/0711.1954}{arXiv:0711.1954 [hep-th]}%
  \bibAnnoteFile{NoStop}{Marino:2007te}%
\bibitem{Marino:2008vx}%
  \BibitemOpen
  \bibfield{author}{%
  \bibinfo {author} {\bibfnamefont{Marcos}\ \bibnamefont{Marino}}, \bibinfo
  {author} {\bibfnamefont{Ricardo}\ \bibnamefont{Schiappa}},\ and\ \bibinfo
  {author} {\bibfnamefont{Marlene}\ \bibnamefont{Weiss}},\ }%
  \bibfield{title}{%
  \enquote{\bibinfo {title} {{Multi-Instantons and Multi-Cuts}},}\ }%
  \bibfield{journal}{%
  \Doi{10.1063/1.3097755}{\bibinfo {journal} {J. Math. Phys.}}\ }%
  \textbf{\bibinfo {volume} {50}},\ \bibinfo {pages} {052301} (\bibinfo {year}
  {2009}),\ \Eprint{http://arxiv.org/abs/0809.2619}{arXiv:0809.2619 [hep-th]}%
  \bibAnnoteFile{NoStop}{Marino:2008vx}%
\bibitem{sss}%
  \BibitemOpen
  \bibfield{author}{%
  \bibinfo {author} {\bibfnamefont{Phil}\ \bibnamefont{Saad}}, \bibinfo
  {author} {\bibfnamefont{Stephen~H.}\ \bibnamefont{Shenker}},\ and\ \bibinfo
  {author} {\bibfnamefont{Douglas}\ \bibnamefont{Stanford}},\ }%
  \bibfield{title}{%
  \enquote{\bibinfo {title} {{JT gravity as a matrix integral}},}\ }%
   (\bibinfo {month} {3}\ \bibinfo {year} {2019}),\
  \Eprint{http://arxiv.org/abs/1903.11115}{arXiv:1903.11115 [hep-th]}%
  \bibAnnoteFile{NoStop}{sss}%
\bibitem{Marino:2012zq}%
  \BibitemOpen
  \bibfield{author}{%
  \bibinfo {author} {\bibfnamefont{Marcos}\ \bibnamefont{Mari\~no}},\ }%
  \bibfield{title}{%
  \enquote{\bibinfo {title} {{Lectures on non-perturbative effects in large $N$
  gauge theories, matrix models and strings}},}\ }%
  \bibfield{journal}{%
  \Doi{10.1002/prop.201400005}{\bibinfo {journal} {Fortsch. Phys.}}\ }%
  \textbf{\bibinfo {volume} {62}},\ \bibinfo {pages} {455--540} (\bibinfo
  {year} {2014}),\ \Eprint{http://arxiv.org/abs/1206.6272}{arXiv:1206.6272
  [hep-th]}%
  \bibAnnoteFile{NoStop}{Marino:2012zq}%
\bibitem{Dunne:2015eaa}%
  \BibitemOpen
  \bibfield{author}{%
  \bibinfo {author} {\bibfnamefont{Gerald~V.}\ \bibnamefont{Dunne}}\ and\
  \bibinfo {author} {\bibfnamefont{Mithat}\ \bibnamefont{\"Unsal}},\ }%
  \bibfield{title}{%
  \enquote{\bibinfo {title} {{What is QFT? Resurgent trans-series, Lefschetz
  thimbles, and new exact saddles}},}\ }%
  \bibfield{journal}{%
  \Doi{10.22323/1.251.0010}{\bibinfo {journal} {PoS}}\ }%
  \textbf{\bibinfo {volume} {LATTICE2015}},\ \bibinfo {pages} {010} (\bibinfo
  {year} {2016}),\ \Eprint{http://arxiv.org/abs/1511.05977}{arXiv:1511.05977
  [hep-lat]}%
  \bibAnnoteFile{NoStop}{Dunne:2015eaa}%
\bibitem{Aniceto:2018bis}%
  \BibitemOpen
  \bibfield{author}{%
  \bibinfo {author} {\bibfnamefont{In\^es}\ \bibnamefont{Aniceto}}, \bibinfo
  {author} {\bibfnamefont{Gokce}\ \bibnamefont{Basar}},\ and\ \bibinfo {author}
  {\bibfnamefont{Ricardo}\ \bibnamefont{Schiappa}},\ }%
  \bibfield{title}{%
  \enquote{\bibinfo {title} {{A Primer on Resurgent Transseries and Their
  Asymptotics}},}\ }%
  \bibfield{journal}{%
  \Doi{10.1016/j.physrep.2019.02.003}{\bibinfo {journal} {Phys. Rept.}}\ }%
  \textbf{\bibinfo {volume} {809}},\ \bibinfo {pages} {1--135} (\bibinfo {year}
  {2019}),\ \Eprint{http://arxiv.org/abs/1802.10441}{arXiv:1802.10441
  [hep-th]}%
  \bibAnnoteFile{NoStop}{Aniceto:2018bis}%
\bibitem{Sen:1999xm}%
  \BibitemOpen
  \bibfield{author}{%
  \bibinfo {author} {\bibfnamefont{Ashoke}\ \bibnamefont{Sen}},\ }%
  \bibfield{title}{%
  \enquote{\bibinfo {title} {{Universality of the tachyon potential}},}\ }%
  \bibfield{journal}{%
  \Doi{10.1088/1126-6708/1999/12/027}{\bibinfo {journal} {JHEP}}\ }%
  \textbf{\bibinfo {volume} {12}},\ \bibinfo {pages} {027} (\bibinfo {year}
  {1999}),\ \Eprint{http://arxiv.org/abs/hep-th/9911116}{arXiv:hep-th/9911116}%
  \bibAnnoteFile{NoStop}{Sen:1999xm}%
\bibitem{Sen:1999nx}%
  \BibitemOpen
  \bibfield{author}{%
  \bibinfo {author} {\bibfnamefont{Ashoke}\ \bibnamefont{Sen}}\ and\ \bibinfo
  {author} {\bibfnamefont{Barton}\ \bibnamefont{Zwiebach}},\ }%
  \bibfield{title}{%
  \enquote{\bibinfo {title} {{Tachyon condensation in string field theory}},}\
  }%
  \bibfield{journal}{%
  \Doi{10.1088/1126-6708/2000/03/002}{\bibinfo {journal} {JHEP}}\ }%
  \textbf{\bibinfo {volume} {03}},\ \bibinfo {pages} {002} (\bibinfo {year}
  {2000}),\ \Eprint{http://arxiv.org/abs/hep-th/9912249}{arXiv:hep-th/9912249}%
  \bibAnnoteFile{NoStop}{Sen:1999nx}%
\bibitem{Schnabl:2005gv}%
  \BibitemOpen
  \bibfield{author}{%
  \bibinfo {author} {\bibfnamefont{Martin}\ \bibnamefont{Schnabl}},\ }%
  \bibfield{title}{%
  \enquote{\bibinfo {title} {{Analytic solution for tachyon condensation in
  open string field theory}},}\ }%
  \bibfield{journal}{%
  \Doi{10.4310/ATMP.2006.v10.n4.a1}{\bibinfo {journal} {Adv. Theor. Math.
  Phys.}}\ }%
  \textbf{\bibinfo {volume} {10}},\ \bibinfo {pages} {433--501} (\bibinfo
  {year} {2006}),\
  \Eprint{http://arxiv.org/abs/hep-th/0511286}{arXiv:hep-th/0511286}%
  \bibAnnoteFile{NoStop}{Schnabl:2005gv}%
\bibitem{Erler:2019fye}%
  \BibitemOpen
  \bibfield{author}{%
  \bibinfo {author} {\bibfnamefont{Theodore}\ \bibnamefont{Erler}}\ and\
  \bibinfo {author} {\bibfnamefont{Carlo}\ \bibnamefont{Maccaferri}},\ }%
  \bibfield{title}{%
  \enquote{\bibinfo {title} {{String field theory solution for any open string
  background. Part II}},}\ }%
  \bibfield{journal}{%
  \Doi{10.1007/JHEP01(2020)021}{\bibinfo {journal} {JHEP}}\ }%
  \textbf{\bibinfo {volume} {01}},\ \bibinfo {pages} {021} (\bibinfo {year}
  {2020}),\ \Eprint{http://arxiv.org/abs/1909.11675}{arXiv:1909.11675
  [hep-th]}%
  \bibAnnoteFile{NoStop}{Erler:2019fye}%
\bibitem{Sen:1998tt}%
  \BibitemOpen
  \bibfield{author}{%
  \bibinfo {author} {\bibfnamefont{Ashoke}\ \bibnamefont{Sen}},\ }%
  \bibfield{title}{%
  \enquote{\bibinfo {title} {{SO(32) spinors of type I and other solitons on
  brane - anti-brane pair}},}\ }%
  \bibfield{journal}{%
  \Doi{10.1088/1126-6708/1998/09/023}{\bibinfo {journal} {JHEP}}\ }%
  \textbf{\bibinfo {volume} {09}},\ \bibinfo {pages} {023} (\bibinfo {year}
  {1998}),\ \Eprint{http://arxiv.org/abs/hep-th/9808141}{arXiv:hep-th/9808141}%
  \bibAnnoteFile{NoStop}{Sen:1998tt}%
\bibitem{Witten:1998cd}%
  \BibitemOpen
  \bibfield{author}{%
  \bibinfo {author} {\bibfnamefont{Edward}\ \bibnamefont{Witten}},\ }%
  \bibfield{title}{%
  \enquote{\bibinfo {title} {{D-branes and K theory}},}\ }%
  \bibfield{journal}{%
  \Doi{10.1088/1126-6708/1998/12/019}{\bibinfo {journal} {JHEP}}\ }%
  \textbf{\bibinfo {volume} {12}},\ \bibinfo {pages} {019} (\bibinfo {year}
  {1998}),\ \Eprint{http://arxiv.org/abs/hep-th/9810188}{arXiv:hep-th/9810188}%
  \bibAnnoteFile{NoStop}{Witten:1998cd}%
\bibitem{Berkovits:2000hf}%
  \BibitemOpen
  \bibfield{author}{%
  \bibinfo {author} {\bibfnamefont{Nathan}\ \bibnamefont{Berkovits}}, \bibinfo
  {author} {\bibfnamefont{Ashoke}\ \bibnamefont{Sen}},\ and\ \bibinfo {author}
  {\bibfnamefont{Barton}\ \bibnamefont{Zwiebach}},\ }%
  \bibfield{title}{%
  \enquote{\bibinfo {title} {{Tachyon condensation in superstring field
  theory}},}\ }%
  \bibfield{journal}{%
  \Doi{10.1016/S0550-3213(00)00501-0}{\bibinfo {journal} {Nucl. Phys. B}}\ }%
  \textbf{\bibinfo {volume} {587}},\ \bibinfo {pages} {147--178} (\bibinfo
  {year} {2000}),\
  \Eprint{http://arxiv.org/abs/hep-th/0002211}{arXiv:hep-th/0002211}%
  \bibAnnoteFile{NoStop}{Berkovits:2000hf}%
\bibitem{Kazakov:2004du}%
  \BibitemOpen
  \bibfield{author}{%
  \bibinfo {author} {\bibfnamefont{Vladimir~A.}\ \bibnamefont{Kazakov}}\ and\
  \bibinfo {author} {\bibfnamefont{Ivan~K.}\ \bibnamefont{Kostov}},\ }%
  \enquote{\bibinfo {title} {{Instantons in noncritical strings from the two
  matrix model}},}\ in\ \Doi{10.1142/9789812775344_0045}{\emph{\bibinfo
  {booktitle} {{From Fields to Strings: Circumnavigating Theoretical Physics: A
  Conference in Tribute to Ian Kogan}}}}\ (\bibinfo {year} {2004})\ pp.\
  \bibinfo {pages} {1864--1894},\
  \Eprint{http://arxiv.org/abs/hep-th/0403152}{arXiv:hep-th/0403152}%
  \bibAnnoteFile{NoStop}{Kazakov:2004du}%
\bibitem{Schiappa:2013opa}%
  \BibitemOpen
  \bibfield{author}{%
  \bibinfo {author} {\bibfnamefont{Ricardo}\ \bibnamefont{Schiappa}}\ and\
  \bibinfo {author} {\bibfnamefont{Ricardo}\ \bibnamefont{Vaz}},\ }%
  \bibfield{title}{%
  \enquote{\bibinfo {title} {{The Resurgence of Instantons: Multi-Cut Stokes
  Phases and the Painleve II Equation}},}\ }%
  \bibfield{journal}{%
  \Doi{10.1007/s00220-014-2028-7}{\bibinfo {journal} {Commun. Math. Phys.}}\ }%
  \textbf{\bibinfo {volume} {330}},\ \bibinfo {pages} {655--721} (\bibinfo
  {year} {2014}),\ \Eprint{http://arxiv.org/abs/1302.5138}{arXiv:1302.5138
  [hep-th]}%
  \bibAnnoteFile{NoStop}{Schiappa:2013opa}%
\bibitem{Sen:2021jbr}%
  \BibitemOpen
  \bibfield{author}{%
  \bibinfo {author} {\bibfnamefont{Ashoke}\ \bibnamefont{Sen}},\ }%
  \bibfield{title}{%
  \enquote{\bibinfo {title} {{Muti-instanton amplitudes in type IIB string
  theory}},}\ }%
  \bibfield{journal}{%
  \Doi{10.1007/JHEP12(2021)065}{\bibinfo {journal} {JHEP}}\ }%
  \textbf{\bibinfo {volume} {12}},\ \bibinfo {pages} {065} (\bibinfo {year}
  {2021}),\ \Eprint{http://arxiv.org/abs/2104.15110}{arXiv:2104.15110
  [hep-th]}%
  \bibAnnoteFile{NoStop}{Sen:2021jbr}%
\bibitem{Cardy89}%
  \BibitemOpen
  \bibfield{author}{%
  \bibinfo {author} {\bibfnamefont{John~L.}\ \bibnamefont{Cardy}},\ }%
  \bibfield{title}{%
  \enquote{\bibinfo {title} {{Boundary Conditions, Fusion Rules and the
  Verlinde Formula}},}\ }%
  \bibfield{journal}{%
  \Doi{10.1016/0550-3213(89)90521-X}{\bibinfo {journal} {Nucl. Phys. B}}\ }%
  \textbf{\bibinfo {volume} {324}},\ \bibinfo {pages} {581--596} (\bibinfo
  {year} {1989})%
  \bibAnnoteFile{NoStop}{Cardy89}%
\bibitem{Seiberg:2003nm}%
  \BibitemOpen
  \bibfield{author}{%
  \bibinfo {author} {\bibfnamefont{Nathan}\ \bibnamefont{Seiberg}}\ and\
  \bibinfo {author} {\bibfnamefont{David}\ \bibnamefont{Shih}},\ }%
  \bibfield{title}{%
  \enquote{\bibinfo {title} {{Branes, rings and matrix models in minimal
  (super)string theory}},}\ }%
  \bibfield{journal}{%
  \Doi{10.1088/1126-6708/2004/02/021}{\bibinfo {journal} {JHEP}}\ }%
  \textbf{\bibinfo {volume} {02}},\ \bibinfo {pages} {021} (\bibinfo {year}
  {2004}),\ \Eprint{http://arxiv.org/abs/hep-th/0312170}{arXiv:hep-th/0312170}%
  \bibAnnoteFile{NoStop}{Seiberg:2003nm}%
\bibitem{Douglas:1990pt}%
  \BibitemOpen
  \bibfield{author}{%
  \bibinfo {author} {\bibfnamefont{Michael~R.}\ \bibnamefont{Douglas}},\ }%
  \enquote{\bibinfo {title} {{The Two matrix model}},}\ in\ \emph{\bibinfo
  {booktitle} {{Cargese Study Institute: Random Surfaces, Quantum Gravity and
  Strings}}}\ (\bibinfo {year} {1990})%
  \bibAnnoteFile{NoStop}{Douglas:1990pt}%
\bibitem{Daul:1993bg}%
  \BibitemOpen
  \bibfield{author}{%
  \bibinfo {author} {\bibfnamefont{J.~M.}\ \bibnamefont{Daul}}, \bibinfo
  {author} {\bibfnamefont{V.~A.}\ \bibnamefont{Kazakov}},\ and\ \bibinfo
  {author} {\bibfnamefont{I.~K.}\ \bibnamefont{Kostov}},\ }%
  \bibfield{title}{%
  \enquote{\bibinfo {title} {{Rational theories of 2-D gravity from the two
  matrix model}},}\ }%
  \bibfield{journal}{%
  \Doi{10.1016/0550-3213(93)90582-A}{\bibinfo {journal} {Nucl. Phys. B}}\ }%
  \textbf{\bibinfo {volume} {409}},\ \bibinfo {pages} {311--338} (\bibinfo
  {year} {1993}),\
  \Eprint{http://arxiv.org/abs/hep-th/9303093}{arXiv:hep-th/9303093}%
  \bibAnnoteFile{NoStop}{Daul:1993bg}%
\bibitem{DiFrancesco:1997nk}%
  \BibitemOpen
  \bibfield{author}{%
  \bibinfo {author} {\bibfnamefont{P.}~\bibnamefont{Di~Francesco}}, \bibinfo
  {author} {\bibfnamefont{P.}~\bibnamefont{Mathieu}},\ and\ \bibinfo {author}
  {\bibfnamefont{D.}~\bibnamefont{Senechal}},\ }%
  \Doi{10.1007/978-1-4612-2256-9}{\emph{\bibinfo {title} {{Conformal Field
  Theory}}}},\ Graduate Texts in Contemporary Physics\ (\bibinfo {publisher}
  {Springer-Verlag},\ \bibinfo {address} {New York},\ \bibinfo {year} {1997})\
  ISBN \bibinfo {isbn} {978-0-387-94785-3, 978-1-4612-7475-9}%
  \bibAnnoteFile{NoStop}{DiFrancesco:1997nk}%
\bibitem{Polchinski:1998rq}%
  \BibitemOpen
  \bibfield{author}{%
  \bibinfo {author} {\bibfnamefont{J.}~\bibnamefont{Polchinski}},\ }%
  \Doi{10.1017/CBO9780511816079}{\emph{\bibinfo {title} {{String theory. Vol.
  1: An introduction to the bosonic string}}}},\ Cambridge Monographs on
  Mathematical Physics\ (\bibinfo {publisher} {Cambridge University Press},\
  \bibinfo {year} {2007})%
  \bibAnnoteFile{NoStop}{Polchinski:1998rq}%
\bibitem{Witten:1985cc}%
  \BibitemOpen
  \bibfield{author}{%
  \bibinfo {author} {\bibfnamefont{Edward}\ \bibnamefont{Witten}},\ }%
  \bibfield{title}{%
  \enquote{\bibinfo {title} {{Noncommutative Geometry and String Field
  Theory}},}\ }%
  \bibfield{journal}{%
  \Doi{10.1016/0550-3213(86)90155-0}{\bibinfo {journal} {Nucl. Phys. B}}\ }%
  \textbf{\bibinfo {volume} {268}},\ \bibinfo {pages} {253--294} (\bibinfo
  {year} {1986})%
  \bibAnnoteFile{NoStop}{Witten:1985cc}%
\bibitem{marino2005chern}%
  \BibitemOpen
  \bibfield{author}{%
  \bibinfo {author} {\bibfnamefont{Marcos}\ \bibnamefont{Marino}},\ }%
  \emph{\bibinfo {title} {Chern-Simons Theory, Matrix Models, and Topological
  Strings}},\ International series of monographs on physics\ (\bibinfo
  {publisher} {Clarendon Press},\ \bibinfo {year} {2005})\ ISBN \bibinfo {isbn}
  {9780198568490}%
  \bibAnnoteFile{NoStop}{marino2005chern}%
\bibitem{Brezin:1977sv}%
  \BibitemOpen
  \bibfield{author}{%
  \bibinfo {author} {\bibfnamefont{E.}~\bibnamefont{Brezin}}, \bibinfo {author}
  {\bibfnamefont{C.}~\bibnamefont{Itzykson}}, \bibinfo {author}
  {\bibfnamefont{G.}~\bibnamefont{Parisi}},\ and\ \bibinfo {author}
  {\bibfnamefont{J.~B.}\ \bibnamefont{Zuber}},\ }%
  \bibfield{title}{%
  \enquote{\bibinfo {title} {{Planar Diagrams}},}\ }%
  \bibfield{journal}{%
  \Doi{10.1007/BF01614153}{\bibinfo {journal} {Commun. Math. Phys.}}\ }%
  \textbf{\bibinfo {volume} {59}},\ \bibinfo {pages} {35} (\bibinfo {year}
  {1978})%
  \bibAnnoteFile{NoStop}{Brezin:1977sv}%
\bibitem{Bessis:1980ss}%
  \BibitemOpen
  \bibfield{author}{%
  \bibinfo {author} {\bibfnamefont{D.}~\bibnamefont{Bessis}}, \bibinfo {author}
  {\bibfnamefont{C.}~\bibnamefont{Itzykson}},\ and\ \bibinfo {author}
  {\bibfnamefont{J.~B.}\ \bibnamefont{Zuber}},\ }%
  \bibfield{title}{%
  \enquote{\bibinfo {title} {{Quantum field theory techniques in graphical
  enumeration}},}\ }%
  \bibfield{journal}{%
  \Doi{10.1016/0196-8858(80)90008-1}{\bibinfo {journal} {Adv. Appl. Math.}}\ }%
  \textbf{\bibinfo {volume} {1}},\ \bibinfo {pages} {109--157} (\bibinfo {year}
  {1980})%
  \bibAnnoteFile{NoStop}{Bessis:1980ss}%
\bibitem{ercolani2003asymptotics}%
  \BibitemOpen
  \bibfield{author}{%
  \bibinfo {author} {\bibfnamefont{Nicholas~M}\ \bibnamefont{Ercolani}}\ and\
  \bibinfo {author} {\bibfnamefont{KDT-R}\ \bibnamefont{McLaughlin}},\ }%
  \bibfield{title}{%
  \enquote{\bibinfo {title} {Asymptotics of the partition function for random
  matrices via riemann-hilbert techniques and applications to graphical
  enumeration},}\ }%
  \bibfield{journal}{%
  \bibinfo {journal} {International Mathematics Research Notices}\ }%
  \textbf{\bibinfo {volume} {2003}},\ \bibinfo {pages} {755--820} (\bibinfo
  {year} {2003})%
  \bibAnnoteFile{NoStop}{ercolani2003asymptotics}%
\bibitem{Moore:1991ir}%
  \BibitemOpen
  \bibfield{author}{%
  \bibinfo {author} {\bibfnamefont{Gregory~W.}\ \bibnamefont{Moore}}, \bibinfo
  {author} {\bibfnamefont{Nathan}\ \bibnamefont{Seiberg}},\ and\ \bibinfo
  {author} {\bibfnamefont{Matthias}\ \bibnamefont{Staudacher}},\ }%
  \bibfield{title}{%
  \enquote{\bibinfo {title} {{From loops to states in 2-D quantum gravity}},}\
  }%
  \bibfield{journal}{%
  \Doi{10.1016/0550-3213(91)90548-C}{\bibinfo {journal} {Nucl. Phys. B}}\ }%
  \textbf{\bibinfo {volume} {362}},\ \bibinfo {pages} {665--709} (\bibinfo
  {year} {1991})%
  \bibAnnoteFile{NoStop}{Moore:1991ir}%
\bibitem{msy}%
  \BibitemOpen
  \bibfield{author}{%
  \bibinfo {author} {\bibfnamefont{Raghu}\ \bibnamefont{Mahajan}}, \bibinfo
  {author} {\bibfnamefont{Douglas}\ \bibnamefont{Stanford}},\ and\ \bibinfo
  {author} {\bibfnamefont{Cynthia}\ \bibnamefont{Yan}},\ }%
  \bibfield{title}{%
  \enquote{\bibinfo {title} {{Sphere and disk partition functions in Liouville
  and in matrix integrals}},}\ }%
   (\bibinfo {month} {7}\ \bibinfo {year} {2021}),\
  \Eprint{http://arxiv.org/abs/2107.01172}{arXiv:2107.01172 [hep-th]}%
  \bibAnnoteFile{NoStop}{msy}%
\bibitem{akk}%
  \BibitemOpen
  \bibfield{author}{%
  \bibinfo {author} {\bibfnamefont{Sergei~Yu.}\ \bibnamefont{Alexandrov}},
  \bibinfo {author} {\bibfnamefont{Vladimir~A.}\ \bibnamefont{Kazakov}},\ and\
  \bibinfo {author} {\bibfnamefont{David}\ \bibnamefont{Kutasov}},\ }%
  \bibfield{title}{%
  \enquote{\bibinfo {title} {{Nonperturbative effects in matrix models and
  D-branes}},}\ }%
  \bibfield{journal}{%
  \Doi{10.1088/1126-6708/2003/09/057}{\bibinfo {journal} {JHEP}}\ }%
  \textbf{\bibinfo {volume} {09}},\ \bibinfo {pages} {057} (\bibinfo {year}
  {2003}),\ \Eprint{http://arxiv.org/abs/hep-th/0306177}{arXiv:hep-th/0306177}%
  \bibAnnoteFile{NoStop}{akk}%
\bibitem{Harish-Chandra:1957dhy}%
  \BibitemOpen
  \bibfield{author}{%
  \bibinfo {author} {\bibnamefont{Harish-Chandra}},\ }%
  \bibfield{title}{%
  \enquote{\bibinfo {title} {{Differential Operators on a Semisimple Lie
  Algebra}},}\ }%
  \bibfield{journal}{%
  \Doi{10.2307/2372387}{\bibinfo {journal} {Am. J. Math.}}\ }%
  \textbf{\bibinfo {volume} {79}},\ \bibinfo {pages} {87} (\bibinfo {year}
  {1957})%
  \bibAnnoteFile{NoStop}{Harish-Chandra:1957dhy}%
\bibitem{Itzykson:1979fi}%
  \BibitemOpen
  \bibfield{author}{%
  \bibinfo {author} {\bibfnamefont{C.}~\bibnamefont{Itzykson}}\ and\ \bibinfo
  {author} {\bibfnamefont{J.~B.}\ \bibnamefont{Zuber}},\ }%
  \bibfield{title}{%
  \enquote{\bibinfo {title} {{The Planar Approximation. 2.}}.}\ }%
  \bibfield{journal}{%
  \Doi{10.1063/1.524438}{\bibinfo {journal} {J. Math. Phys.}}\ }%
  \textbf{\bibinfo {volume} {21}},\ \bibinfo {pages} {411} (\bibinfo {year}
  {1980})%
  \bibAnnoteFile{NoStop}{Itzykson:1979fi}%
\bibitem{Zinn-Justin:2002rai}%
  \BibitemOpen
  \bibfield{author}{%
  \bibinfo {author} {\bibfnamefont{Paul}\ \bibnamefont{Zinn-Justin}}\ and\
  \bibinfo {author} {\bibfnamefont{J.~B.}\ \bibnamefont{Zuber}},\ }%
  \bibfield{title}{%
  \enquote{\bibinfo {title} {{On some integrals over the U(N) unitary group and
  their large N limit}},}\ }%
  \bibfield{journal}{%
  \Doi{10.1088/0305-4470/36/12/318}{\bibinfo {journal} {J. Phys. A}}\ }%
  \textbf{\bibinfo {volume} {36}},\ \bibinfo {pages} {3173--3194} (\bibinfo
  {year} {2003}),\
  \Eprint{http://arxiv.org/abs/math-ph/0209019}{arXiv:math-ph/0209019}%
  \bibAnnoteFile{NoStop}{Zinn-Justin:2002rai}%
\bibitem{Mehta:1981xt}%
  \BibitemOpen
  \bibfield{author}{%
  \bibinfo {author} {\bibfnamefont{M.~L.}\ \bibnamefont{Mehta}},\ }%
  \bibfield{title}{%
  \enquote{\bibinfo {title} {{A Method of Integration Over Matrix
  Variables}},}\ }%
  \bibfield{journal}{%
  \Doi{10.1007/BF01208498}{\bibinfo {journal} {Commun. Math. Phys.}}\ }%
  \textbf{\bibinfo {volume} {79}},\ \bibinfo {pages} {327--340} (\bibinfo
  {year} {1981})%
  \bibAnnoteFile{NoStop}{Mehta:1981xt}%
\bibitem{Ambjorn:1992gw}%
  \BibitemOpen
  \bibfield{author}{%
  \bibinfo {author} {\bibfnamefont{Jan}\ \bibnamefont{Ambj\o{}rn}}, \bibinfo
  {author} {\bibfnamefont{L.}~\bibnamefont{Chekhov}}, \bibinfo {author}
  {\bibfnamefont{C.~F.}\ \bibnamefont{Kristjansen}},\ and\ \bibinfo {author}
  {\bibfnamefont{Yu.}\ \bibnamefont{Makeenko}},\ }%
  \bibfield{title}{%
  \enquote{\bibinfo {title} {{Matrix model calculations beyond the spherical
  limit}},}\ }%
  \bibfield{journal}{%
  \Doi{10.1016/0550-3213(93)90476-6}{\bibinfo {journal} {Nucl. Phys. B}}\ }%
  \textbf{\bibinfo {volume} {404}},\ \bibinfo {pages} {127--172} (\bibinfo
  {year} {1993}),\ \bibinfo {note} {[Erratum: Nucl.Phys.B 449, 681--681
  (1995)]},\
  \Eprint{http://arxiv.org/abs/hep-th/9302014}{arXiv:hep-th/9302014}%
  \bibAnnoteFile{NoStop}{Ambjorn:1992gw}%
\bibitem{Ambjorn:1990ji}%
  \BibitemOpen
  \bibfield{author}{%
  \bibinfo {author} {\bibfnamefont{Jan}\ \bibnamefont{Ambjorn}}, \bibinfo
  {author} {\bibfnamefont{J.}~\bibnamefont{Jurkiewicz}},\ and\ \bibinfo
  {author} {\bibfnamefont{Yu.~M.}\ \bibnamefont{Makeenko}},\ }%
  \bibfield{title}{%
  \enquote{\bibinfo {title} {{Multiloop correlators for two-dimensional quantum
  gravity}},}\ }%
  \bibfield{journal}{%
  \Doi{10.1016/0370-2693(90)90790-D}{\bibinfo {journal} {Phys. Lett. B}}\ }%
  \textbf{\bibinfo {volume} {251}},\ \bibinfo {pages} {517--524} (\bibinfo
  {year} {1990})%
  \bibAnnoteFile{NoStop}{Ambjorn:1990ji}%
\bibitem{Eynard:2015aea}%
  \BibitemOpen
  \bibfield{author}{%
  \bibinfo {author} {\bibfnamefont{Bertrand}\ \bibnamefont{Eynard}}, \bibinfo
  {author} {\bibfnamefont{Taro}\ \bibnamefont{Kimura}},\ and\ \bibinfo {author}
  {\bibfnamefont{Sylvain}\ \bibnamefont{Ribault}},\ }%
  \bibfield{title}{%
  \enquote{\bibinfo {title} {{Random matrices}},}\ }%
   (\bibinfo {month} {10}\ \bibinfo {year} {2015}),\
  \Eprint{http://arxiv.org/abs/1510.04430}{arXiv:1510.04430 [math-ph]}%
  \bibAnnoteFile{NoStop}{Eynard:2015aea}%
\bibitem{Eynard:2002kg}%
  \BibitemOpen
  \bibfield{author}{%
  \bibinfo {author} {\bibfnamefont{B.}~\bibnamefont{Eynard}},\ }%
  \bibfield{title}{%
  \enquote{\bibinfo {title} {{Large N expansion of the 2 matrix model}},}\ }%
  \bibfield{journal}{%
  \Doi{10.1088/1126-6708/2003/01/051}{\bibinfo {journal} {JHEP}}\ }%
  \textbf{\bibinfo {volume} {01}},\ \bibinfo {pages} {051} (\bibinfo {year}
  {2003}),\ \Eprint{http://arxiv.org/abs/hep-th/0210047}{arXiv:hep-th/0210047}%
  \bibAnnoteFile{NoStop}{Eynard:2002kg}%
\bibitem{Bertola:2003rp}%
  \BibitemOpen
  \bibfield{author}{%
  \bibinfo {author} {\bibfnamefont{M.}~\bibnamefont{Bertola}},\ }%
  \bibfield{title}{%
  \enquote{\bibinfo {title} {{Free energy of the two matrix model / dToda tau
  function}},}\ }%
  \bibfield{journal}{%
  \Doi{10.1016/j.nuclphysb.2003.07.029}{\bibinfo {journal} {Nucl. Phys. B}}\ }%
  \textbf{\bibinfo {volume} {669}},\ \bibinfo {pages} {435--461} (\bibinfo
  {year} {2003}),\
  \Eprint{http://arxiv.org/abs/hep-th/0306184}{arXiv:hep-th/0306184}%
  \bibAnnoteFile{NoStop}{Bertola:2003rp}%
\end{thebibliography}%
\end{document}